\documentclass[12pt,letterpaper]{article}
\usepackage{setspace}
\usepackage{graphicx}
\begin{document}
\normalsize

\newcommand{\eVq}{\ensuremath{\text{eV}^2}}
\newcommand{\Dmq}{\Delta m^2}
\newcommand{\Dlt}{\Delta\delta}
\newcommand{\Eps}{{\varepsilon}}
\newcommand{\Epp}{{\varepsilon'}}
\newcommand\newvskip{\newline\vspace{1pt}}
\newcommand\etal{{\it et al.}}
\newcommand\stf{$S_{34}(0)$~}
\newcommand\sos{$S_{17}(0)$~}
\newcommand\xsos{$S_{17}$~}
\newcommand\bepg{$^7Be(p,\gamma)^8B$}
\newcommand\xbe{$^7Be$}
\def\ra{\rightarrow} 
\newcommand{\beq}{\begin{equation}}
\newcommand{\eeq}{\end{equation}}
\def\gs{\mathrel{ \rlap{\raise
0.511ex \hbox{$>$}}{\lower 0.511ex \hbox{$\sim$}}}} \def\ls{\mathrel{
\rlap{\raise 0.511ex \hbox{$<$}}{\lower 0.511ex \hbox{$\sim$}}}}
\newcommand{\obb}{0\mbox{$\nu\beta\beta$}}
\newcommand{\onbb}{neutrinoless double beta decay}
\newcommand{\Slash}[1]{\mbox{$#1\hspace{-.6em}/$}}
\newcommand{\ba}{\begin{array}{c}}
\newcommand{\baz}{\begin{array}{cc}}
\newcommand{\bad}{\begin{array}{ccc}}
\newcommand{\bea}{\begin{equation} \begin{array}{c}}
\newcommand{\eea}{ \end{array} \end{equation}}
\newcommand{\ea}{\end{array}} \newcommand{\D}{\displaystyle}
\newcommand{\dms}{\mbox{$\Delta m^2_{\odot}$ }}
\newcommand{\dma}{\mbox{$\Delta m^2_{\rm A}$ }}
\newcommand{\imeff}{\mbox{$\langle \frac{\D 1}{\D m} \rangle$ }}
\newcommand{\mab}{\mbox{$\langle m_{\alpha \beta} \rangle $}}
\newcommand{\tm}{\mbox{$\tilde{m}$}} \def\ra{\rightarrow}
\newcommand{\ppp}{\mbox{$(+++)$}} \newcommand{\pmm}{\mbox{$(+--)$}}
\newcommand{\mpm}{\mbox{$(-+-)$}} \newcommand{\mmp}{\mbox{$(--+)$}}
%
%
\renewcommand{\floatpagefraction}{1} \renewcommand{\textfraction}{0}
\renewcommand{\topfraction}{1} \renewcommand{\bottomfraction}{0.5}
\textwidth 16.5cm \textheight 23.0cm \setlength{\marginparwidth}{0cm}
\setlength{\marginparsep}{0cm} \setlength{\topmargin}{-1cm}
\setlength{\evensidemargin}{-0.4cm} \setlength{\oddsidemargin}{-0.4cm}

\def\gtap{\mathrel{ \rlap{\raise 0.511ex \hbox{$>$}}{\lower 0.511ex
   \hbox{$\sim$}}}} 
\def\ltap{\mathrel{ \rlap{\raise 0.511ex
   \hbox{$<$}}{\lower 0.511ex \hbox{$\sim$}}}}
   \newcommand{\deltaatm}{\mbox{$\Delta m^2_{23}$}}
   \newcommand{\deltasol}{\mbox{$ \Delta m^2_{21}$}}
   \newcommand{\utre}{\mbox{$|U_{\mathrm{e} 3}|$}}
   \newcommand{\betabeta}{\mbox{$(\beta \beta)_{0 \nu} $}}
   \newcommand{\meffih}{\mbox{$ \left|< \! m \! >
         \right|^{\mathrm{IH}}$}} 
   \newcommand{\meffqd}{\mbox{$ \left|< \! m \! >
         \right|^{\mathrm{QD}}$}} 
   \newcommand{\meff}{\mbox{$\left| < \! m \! > \right|$}}
   \newcommand{\meffexp}{\mbox{$(\left| < \! m \! > 
         \right|_{\mbox{}_{\mathrm{exp}}})_{\mbox{}_{\mathrm{MIN}}}~$}}
   \newcommand{\hbeta}{$\mbox{}^3 {\rm H}$ $\beta$-decay }
   \newcommand{\eV}{\mbox{$ \ \mathrm{eV}$}}
   \newcommand{\deltatre}{\mbox{$ \Delta m^2_{32} \ $}}
   \newcommand{\deltadue}{\mbox{$ \Delta m^2_{21} \ $}}
   \newcommand{\ueuno}{\mbox{$ \ |U_{\mathrm{e} 1}|^2 \ $}}
   \newcommand{\uedue}{\mbox{$ \ |U_{\mathrm{e} 2}|^2 \ $}}
   \newcommand{\uetre}{\mbox{$ \ |U_{\mathrm{e} 3}|^2 \ $}}
   \newcommand{\me}{\mbox{$ m_{\bar{\nu}_{e}}$}}
\newcommand{\am}{\alpha}
\newcommand{\an}{\beta}
\newcommand{\dmamax}{\mbox{$(\Delta m^2_{\mathrm{atm}})_{ \!
\mbox{}_{\mathrm{MAX}}} \ $}} \newcommand{\dmamin}{\mbox{$(\Delta
m^2_{\mathrm{atm}})_{ \! \mbox{}_{\mathrm{MIN}}} \ $}}
\newcommand{\deltasolmax}{\mbox{$ (\Delta m^2_{\odot})_{ \!
\mbox{}_{\mathrm{MAX}}} \ $}} \newcommand{\deltasolmin}{\mbox{$(\Delta
m^2_{\odot})_{ \! \mbox{}_{\mathrm{MIN}}} \ $}}
\newcommand{\utremax}{\mbox{$|U_{\mathrm{e} 3}|^2_{ \!
\mbox{}_{\mathrm{MAX}}}$ }}
\newcommand{\utremin}{\mbox{$|U_{\mathrm{e} 3}|^2_{ \!
\mbox{}_{\mathrm{MIN}}} $ }} \newcommand{\utretilda}{\mbox{
$\widehat{|U_{\mathrm{e} 3}|^2}$}}
\newcommand{\uuno}{\mbox{$|U_{\mathrm{e} 1}|^2$}}
\newcommand{\uunomax}{\mbox{$|U_{\mathrm{e} 1}|^2_{ \!
\mbox{}_{\mathrm{MAX}}}$ }}
\newcommand{\uunomin}{\mbox{$|U_{\mathrm{e} 1}|^2_{ \!
\mbox{}_{\mathrm{MIN}}}$ }} \newcommand{\ts}{\mbox{$\tan^2
\theta_\odot$}}

\hyphenation{par-ti-cu-lar} \hyphenation{ex-pe-ri-men-tal}
\hyphenation{dif-fe-rent} \hyphenation{bet-we-en}
\hyphenation{mo-du-lus}


\title{Report of the Solar and Atmospheric Neutrino Experiments Working Group of the APS
Multidivisional Neutrino Study\\
}
\author{H. Back, J.N. Bahcall, J. Bernabeu, M. G. Boulay, T. Bowles, \\
F. Calaprice, A. Champagne, M. Gai, C. Galbiati, H. Gallagher, \\
C. Gonzalez-Garcia, R.L. Hahn, K.M. Heeger, A. Hime, C.K. Jung, J.R.Klein, \\
M. Koike, R. Lanou, J.G. Learned, K. T. Lesko, J.  Losecco, M. Maltoni, A. Mann, \\
D. McKinsey, S. Palomares-Ruiz, C.Pe\~{n}a-Garay, S.T. Petcov, A. Piepke, \\
M.Pitt, R. Raghavan, R.G.H. Robertson, K. Scholberg, \\
H. W. Sobel, T. Takeuchi, R. Vogelaar, L.  Wolfenstein \\ }

\maketitle

\pagebreak

\tableofcontents

\pagebreak

\section{Executive Priority Summary}

The highest priority of the Solar and Atmospheric Neutrino Experiment
Working Group is the development of a real-time, precision experiment
that measures the $pp$ solar neutrino flux.  A measurement of the $pp$
solar neutrino flux, in comparison with the existing precision measurements
of the high energy $^8$B neutrino flux, will demonstrate the transition
between vacuum and matter-dominated oscillations, thereby quantitatively
testing a fundamental prediction of the standard scenario of neutrino flavor
transformation. The initial solar neutrino beam is pure $\nu_e$, which
also permits sensitive tests for sterile neutrinos.  The $pp$ experiment
will also permit a significantly improved determination of $\theta_{12}$
and, together with other solar neutrino measurements, either a measurement
of $\theta_{13}$ or a constraint a factor of two lower than existing bounds.

	In combination with the essential pre-requisite experiments that
will measure the $^7$Be solar neutrino flux with a precision of 5\%, a
measurement of the $pp$ solar neutrino flux will constitute a sensitive test
for non-standard energy generation mechanisms within the Sun.  The Standard
Solar Model predicts that the $pp$ and $^7$Be neutrinos together constitute
more than 98\% of the solar neutrino flux. The comparison of the solar
luminosity measured via neutrinos to that measured via photons will test
for any unknown energy generation mechanisms within the nearest star.
A precise measurement of the $pp$ neutrino flux (predicted to be 92\% of
the total flux) will also test stringently the theory of stellar evolution
since the Standard Solar Model predicts the $pp$ flux with a theoretical
uncertainty of 1\%.

	We also find that an atmospheric neutrino experiment capable of
resolving the mass hierarchy is a high priority.  Atmospheric neutrino
experiments may be the only alternative to very long baseline accelerator
experiments as a way of resolving this fundamental question. Such an
experiment could be a very large scale water Cerenkov detector, or
a magnetized detector with flavor and antiflavor sensitivity.

	Additional priorities are nuclear physics measurements
which will reduce the uncertainties in the predictions of the Standard Solar
Model, and similar supporting measurements for atmospheric neutrinos (cosmic
ray fluxes, magnetic fields, etc.).  We note as well that the detectors
for both solar and atmospheric neutrino measurements can serve as
multipurpose detectors, with capabilities of discovering dark matter,
relic supernova neutrinos, proton decay, or as targets for long baseline 
accelerator neutrino experiments.

\pagebreak

	Figure~\ref{fig:timeline} shows a potential timeline for
these experiments.
\begin{figure}[h!]
\begin{center}
\includegraphics[width=1.0\textwidth]{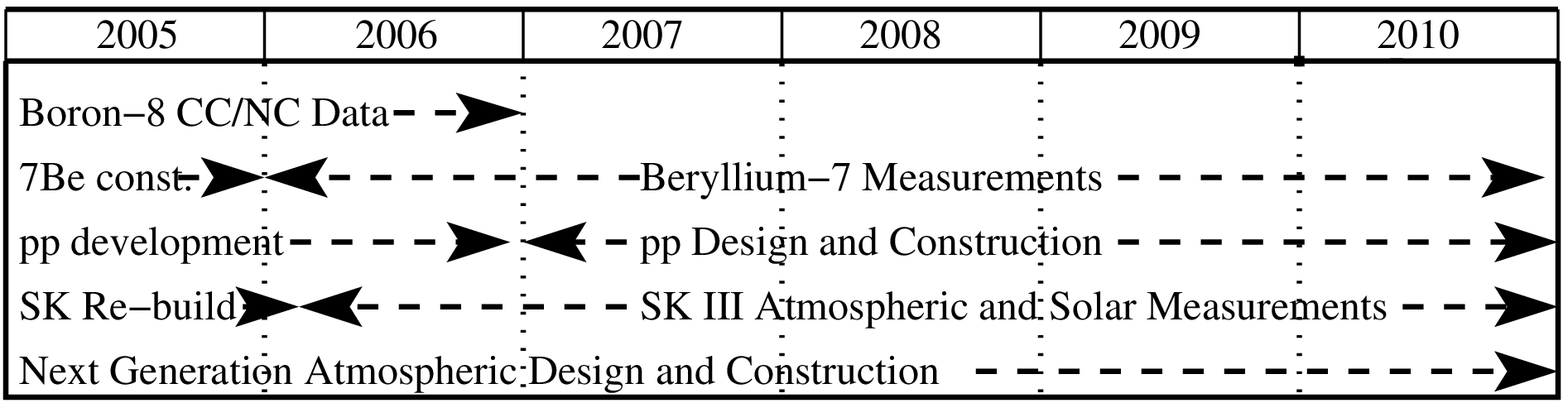}
\caption{Timeline for future solar and atmospheric neutrino experiments.
\label{fig:timeline}}
\end{center}
\end{figure}

\section{Introduction}
\label{sec:intro}

\subsection{Discovery Potential}

	Both the first evidence and the first discoveries of neutrino
flavor transformation have come from experiments which use neutrino beams
provided by Nature.  These discoveries were remarkable not only because
they were unexpected---they were discoveries in the purest sense---but
that they were made initially by experiments designed to do different
physics.  Ray Davis's solar neutrino experiment~\cite{davis} was created
to study solar astrophysics, not the particle physics of neutrinos.
The IMB~\cite{IMB1,IMB2} and Kamiokande~\cite{KIIatm} experiments were hoping
to observe proton decay, rather than study the (ostensibly relatively
uninteresting) atmospheric neutrino flux.  That these experiments and their
successors~\cite{KIIsol,SAGE,GALLEX,sksol,snocces,GNO} have had such a great
impact upon our view of neutrinos and the Standard Model underscores two of
the most important motivations for continuing current and creating future
solar and atmospheric neutrino experiments: they are naturally sensitive to
a broad range of physics (beyond even neutrino physics), and they therefore
have a great potential for the discovery of what is truly new and unexpected.

	The fact that solar and atmospheric neutrino experiments use
naturally created neutrino beams raises the third important motivation---the
beams themselves are intrinsically interesting.  Studying atmospheric
neutrinos can tell us about the primary cosmic ray flux, and at high energies
it may bring us information about astrophysical sources of neutrinos (see
Report of Astrophysics Working Group) or perhaps even something about
particle interactions in regimes still inaccessible to accelerators.  For
solar neutrinos, the interest of the beam is even greater: as the only
particles which can travel undisturbed from the solar core to us, neutrinos
tell us details about the inner workings of the Sun.  The recent striking
confirmation~\cite{snocces,snoccnc,snodn,snosalt} of the predictions of the
Standard Solar Model~\cite{BP04} (SSM) are virtually the tip of the iceberg:
we have not yet examined in an exclusive way more than 99\% of the solar
neutrino flux.  The discovery and understanding of neutrino flavor
transformation now allows us to return to the original solar neutrino
project---using neutrinos to understand the Sun.

	The fourth and perhaps strongest motivation for solar and atmospheric
neutrino experiments is that they have a vital role yet to play in exploring
the new physics of neutrinos.  The beams used in these experiments give
them unique sensitivity to some of the most interesting new phenomena.
The solar beam is energetically broadband, free of flavor backgrounds,
and passes through quantities of matter obviously unavaible to terrestrial
experiments.  The atmospheric beam is also broadband, but unlike the solar
beam it has the additional advantage of a baseline which varies from tens
of kilometers to many thousands.  

\subsection{Primary Physics Goals}

	In the work described here, we have chosen to focus on the following 
primary physics questions:
\begin{itemize}

\item {\it Is our model of neutrino mixing and oscillation complete, or are there other mechanisms at work?}

	To test the oscillation model, we must search for sub-dominant
effects such as non-standard interactions, make precision comparisons
to the measurements of other experiments in different regimes, and
verify the predictions of both the matter effect and vacuum oscillation.
The breadth of the energy spectrum, the extremely long baselines, and the
matter densities traversed by solar and atmospheric neutrinos make them
very different than terrestrial experiments, and hence measurements in all
three mixing sectors---including limits on $\theta_{13}$---can be compared
to terrestrial measurements and thus potentially uncover new physics.

\item {\it Is nuclear fusion the only source of the Sun's energy, and is it 
a steady state system?}

	Comparison of the total energy output of the Sun measured in
neutrinos must agree with the total measured in photons, if nuclear
fusion is the only energy generation mechanism at work.  In addition,
the comparison of neutrino to photon luminosities will tell us whether
the Sun is in an approximately steady state by telling us whether the
rate of energy generation in the core is equal to that radiated through
the solar surface---the heat and light we see today at the solar surface
was created in the interior $\sim$ 40,000 years ago, while the neutrinos
are just over eight minutes old.

\item {\it What is the correct hierarchical ordering of the neutrino masses?}

	Atmospheric neutrinos which pass through the Earth's core and mantle
will have their transformation altered due to the matter effect, dependent
upon the sign of the $\Delta m^2_{13}$ mass difference.  Future large scale
water Cerenkov experiments may be able to observe this difference in the ratio
of $\mu$-like to $e$-like neutrino interactions, while magnetized atmospheric
neutrino experiments may be able to see the effect simply by comparing the
number of detected $\nu_{\mu}$ to $\bar{\nu_{\mu}}$ events.

%

\end{itemize}

\section{The Standard Solar Model and Solar Neutrino Experiments}
\label{sec:ssmexp}

	The forty-year effort which began as a way to understand the Sun's
neutrino production ultimately taught us two remarkable things: that the
Sun's neutrinos are changing flavor between their creation in the solar
interior and their detection on Earth, and that the Standard Solar Model's 
predictions of the $^8$B flux of neutrinos was accurate to a degree well 
within its theoretical uncertainties.

	Figure~\ref{fig:ssmspec} summarizes the Standard Solar Model's
predictions for the neutrino fluxes and spectra.  In Figure~\ref{fig:ssmspec}
the neutrinos labeled $pp$, $pep$, $^7$Be, $^8$B, and {\em hep} belong to the
`$pp$-chain' which for a star like the Sun dominates over those from the
CNO cycle.  Of the neutrinos in the $pp$ chain, those from the initial
reaction $p + p \rightarrow d + e^+ + \nu_e$ make up over 92\% of the
entire solar flux.

\begin{figure}[h!]
\begin{center}
\includegraphics[angle=-90, width=0.7\textwidth]{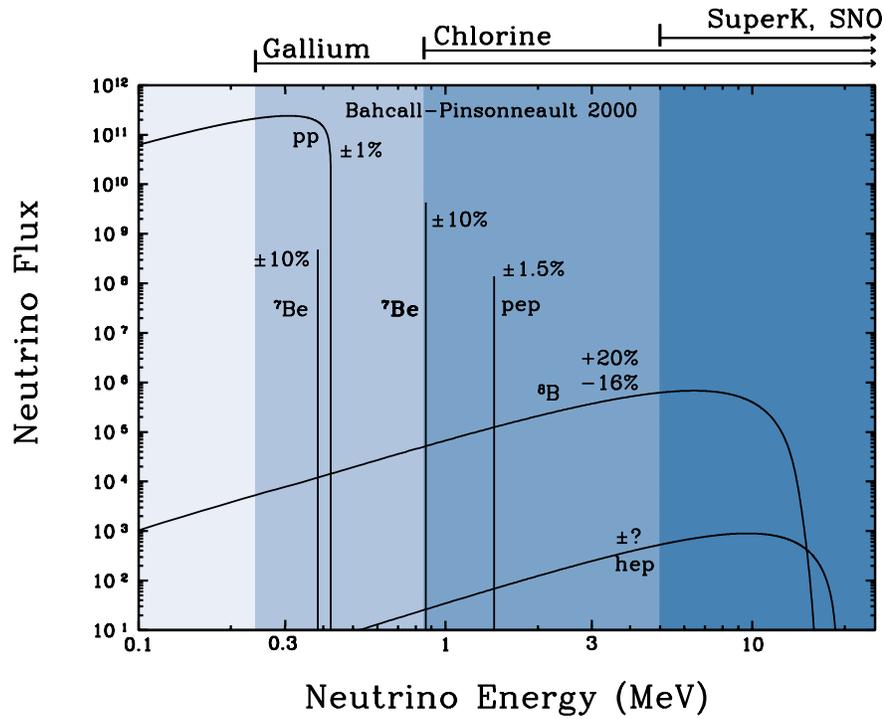}
\caption{Differential Standard Solar Model neutrino fluxes~\cite{jnbweb}.\label{fig:ssmspec}}
\end{center}
\end{figure}
\begin{figure}[h!]
\begin{center}
\includegraphics[angle=-90, width=0.7\textwidth]{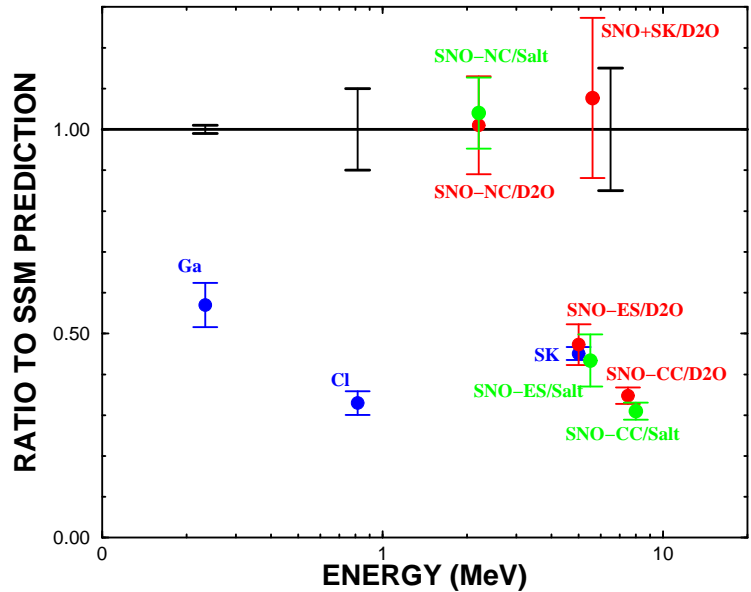}
\caption{Comparison of measurements to Standard Solar Model predictions.  \label{fig:ssmexp}}
\end{center}
\end{figure}

	Figure~\ref{fig:ssmexp} shows the past forty years of measurements of 
the solar neutrino fluxes.  In the figure, the measurements are plotted 
in terms of their respective neutrino energy thresholds.  The experiments
divide into two classes: radiochemical experiments like the original Davis
Chlorine detector, and real-time experiments like Super-Kamiokande and SNO.

The radiochemical experiments do not provide any direct spectral information
about the solar fluxes, but rather make inclusive measurements of all
neutrino sources above their particular reaction threshold.  For the Gallium
based experiments such as SAGE~\cite{SAGE} and Gallex~\cite{GALLEX}, this
sensitivity extends all the way down to the $pp$ neutrinos but includes
all neutrinos above the threshold of 0.233~MeV (even neutrinos of the CNO
cycle should they exist).  The Chlorine threshold is above that of the
$pp$ neutrinos, but is sensitive to the neutrinos from $^7$Be and $^8$B,
as well as potential CNO neutrinos.  For all radiochemical experiments,
the interpretation of the observed rates as measurements of neutrino
mixing assume that the Standard Solar Model calculated fluxes are correct
within their theoretical uncertainties.  In addition, the best values of
the mixing parameters are obtained when the `luminosity constraint' is
imposed, requiring the sum of all the energy radiated by the Sun through
neutrinos to agree with that radiated through photons.

To date, the real-time experiments have all been water Cerenkov experiments.
As such, their neutrino energy thresholds are relatively high, and they are
sensitive exclusively to the neutrinos from the solar $^8$B reaction (if the 
flux of neutrinos from the {\em hep} reaction were high enough they would also be 
included in the measurements).   This exclusivity has had a great advantage:
comparison of the number of neutrinos measured through the charged current (CC)
reaction in SNO's heavy water ($\nu_e + d \rightarrow e^- + p + p$) to that
measured via the elastic scattering (ES) of electrons in Super-Kamiokande's
light water  ($\nu + e^- \rightarrow \nu + e^-$) allowed the first
model-independent demonstration of the transformation of solar
neutrinos~\cite{sksol,snocces}.  SNO's subsequent measurement of the rate of 
neutral current (NC) events in D$_2$O ($\nu + d \rightarrow \nu + n+ p $) provided
the first direct measurement of the total active $^8$B flux~\cite{snoccnc}.  
In both cases---the combination of the SNO and Super-Kamiokande measurements
as well as the SNO NC measurement, the measurements of the $^8$B flux were in 
excellent agreement with the predictions of the Standard Solar Model for
that flux.  The SNO measurements therefore allow measurements of neutrino
mixing parameters without any reliance upon the predicted Standard Solar Model
$^8$B neutrino flux.

	The real-time experiments also allow searches for time-dependent
variations (such as a Day/Night asymmetry) and comparisons of the observed
recoil electron spectrum to expectations for the $^8$B neutrinos. As of yet,
no significant asymmetry or distortion of the spectrum has been observed.

	With the integral measurements of the radiochemical experiments, 
the differential real-time exclusive measurements of the water Cerenkov experiments, and
the fluxes from the Standard Solar Model for all but the $^8$B neutrinos, the
allowed region of mixing parameters is restricted to the large mixing angle
region (LMA).  Figure~\ref{fig:allowed} shows this allowed region, for all
solar neutrino data.
\begin{figure}[t]
\begin{center}
\includegraphics[width=0.8\textwidth]{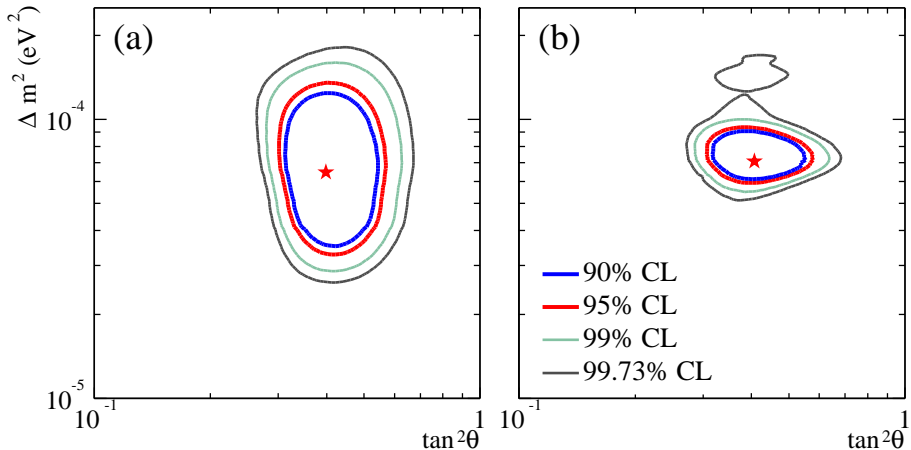}
\caption{Measurements of the mixing parameters for (a) all solar data and
(b) solar+KamLAND data~\cite{snosalt}, before the new results from
the Neutrino 2004 conference.\label{fig:allowed}}
\end{center}
\end{figure}

\subsection{Testing the Model of the Sun}
\label{sec:ssmtest}

	The idea that the Sun generates power through nuclear fusion in its
core was first suggested in 1919 by Sir Arthur Eddington,
who pointed out that the nuclear energy stored in the Sun is ``well-nigh
inexhaustible'', and therefore could explain the apparent age of the
Solar System.  Hans Bethe developed the first detailed model of stellar
fusion, in which the CNO cycle was thought to be the dominant process.

	Despite the obvious appeal of the theory, simple observations of
the solar luminosity are not enough to demonstrate that nuclear fusion is,
in fact, the solar energy source.  As John Bahcall wrote in 1964:``No
\underline{direct} evidence for the existence of nuclear reactions in
the interiors of stars has yet been obtained...Only neutrinos, with their
extremely small interaction cross sections, can enable us to \underline{
see into the interior of a star} and thus verify directly the hypothesis of
nuclear energy generation in stars.''~\cite{bahcall64}.  The idea only became
feasible when Bahcall and Davis showed that a reasonably-sized Chlorine
detector could observe the neutrinos at $^7$Be energies and higher.

	No one anticipated that it would take nearly four decades and eight
different experiments before Bahcall and Davis's original idea of testing
the model of the Sun in detail could become a reality.  With the measurements of
SNO and the KamLAND reactor experiment, the problem of neutrino mixing
can now be decoupled from the study of neutrinos as the signature of solar
energy generation.

	What we know: the Standard Solar Model correctly predicts the
flux of $^8$B neutrinos measured by SNO, and that globally fitting all
the solar neutrino data (and the data from the KamLAND reactor experiment)
for the neutrino fluxes and mixing parameters, provides good agreement
with the Standard Solar Model.	Table~\ref{tbl:gfit_roadmap}, reproduced
here from Ref.~\cite{roadmap}, shows the resultant mixing angle and the
ratio of the fitted fluxes to the predictions of the SSM, that is
\[ f = \frac{\phi_{\rm fit}}{\phi_{\rm SSM}}. \]
\begin{table}[t]
\begin{center}
\begin{tabular}{lcccc}
\hline \noalign{\smallskip} Analysis &$\tan^2\theta_{12}$&$f_B$& $f_{Be}$ & $f_{pp}$ \\
\hline \noalign{\smallskip}\noalign{\smallskip} A & $0.45^{+0.04}_{-0.06}$ ($^{+0.24}_{-0.16}$) &
$0.99^{+0.05}_{-0.03}$ ($^{+0.14}_{-0.13}$) & $0.13^{+0.41}_{-0.13}$ ($^{+1.27}_{-0.13}$)
& $1.38^{+0.18}_{-0.25}$ ($^{+0.47}_{-0.75}$)\\
\noalign{\smallskip}\noalign{\smallskip} A + lum & $0.40^{+0.06}_{-0.04}$ ($^{+0.23}_{-0.12}$) &
$1.02^{+0.03}_{-0.05}$ ($^{+0.12}_{-0.14}$) & $0.58^{+0.26}_{-0.25}$ ($^{+0.81}_{-0.58}$)
& $1.03^{+0.02}_{-0.02}$ ($^{+0.05}_{-0.06}$)\\
\noalign{\smallskip}\noalign{\smallskip} B + lum & $0.41^{+0.05}_{-0.05}$ ($^{+0.22}_{-0.13}$) &
$1.01^{+0.04}_{-0.04}$ ($^{+0.13}_{-0.13}$)& $0.93^{+0.25}_{-0.63}$ ($^{+0.80}_{-0.93}$)
& $1.02^{+0.02}_{-0.02}$ ($^{+0.06}_{-0.06}$)\\
\noalign{\smallskip} \hline
\end{tabular}
\caption{Allowed neutrino parameters with free $p-p$, $^7$Be, and $^8$B solar neutrino fluxes: with and without luminosity constraint, from Ref.~\cite{roadmap}. 
For all cases presented in this table, $\Delta m^2 =
7.3^{+0.4}_{-0.6}\times 10^{-5} {\rm eV^2} $ The results given here were obtained using all the currently
available data from the solar~~\cite{GALLEX,sksol,snoccnc,snodn,GNO,snosalt,chlorine,sage02} and
KamLAND~\cite{kamland} neutrino experiments.  All other (much less important) solar neutrino
fluxes are assumed to have the standard solar model (BP00) predicted values and uncertainties. \label{tbl:gfit_roadmap}}
\end{center}
\end{table}
	The top row of Table~\ref{tbl:gfit_roadmap} shows the fluxes without
the luminosity constraint imposed, and we can see that the best fit $pp$ and
$^7$Be neutrino fluxes have very large uncertainties and in fact do not
agree with their SSM values (nor do they even obey the luminosity constraint
itself).  The $^8$B flux, which is constrained by the SNO neutral current
measurements, stays close to its SSM value even if the luminosity constraint
it not imposed.  The second row of Table~\ref{tbl:gfit_roadmap} is the same fit
as in the first row, but now with the luminosity constraint imposed, and 
the third row is the same as the second, but with the CNO neutrino fluxes 
treated as free parameters.  The important points to take from 
Table~\ref{tbl:gfit_roadmap} are: 
\begin{itemize}
\item Without the luminosity constraint, the $pp$ and $^7$Be fluxes are
very poorly known, and the luminosity constraint is violated
\item Even with the luminosity constraint, the $^7$Be flux is still very
poorly determined, with uncertainties as large as 40\%
\item With the luminosity constraint, the $pp$ flux is known 
with a precision, $\pm 2$\%, comparable to but still larger than the 
theoretical uncertainty, $\pm 1$\% in the SSM prediction.
\end{itemize}

	If we are to test the Standard Solar Model further, we therefore
first need to measure the $^7$Be neutrinos.  The planned measurements (see
Section~\ref{sec:7be}) are likely to improve knowledge of this flux 
significantly over what is now known.  The measurements of the
$^7$Be flux can also give us direct information about some of the critical
parts of the Standard Solar Model, such as the ratio of rates of $^3$He+$^4$He
to $^3$He+$^3$He~\cite{jnbhist}.  In addition, a measurement of $^7$Be
will improve the determination of the $pp$ neutrino flux from the Gallium
experiments~\cite{sage02,GALLEX} to which both $pp$ and $^7$Be neutrinos
contribute to the rate.  With the luminosity constraint, the $pp$ flux
will be determined with a precision 2 to 4 times better than presently known
(2$\%$), and test the precise prediction of the $pp$ SSM
(1$\%$)~\cite{roadmap}.

	An exclusive, real-time measurement of the $pp$ flux can provide us
with an even more general test of the Standard Solar Model.  In combination
with the planned (and necessary) $^7$Be measurements and the existing
$^8$B measurements, a $pp$ measurement will allow a precise test of the
luminosity constraint itself, by comparing the inferred luminosity based
on the neutrino fluxes with the observed photon luminosity.  Such a test
will tell us whether there are any energy generation mechanisms beyond
nuclear fusion.  In addition, we will learn whether the Sun is in a steady
state, because the neutrino luminosity tells us how it burns today, while
the photons tell us how it burned over 40,000 years ago.  The current
comparison of these luminosities is not very precise~\cite{roadmap}:
\begin{equation}
 \frac{L_\odot{\rm (neutrino-inferred)}}{L_\odot}~=~1.4^{+0.2}_{-0.3} \left(^{+0.7}_{-0.6}\right).
\label{eq:lnuoverlphoton}
\end{equation}

	We see that, at 3$\sigma$, the inferred luminosity can be 2.1 times
larger than the measured photon luminosity, or 0.8 times smaller.  The fact
that the solar neutrino flux is overwhelmingly $pp$ neutrinos means that
the precision of this comparison approximately scales with the precision
of a measurement of the $pp$ flux---a measurement with a precision of 5\%
will reduce the uncertainties on this comparison to $\sim$4\%.

We also note that, although not explicitly listed in
Table~\ref{tbl:gfit_roadmap}, a measurement of the flux of neutrinos from
the $pep$ reaction can provide much of the same information as the $pp$
measurements, if we are willing to make the Standard Solar Model assumption
that the rates of the two reactions are strongly coupled.

\subsection{Testing the Neutrino Oscillation Model}
\label{sec:soltests}

	  The idea that the Standard Model `accommodates' the new found
neutrino properties must recognize that the oscillation model of neutrino
flavor transformation is just that---a model---and until we test that model
with the kind of precision with which we have explored the rest of particle
physics, we do not know whether it is in fact a satisfactory description
of neutrinos.  Even if we accept that the combination of the atmospheric
and the solar results taken together are compelling evidence that flavor
transformation in the neutrino sector is explained by the additional seven
new Standard Model parameters, we as yet have no experimental evidence that
the mixing involves three flavors in the way it does in the quark sector.
We even have evidence to the contrary---the results of the LSND experiment,
in combination with the results in the solar and atmospheric sector, point
to either the existence of a fourth family or perhaps even stranger physics,
such as a violation of CPT symmetry.

	To test the model, therefore, we need to look directly for evidence
of sub-dominant effects (Section~\ref{sec:npsol}), verify some of the basic
predictions of the model like the matter effect (Section~\ref{sec:msw}),
and compare the predictions of the model across different physical
regimes (Sections~\ref{sec:expcomp12} and~\ref{sec:expcomp13}).  The luminosity
test described in the previous section (Section~\ref{sec:ssmtest}) is its own
global test of neutrino properties.  For example, were the neutrino luminosity to 
fall substantially short of the total luminosity, it could be evidence of energy 
loss to sterile neutrino species.

\subsubsection{Other Transformation Hypotheses}
\label{sec:npsol}
Neutrino masses and mixing are not the only mechanism for neutrino flavor
oscillations. 
They can also be generated
by a variety of forms of nonstandard neutrino interactions or
properties. In general these alternative mechanisms share a common
feature: they require the existence of an interaction (other than the
neutrino mass terms) that can mix neutrino flavours. 
Among others this effect can arise due to:\\

\begin{tabular} {ll}
Violation of Equivalence Principle (VEP)~\cite{VEP}:& \\
(non universal coupling of neutrinos {$\gamma_1\neq\gamma_2$} & 
\\to gravitational potential {$\phi$}) & 
${\lambda}= \frac{\pi} {E |\phi| \delta\gamma }$ \\[+0.2cm]
Violation of Lorentz Invariance (VLI)~\cite{VLI}: &\\
(non universal asymptotic velocity of neutrinos $v_1\neq v_2$) 
& $\lambda=\frac{2\pi}{E \delta v}$ 
\\[+0.2cm]
Non universal couplings of neutrinos q
$k_1\neq k_2$ & \\
to gravitational 
torsion strength $Q$~\cite{torsion} 
&  $\lambda=\frac{2\pi}{Q \delta k }$\\ [+0.2cm]
Violation of Lorentz Invariance (VLI) & \\
due to CPT violating terms~\cite{VLICPT} 
$\bar{\nu}_L^\alpha {b_\mu^{\alpha\beta}} \gamma_\mu \nu_L^\beta$ 
& ${\lambda}=\pm 
\frac{2\pi}{{\delta b}}$ \\[+0.2cm]
Non-standard $\nu$ interactions in matter~\cite{NSI}: & \\
{$G_F { \varepsilon_{\alpha \beta}} 
(\overline{\nu_\alpha} \gamma^\mu \nu_\beta)(\overline f \gamma_\mu f)$}
& ${\lambda}= \frac
{2\pi} {2 \sqrt{2} G_f \; N_f\; 
\sqrt{\varepsilon_{\alpha\beta}^2+
(\varepsilon_{\alpha\alpha}-\varepsilon_{\beta\beta})^2/4}}$ 
\end{tabular}
\vskip 0.2cm
where $\lambda$ is the oscillation length.

From the point of view of neutrino
oscillation phenomenology, the most relevant feature of these
scenarios is that, in general, they imply a departure from the
$E^{1}$ ($\lambda =\frac{4\pi E}{\Delta m^2}$)
energy dependence of the oscillation wavelength.

Some of these scenarios have been invoked in the literature as 
explanations for the solar neutrino data alternative to mass oscillations.
Prior to the arrival of KamLAND, some of them could
still provide a good fit~\cite{solnp} to the data.

The observation of oscillations in KamLAND with parameters which are
consistent with solar LMA oscillations clearly rules out these mechanisms
as dominant source of the solar neutrino flavor transitions. 
However they may still exist at the sub-dominant level. This raises 
the question of to what point the possible presence of
these forms of new physics (NP), even if sub-dominant, can be constrained 
by the analysis of solar and atmospheric data. Or, conversely, 
to what level our present determination of the 
neutrino masses and mixing is robust
under the presence of phenomenologically allowed NP effects.

At present, there is no general analysis in the literature 
which answers these questions quantitatively. However one may argue 
that (unlike for atmospheric neutrinos), existing data on 
solar neutrinos by itself is 
unlikely to provide strong constraints on these forms of NP.
Therefore, as long as the KamLAND data is not affected by these NP effects,
there should be still more room for these effects in the analysis 
of the solar+KamLAND than there is in the corresponding analysis of
atmospheric data. The reason for this is the scarce information 
from solar neutrino data on the energy dependence of the $\nu_e$
survival probability $P_{ee}$, as
illustrated in Fig.\ref{solpee} where we show the
results of a fit to the observed solar rates in terms of the 
averaged $P_{ee}$ for three energy regions of the solar neutrino 
spectrum  (from Ref.\cite{barger}). 
\begin{figure}
\begin{center}
\includegraphics[width=4.in]{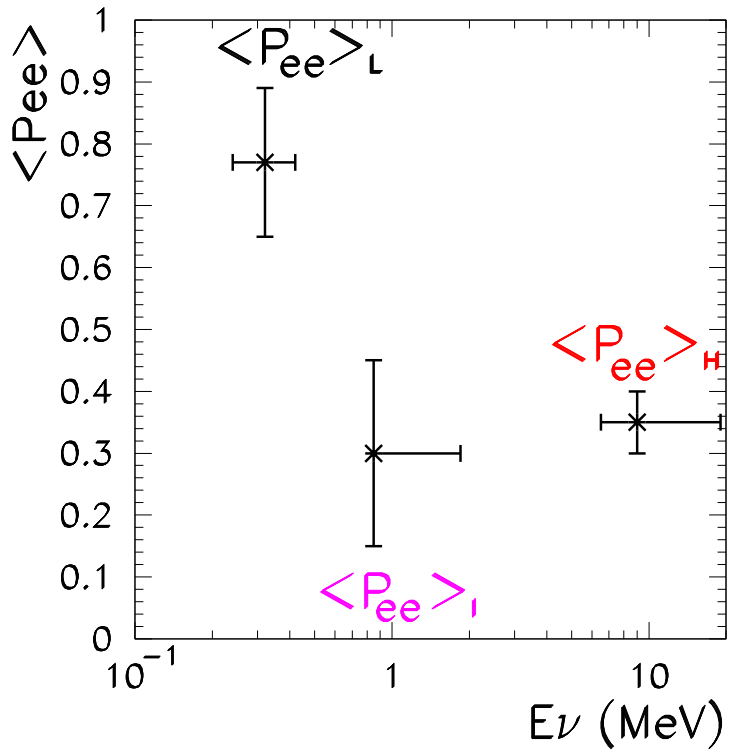} 
\caption{Reconstructed values of the 
survival probability of solar neutrinos in different energy ranges 
from a fit to the observed
rates. \label{solpee}}
\end{center}
\end{figure}

One illustrative example of this conclusion can be found in Ref.~\cite{carlos,orlando}. The authors of these works 
find that the inclusion of somewhat large but still allowed non-standard
neutrino interactions affecting the propagation of neutrinos 
in the Sun and Earth matter  can shift the allowed region of oscillation 
parameters in the solar+KamLAND analysis 
to lower $\Delta m^2$ values without spoiling the quality of 
the fit. 

	Recently, there has been a suggestion that the mass varying neutrino
(MaVaNs) hypothesis, put forward as an explanation of the 
of the origin of the Dark Energy and the coincidence of its magnitude
with the neutrino mass splittings, may produce matter effects which
will alter the solar and atmospheric neutrino oscillations~\cite{mavans}.
This hypothesis can be made to fit simultaneously the solar, atmospheric, and 
LSND results.

\subsubsection{MSW Effect}
\label{sec:msw}

	One of the predictions of the neutrino oscillation model is that
matter can strongly affect the neutrino survival probability (the `MSW
effect'~\cite{msw}).  The effect arises because matter is made out
of first generation material---the $\nu_e$'s interact with electrons
via both charged current and neutral current channels, while at solar
neutrino energies the other active flavor eigenstates have only neutral
current interactions.	The resultant difference in the forward scattering
amplitudes makes the matter of the Sun birefringent to neutrinos, and the
oscillation already caused by the neutrino mass differences can be enhanced
by this additional dispersion.   Beyond being a confirmation of our new
model of neutrinos, the MSW effect is a beautiful phenomenon in its own
right: as the neutrinos propagate from solar center to surface, the Sun's
changing density alters the effective mixing angles in an energy-dependent
way, leaving its quantum mechanical imprint for us to observe on Earth.

The effective Hamiltonian for two-neutrino propagation in matter can be written
conveniently in the
familiar
form~\cite{msw,bethe,Mikheev:ik,messiah,neutrinoastrophysics,conchayossi}

\begin{equation}
H ~=~ \left ( \begin{array}{cc} \frac{\Delta m^2 cos 2 \theta_{12}}{4 E}-
\frac{\sqrt{2}G_{\rm F} n_{\rm
e}}{2}&
\frac{\Delta m^2 sin2 \theta_{12}}{2 E}\\
 \frac{\Delta m^2 sin2 \theta_{12}}{2 E} &
  -\frac{\Delta m^2 cos 2 \theta_{12}}{4 E}+ \frac{\sqrt{2}G_{\rm F} n_{\rm
  e}}{2}\end{array}\right) \, .
   \label{eq:hamiltonian}
\end{equation}
Here $\Delta m^2$ and $\theta_{12}$ are, respectively, the difference in the
squares of the masses of the
two neutrinos and the vacuum mixing angle, $E$ is the energy of the neutrino,
$G_{\rm F}$ is the Fermi
coupling constant, and $n_{\rm e}$ is the electron number density at the
position at which the
propagating neutrino was produced.

The  relative importance of the MSW matter term and the kinematic vacuum
oscillation term in the
Hamiltonian be parameterized by the quantity, $\beta$, which represents the
ratio of matter to vacuum
effects~\cite{roadmap}. From equation~\ref{eq:hamiltonian} we see that the appropriate ratio is
\begin{equation}
\beta= \frac{2 \sqrt2 G_F n_e E_\nu}{\Delta m^2}\, . \label{eq:defbeta}
\end{equation}
The quantity $\beta$ is the ratio between the oscillation length in matter and
the oscillation length in vacuum. In convenient units,
$\beta$ can be written as
 \begin{equation}
 \beta= 0.22 \, \left[\frac{E_\nu}{1~{\rm MeV}}\right]\, \left[
 \frac{\mu_e\rho}{100~{\rm
 g~cm}^{-3}}\right] \, \left[ \frac{7 \times 10^{-5} eV^2}{\Delta m^2}\right]\,
 ,
 \label{eq:betaconvenient}
 \end{equation}
where $\mu_e$ is the electron mean molecular weight ($\mu_e \approx 0.5(1 +
 X)$, where X is the mass
fraction of hydrogen) and $\rho$ is the total density, both evaluated at the
location where the neutrino
is produced. For the electron density at the center of the standard solar
model, $\beta = 0.22$ for $ E =
1$MeV and $\Delta m^2 =  7\times 10^{-5} {\rm eV^2}$.

There are three explicit signatures of the MSW effect which can be
observed with solar neutrinos.  The first is the `Day/Night' effect in
which $\nu_e$'s which have been transformed by the matter of the Sun into
$\nu_{\mu}$'s and $\nu_{\tau}$'s are changed back to $\nu_e$'s as they pass
through the Earth---the coherent regeneration of $K^0_S$'s is a fair analogy.
As the regeneration is only appreciable for large path lengths, the number
of $\nu_e$'s observed by a detector at night will be larger than
during the day.  

The second signature of the MSW effect is a distortion of the energy
spectrum, in the region of the transition from matter-dominated to
vacuum-dominated oscillations. The energy dependence of the matter mixing
angles and eigenstates leads to energy-dependent survival probabilities which
are different from those for simple vacuum mixing. Figure~\ref{fig:psurvmsw}
shows the turnup in the survival probabilities for some of the mixing
parameters in the LMA region.
\begin{figure}[ht]
\begin{center}
\includegraphics[width=0.7\textwidth]{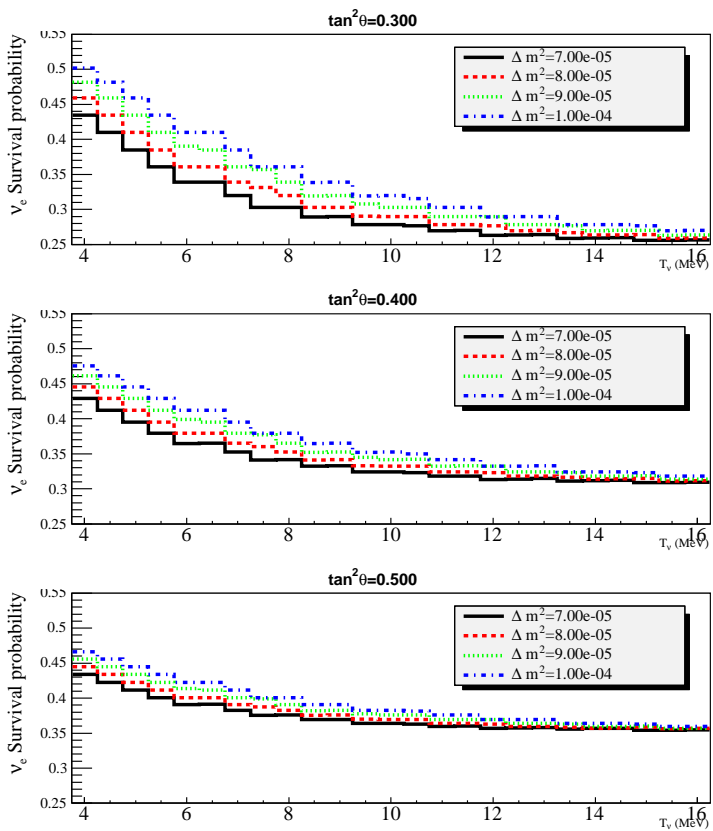}
\caption{$^8$B solar neutrino survival probabilities in the LMA region.
\label{fig:psurvmsw}}
\end{center}
\end{figure}

The third signature is the observation of vacuum-dominated mixing at
low energies~\cite{roadmap}. When the parameter $\beta$ given
in Eqn.~\ref{eq:betaconvenient} is greater than 1, the neutrino
flavor transformation is dominated by matter effects, which occurs
for the highest energy $^8$B neutrinos.  Figure~\ref{fig:vacmswtrans} shows 
the change in survival probability as the neutrino energies are lowered
from the $^8$B energies down to $pp$ energies. The clear transition from
matter-dominated to vacuum-dominated oscillations can be seen, and this
transition region is the same as that shown in Figure~\ref{fig:psurvmsw}.
What Figure~\ref{fig:vacmswtrans} shows is that a demonstration of the
matter effect can be made by comparing the measured survival probability
at high energies to that at low energies.
\begin{figure}[ht]
\begin{center}
\includegraphics[width=0.5\textwidth]{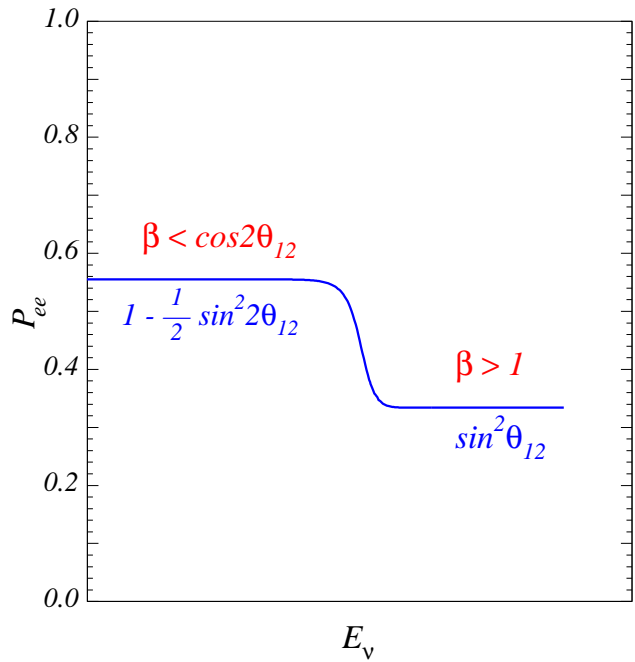}
\caption{Transition between vacuum and matter-dominated flavor transformation,
as a function of energy, from Ref.~\cite{roadmap}.  \label{fig:vacmswtrans}}
\end{center}
\end{figure}

	Based upon the results of the solar neutrino experiments and the
KamLAND experiment, we know that the mixing parameters are in a region
where the MSW effect plays an important role.  As of yet, we have not
directly seen any of its specific signatures.  We conclude that Nature
has been unkind---that the parameters are `unlucky'.  Or perhaps we have
not looked hard enough.

	Below we discuss the prospects for identifying each of these
signatures.
\begin{itemize}

\item Day/Night Asymmetry

	Both the Super-Kamiokande~\cite{superkdn} and SNO~\cite{snodn}
experiments have looked for a Day/Night asymmetry in the flux of $^8$B
solar neutrinos.  A measurement of a Day/Night asymmetry is perhaps the
cleanest of the signatures of the matter effect, because the vast majority
of experimental uncertainties cancel in the asymmetry ratio.  The asymmetry
is a function both of the energy and the zenith angle of the incident
neutrinos, and so often the measurements are published as `Day/Night
spectra', occasionally binned or fit in distributions of zenith angle.


	Currently, only SNO and Super-Kamiokande are operating in a regime
where a Day/Night asymmetry might be observable.  In both cases, however,
the measurements are statistically limited.  Figure~\ref{fig:smirnovdn},
from Ref.~\cite{smirnovdn} shows contours of Day/Night asymmetry expected
for SNO, overlaid with the LMA region of mixing parameters, and we can 
see that the asymmetry is small, even for the lowest allowed $\Delta m^2$
values.
\begin{figure}[ht]
\begin{center}
\includegraphics[width=0.5\textwidth]{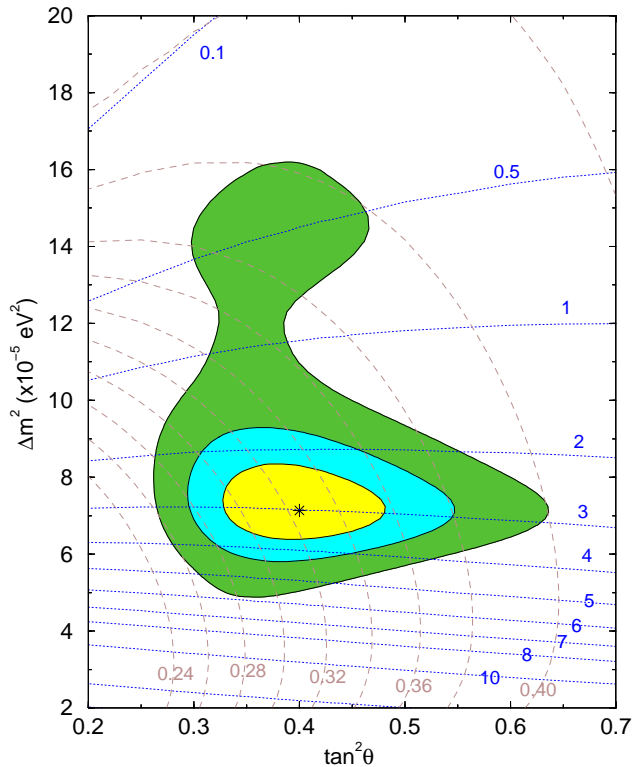}
\caption{Contours of expected Day/Night asymmetry, shown as the horizontal
dotted lines labeled in \%, overlain on the LMA
region~\cite{smirnovdn}.  \label{fig:smirnovdn}}
\end{center}
\end{figure}

 To observe a Day/Night asymmetry with high significance will require
a much larger real-time $^8$B experiment.  Some of the proposals for new
megaton-scale water Cerenkov detectors~\cite{uno,3m} have included
a low background region in the detector whose goal will be to observe the
$^8$B neutrinos.  With a fiducial volume at least seven times that of
Super-Kamiokande, a photocathode coverage of at least 40\%, and an
energy threshold of 6~MeV, a large water Cerenkov detector could see
the expected LMA Day/Night asymmetry of 2\% with a significance
of $\sim 4\sigma$ in roughly 10 years of running~\cite{dnuno}.

\item Spectral Distortion

	To observe the rise in survival probability shown in
Figure~\ref{fig:psurvmsw}, real-time solar neutrino experiments capable of
observing the $^8$B flux are needed.  Both SNO and Super-Kamiokande have
looked for signs of a distortion in the spectrum of observed recoil electrons,
and they do not see any significant effect.  

	 To see the spectral distortion, SNO or Super-K will need to 
lower their energy thresholds---when convolved with the differential cross
sections and the detector energy resolutions, the change in survival
probability does not become noticeable until an electron recoil energy 
below 5~MeV or so.  Investigations into the feasibility of background
reduction in these experiments to see the distortion are underway.

	{\it We note that there are currently no other experiments planned
whose primary goal is to directly probe this region.}  

\item Low E/High E Survival Probability Comparisons

	The Super-Kamiokande and SNO measurements have given us the
survival probabilities for the high energy end of the solar neutrino
spectrum, and so they have mapped out the matter-dominated survival
region shown in the upper end Figure~\ref{fig:vacmswtrans}.  The
Chlorine and Gallium experiments, in combination with the predictions
of the Standard Solar Model, have told us that the survival probability
at low energies looks like the expectation from vacuum-dominated oscillations.
Unfortunately, the integral nature of the low energy experiments means that
they must rely on the assumption that the Standard Solar Model predictions
for the various neutrino sources is correct.  In particular, the inferences
drawn from the radiochemical measurements  assume the neutrino cross
sections can be multiplied by a constant survival probability independent
of energy, and neglect correlations among the systematic uncertainties.
In addition, the CNO neutrinos are typically neglected when
calculating the survival probabilities from these experiments.
As Table~\ref{tbl:gfit_roadmap} shows, the for a fit which includes all
uncertainties, the resulting overall uncertainties are currently too large
for a quantitative test of the MSW scenario. We therefore need exclusive
measurements of the $^7$Be and $pp$ neutrino fluxes to unambiguously
demonstrate the vacuum/matter transition with solar neutrinos.

\end{itemize}

\subsubsection{Precision Comparisons in the (1,2) Sector}
\label{sec:expcomp12}

	The strongest test of our model of neutrino flavor transformation
is to compare the predictions over as wide a physical range as possible.
The model predicts that seven fundamental parameters are all that is needed
to explain every possible observation of neutrino flavor transformation
regardless of lepton number, flavor, energy, baseline, or intervening matter.
As it happens, even fewer parameters are needed to explain the observations
which have been made so far, because the difference in the neutrino masses
and the sizes of the mixing angles are such that most experiments can be 
treated as involving just two flavors.  

	The first precision test across experimental regimes is the
comparison of the measurements of the KamLAND experiment to the solar
neutrino experiments~\cite{kamland}. KamLAND sees a transformation signal
with a range of parameters which include the solar LMA region, yet it
differs in nearly every relevant way from the solar experiments: it looks
at reactor antineutrinos rather than neutrinos; it has a medium baseline
(150~km) rather than the $150 \times 10^6$~km solar baseline; it looks
for disappearance rather than SNO's inclusive appearance; it is sensitive
only to vacuum oscillations rather than matter-enhanced oscillations.
Figure~\ref{fig:kamlma}, from Ref.~\cite{kamland2}, shows the allowed regions
of the mixing parameters determined by KamLAND overlain on the LMA region
determined by the solar experiments.  The fact that there is overlap between
the two regions, and that the best fit point agrees within the measurement
uncertainties, is remarkable confirmation of the oscillation model.
\begin{figure}[ht]
\begin{center}
\includegraphics[width=0.8\textwidth]{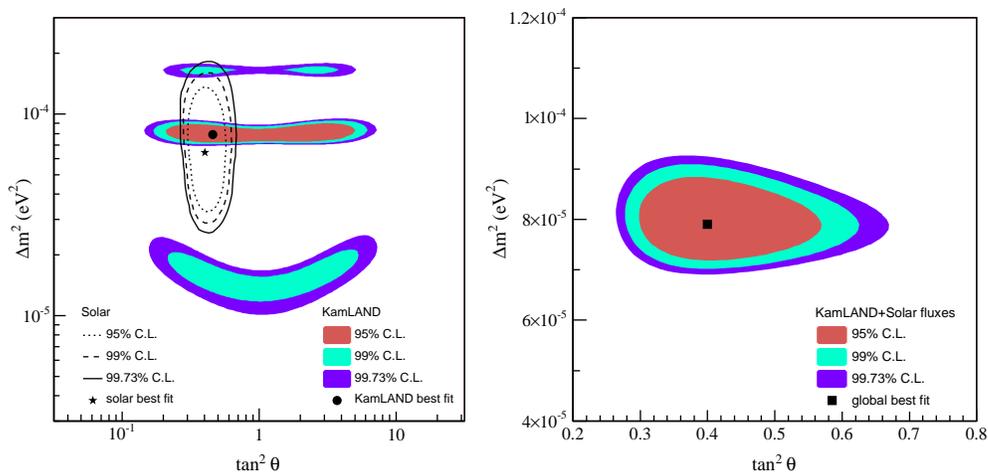}
\caption{Allowed regions of mixing parameters determined by KamLAND, compared
to solar measurements, for the most recent~\cite{kamland2} results.
\label{fig:kamlma}}
\end{center}
\end{figure}

	To go further, and explore some of the possibilities for new physics
described in Section~\ref{sec:npsol},  we need to improve the precision on
the measurements of the mixing parameters in the two regimes.  The most
recent KamLAND results~\cite{kamland2} have improved the statistical
precision of the initial measurements by roughly a factor of four,
eliminating some of the regions in $\Delta m^2_{12}$ which were outside
the region measured by the solar experiments.  The possibility of a more
precise reactor-based (1,2) sector experiment is also being discussed,
perhaps in conjunction with a reactor experiment to measure the value of
$\sin^2 2\theta_{13}$ (see Report of Reactor Working Group).

	SNO will soon publish updated results from
the Phase II (salt) data, which will bring some improvements on the
precision from the solar side.	The next phase of SNO will reduce the
uncertainties on the mixing angle further.  While a $^7$Be measurement
is not expected to improve the measurements of the mixing parameters,
an exclusive measurement of the $pp$ flux (or a measure of the $pep$
flux) can have a substantial effect on the mixing parameters, depending
on the precision of the measurement.  Figure~\ref{fig:solppmix} shows the
improvements on $\tan^2 \theta_{12}$ that could come from a $pp$ measurement,
allowing the $pp$, $^7$Be, $^8$B, and CNO neutrinos as free parameters,
subject only to the luminosity constraint.
\begin{figure}[ht]
\begin{center}
\includegraphics[width=0.5\textwidth]{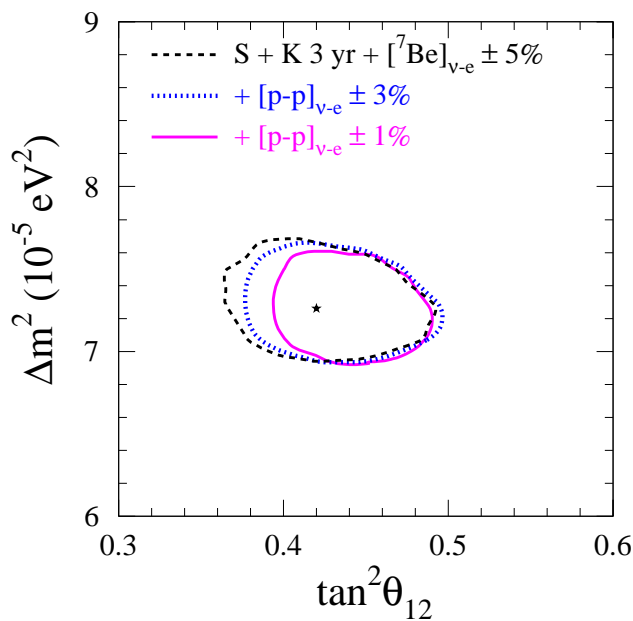}
\caption{Improvements in mixing angle determination for hypothetical
$pp$ experiments, including future KamLAND and $^7$Be measurements. From Ref~\cite{roadmap}.\label{fig:solppmix}}
\end{center}
\end{figure}

\subsubsection{Precision Comparisons of $\theta_{13}$}
\label{sec:expcomp13}

	Like the (1,2) sector, measurements of the (1,3) parameters with
solar neutrino experiments provide tests of the oscillation model in a
very different regime than either reactor or accelerator experiments.
In particular, measurements of $\theta_{13}$ with solar experiments are
essentially independent of the value of $\Delta m^2_{23}$, unlike either
the accelerator or reactor experiments.  At solar neutrino energies,
and the range of allowed values of $\Delta m^2_{13}$, the matter effect
(unfortunately) does not play a significant role in the (1,3) transformation.
However, we still expect to see (1,3) effects due to vacuum mixing.

A global analysis of all available data by Maltoni {\em et al.}
\cite{maltoni} (but without the most recent value for $\Delta m_{23}^2$
or most recent KamLAND measurements~\cite{kamland2}) gives  $\theta_{13} =
4.4_{-4.4}^{+6.3} $ degrees (2 $\sigma$).  The current situation is well
summarized in Figure 8 of \cite{maltoni}, which we reproduce here (Fig.
\ref{maltonifig8}) superimposed with the most recent range for $\Delta
m_{23}^2$.  One can see that near the low end of the mass range the tightest
limits on $\theta_{13}$ are already coming from solar neutrinos and KamLAND.
The relationship between these experiments and $\theta_{13}$ began to be
explored even before results were available from KamLAND \cite{concha}.
\begin{figure}[ht!] \vspace{.2in}
\begin{center}
\includegraphics[width=3in]{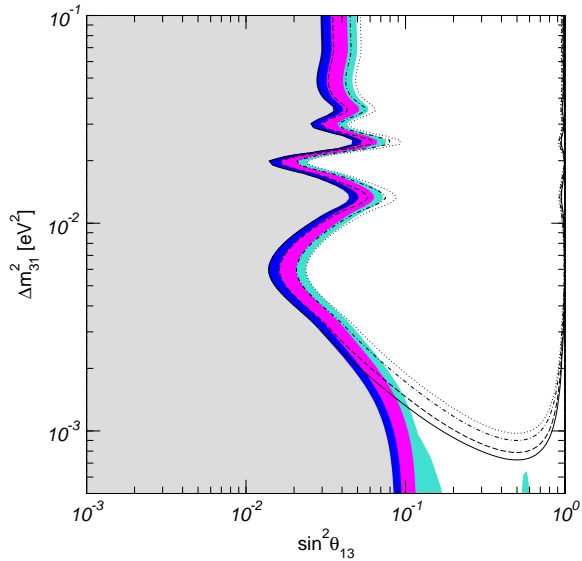}
\end{center}
\caption[amp]{Limits on $\theta_{13}$ from Chooz (lines, 90\%, 95\%, 99\%, and
3$\sigma$), and from Chooz+solar+KamLAND (colored regions) \cite{maltoni}. }
\vspace{.2in}
\label{maltonifig8}
\end{figure}


	Ref.~\cite{hamish} has performed a fit to existing solar neutrino
and KamLAND data, to investigate the effects of new solar measurements on
the limits for $\theta_{13}$, and what follows is described in more detail
there.	The fit includes 5 unknowns, the 3 (total active) solar fluxes
$\phi_1$, $\phi_7$, and $\phi_8$, and two mixing angles, $\theta_{12}$ and
$\theta_{13}$.	The mass-squared difference $\Delta m_{12}^2 $ is fixed by
the ``notch'' in the KamLAND reactor oscillation experiment, and $\Delta
m_{23}^2 $ by the atmospheric neutrino data.  The fit parameters that
are approximately normally distributed are $\phi_1$, $\phi_7$, $\phi_8$,
$\sin^2\theta_{12}$, and $\cos^4\theta_{13}$.

Solar plus KamLAND data already provide some constraint on
$\cos^4\theta_{13}$, with the corresponding angle $\theta_{13} =
7.5_{-7.5}^{+4.8} $ degrees.  The expected statistical improvements from the
KamLAND experiment reduce the overall uncertainties somewhat---in particular
$\theta_{13}$ is non-zero at 1 $\sigma$.  The reason the improvement is
not better is the growth of the correlation coefficient between the mixing
parameters, which is as large as -0.906.   Further improvements cannot be
made without breaking that correlation.

	The way to break the correlation is find a way of measuring the
(1,2) parameters independently from the (1,3) parameters.  Luckily, the
MSW effect, which acts only in the (1,2) sector for solar neutrinos, can
provide this independent measure.  The transition between the vacuum-
and matter-dominated regimes shown in Figure~\ref{fig:vacmswtrans}
shows that better measurements in the low energy regime can provide a
lever arm to distinguish the (1,2) from the (1,3) effects. Unfortunately,
reducing the uncertainties on the Gallium experiments (by, for example,
understanding the cross sections better) does not help very much. The
difficulty is that the $^7$Be flux is not well determined and thus floats
against the survival probability $P_{ee}$. The strong correlation between
the low-energy fluxes $\phi_1$ and $\phi_7$ could in the future be broken
by a $^7$Be experiment, either CC or ES, or by a robust prediction of
$^7$Be within the demonstrably reliable Standard Solar Model \cite{BP04}.
For the latter, a new high-precision determination of the $^3$He($\alpha,
\gamma$)$^7$Be cross section is needed.   For concreteness at this point,
we take a CC experiment with a precision of 5\%. Table \ref{LENS} shows
that the low-energy fluxes are individually determined twice as precisely
and there is some improvement in the separation of the mixing angles. Both
an improved Ga rate and a $^7$Be determination are needed to obtain this
improvement; either by itself is ineffective.

\begin{table}
\caption{Fitted fluxes and mixing parameters under the assumption of a putative CC $^7$Be experiment that measures $\phi_7$ to be 0.511 to an accuracy of 5\%, as well as the KamLAND statistical improvements and the improved SNO CC/NC ratio.}
\medskip
\begin{center}
\begin{tabular}{lccccc} 
\hline
\hline
Parameter & $\phi_1$ & $\phi_7$ & $\phi_8$ & $\sin^2\theta_{12}$ & $\cos^4\theta_{13}$ \\
\hline
Value & 6.00 & 0.525 & $5.33\times 10^{-4}$ & 0.330 & 0.955 \\
1-$\sigma$ error & 0.06 & 0.06 & $0.21\times 10^{-4}$ & 0.025 & 0.025 \\
$\chi^2$ & 3.67 & & & &  \\
\hline
\multicolumn{6}{l}{Correlation Matrix} \\
 $\phi_1$ & 1 & -0.909 & 0.502 & -0.669 &  0.670 \\
$\phi_7$& & 1 & -0.565 & 0.753 & -0.754 \\
$\phi_8$& & & 1 & -0.797 & 0.513 \\
 $\sin^2\theta_{12}$& & & & 1& -0.811 \\
$\cos^4\theta_{13}$& & & & & 1 \\
\hline
\hline\end{tabular}
\end{center}
\label{LENS}
\end{table}
\normalsize

  A factor of two improvement in the precision of the SuperKamiokande solar
 neutrino flux measurement does not significantly improve this separation.
The various scenarios and their effect on the determination of $\theta_{13}$
are summarized in Table \ref{summary}.

\begin{table}
\begin{center}
\caption{Effect of different future advancements on determination of $\theta_{13}$.}
\footnotesize
\begin{tabular}{|l|c|c|c|c|c|c|c|c|c|c|c|c|c|} 
\hline
\hline
SNO CC/NC 5\%	&		&	x	&		&	x	&	x	&	x	&	x	&		&	x	&	x	&	x	\\SNO total 2.5\%	&		&		&		&		&	x	&	x	&		&	x	&		&	x	&	x	\\KamLAND 3 yr	&		&		&	x	&	x	&	x	&	x	&	x	&	x	&	x	&	x	&	x	\\SK 2.5\%	&		&		&		&		&		&		&		&		&		&		&	x	\\Ga 2.3 SNU	&		&		&		&		&		&	x	&		&	x	&	x	&	x	&	x	\\$^7$Be 5\%	&		&		&		&		&		&		&	x	&	x	&	x	&	x	&	x	\\
\hline
$\sigma(\cos^4\theta_{13}$)	&	\footnotesize 0.0548	&	\footnotesize 0.0494	&	\footnotesize 0.0406	&	\footnotesize 0.0359	&	\footnotesize 0.0355	&	\footnotesize 0.0340	&	\footnotesize 0.0329	&	\footnotesize 0.0304	&	\footnotesize 0.0253	&	\footnotesize 0.0252	&	\footnotesize 0.0252	\\
$\Delta\theta_{13}$ (deg) & 10.0 & 9.1 & 8.2 & 7.7 & 7.7& 7.5 & 7.4 & 7.1 & 6.5 & 6.5 & 6.5 \\
\hline
\hline\end{tabular}
\end{center}
\label{summary}
\end{table}

A recent analysis~\cite{postkam} including the most recent KamLAND
data~\cite{kamland2} as well as the K2K results~\cite{K2K}, finds that
$\sin^2 \theta_{13} < 0.048$ at 3$\sigma$, allowing all the neutrino
fluxes to be free.

In summary, solar neutrino experiments and KamLAND provide information about
$\theta_{13}$ that is independent of the Chooz and atmospheric neutrino
determination, and therefore also essentially independent of the value
of $\Delta m_{23}^2$.  Since the solar and KamLAND experiments depend also on
$\theta_{12}$, a means of separating the effects of $\theta_{12}$ and $\theta_{13}$ is needed.
Beyond the existing data, improved separation can be obtained from any pair
of experiments  from the set  consisting of  a $^7$Be experiment or SSM
prediction, SNO CC/NC, and KamLAND rate.  The KamLAND spectral shape plays a
separate but key role in fixing $\Delta m_{12}^2$.  To obtain a significant
improvement in the determination of $\theta_{13}$ requires several improvements
in ongoing experiments; the improvement from any one is generally modest
by itself, but each is needed to make the gains.   If $\theta_{13}$ is about
12 degrees, close to its present upper limit, a 3-$\sigma$ determination
from solar and KamLAND data is possible.  No specific model inputs have
been used in this analysis other than the assumption that the Sun is in
quasi-static equilibrium generating energy by light element fusion.

\subsubsection{Sterile Neutrinos}
\label{sec:sterile}

	As described in Section~\ref{sec:ssmtest}, the precision with which
the flux of the lowest energy neutrinos can be predicted is better than
2\%---as well as most terrestrial reactor or accelerator neutrino fluxes are 
known.  Comparison of the number of low energy neutrinos measured to the
number predicted, including the (now) known mixing effects, can demonstrate
whether there is mixing to sterile neutrino species.  
Based upon existing solar data and the first results of the KamLAND experiment,
the 1$\sigma$ allowed range for the active-sterile admixture is~\cite{roadmap}
\begin{equation}
\sin^2\eta\leq 0.09
\label{eq:etacurrent}
\end{equation}
where $\sin^2\eta$ represents the mixing fraction to sterile states.
Future measurements by KamLAND, SNO, and Super-Kamiokande are not likely to
improve this bound substantially~\cite{roadmap}, nor will future
$^7$Be measurements.  A precision $pp$ experiment could bring the bound down
by as much as 20\%.

An more recent analysis~\cite{postkam}, including new data presetned
at the Neutrino 2004 conference, shows that the limits on a sterile
fraction have not changed much. The best fit is still zero admixture
to sterile.

\subsection{High Energy ($> 5$~MeV) Experiments}

	The highest energy solar neutrinos in the Standard Solar
Model are from the the $^8$B and {\em hep} reactions.  As described
in Section~\ref{sec:ssmexp}, the $^8$B neutrinos have been
observed by the water Cerenkov experiments Kamiokande~II~\cite{KIIsol}
and Super-Kamiokande~\cite{sksol}  via the elastic scattering (ES) of
electrons  ($\nu + e \rightarrow \nu + e$), and in SNO via both the charged
current (CC) ($\nu_e + d \rightarrow e + p + p$) and neutral current (NC)
($\nu + d \rightarrow \nu + p + n$) reactions on deuterium.  The latter
two measurements allowed the first model-independent measurement of solar
neutrino mixing, as well as the first confirmation of a Standard Solar Model
predicted neutrino flux.  To date, the neutrinos from the {\em hep} reaction
have not been observed, though upper limits on the flux have been set, placing
it less than about five times the predicted value of the Standard Solar Model.

	Both Super-Kamiokande and SNO will continue to run over the next few
years.  Currently, the Super-Kamiokande solar neutrino measurements are
limited because the loss of the PMT's has effectively raised the energy
threshold.  When the PMT's are replaced, Super-Kamiokande will resume its
solar neutrino measurements. SNO will complete its final data acquisition phase 
at the end of 2006.

	SNO has just begun a new phase of running, in which discrete, $^3$He
proportional counters have been installed within the heavy water volume. 
The $^3$He counters will allow SNO to measure the number of neutrons created by the
neutral current reaction on an event-by-event basis.  The new measurement of
the NC rate will therefore be systematically independent of the previous SNO
measurements in the pure D$_2$O and salt phases.  In addition, the $^3$He
counters remove neutrons from the events measured with Cerenkov light.  The
combination of these two effects means that in the third phase of SNO, the chance
of observing an MSW-produced spectral distortion is enhanced---the neutrons from
the NC reaction which are effectively a background in the prior SNO phases are
both reduced in number and normalized independently.  With some effort to lower
the analysis threshold by $\sim$ 0.5-1.0~MeV,  it may be possible to observe
a spectral distortion if the mixing parameters lie within the `northwest' quadrant
of the allowed region shown in Figure~\ref{fig:allowed}.  

	The third and final phase of SNO will therefore improve our knowledge of
the mixing parameters through the improved precision of the new measurements, 
allow a more sensitive search for an MSW distortion, and also provide additional
statistics in the search for a Day/Night asymmetry.

	As mentioned in Section~\ref{sec:msw}, there is currently
no experiment planned whose goal is the measurement of the $^8$B
spectrum in the region 1-5 MeV.  However, megaton-scale water Cerenkov
experiments~\cite{uno,hyperk} are being discussed which could observe
the $^8$B and {\em hep} neutrinos.  If built, the enormous statistics
these experiments would have may make it possible to observe even a
small Day/Night effect.  This would be particularly important in the
context of testing the neutrino oscillation model, as we will know
based on KamLAND or future (1,2) sector reactor experiments how large
the Day/Night asymmetry should be.  In addition, a megaton-scale water
Cerenkov experiment may be able to finally see the {\em hep} neutrinos,
thus confirming another piece of the Standard Solar Model.

\subsection{Low Energy ($< 2$~MeV) Experiments}

\subsubsection{$^7$Be}
\label{sec:7be}

	The flux of solar neutrinos from the $^7$Be reaction are the least
well-known based on the measurements to date.  In addition to verifying the
Standard Solar Model, a precision measurement of the $^7$Be neutrinos is
critical to the luminosity test described in Section~\ref{sec:ssmtest}.  
There are currently two experiments which may, in the near future, be able
to measure this flux.  We describe their current status and prospects below.

\begin{itemize}

\item {\bf KamLAND}

The KamLAND detector is a high light yield, high resolution (6.7\%/$\sqrt{\rm
E}$) liquid scintillator which is, in principle, also well suited for the
detection of low energy $^7$Be solar neutrinos. Elastic neutrino-electron
scattering would serve as the detection reaction: 
$\nu_e + e^- \rightarrow  \nu_e + e^-$ 
The interaction of the mono-energetic solar $^7$Be neutrinos (E$_\nu$=862
keV) will result in a Compton-like continuous recoil spectrum with an
endpoint energy of T$_{max}$=665 keV. This detection reaction provides
no signature allowing tagging. Such a measurement therefore has to
be performed in singles counting mode. (For the measurement of reactor
antineutrinos KamLAND makes use of the correlated detection of positrons
and neutrons by utilizing: $\overline{\nu}_e + p \rightarrow e^+ + n$,
which greatly reduces background).  The scintillator and its surrounding
technical components therefore must be of sufficient radio-purity in order
to avoid being overwhelmed by radioactive background. Signal event rates
of about 170 per day can be expected after appropriate fiducial volume cut
(600 tons assumed here). A more restrictive cut can be applied to counter
non-scintillator backgrounds.  This rather substantial rate partially
compensates for the lack of signature compared to the antineutrino detection
where KamLAND detects about one event per 2.7 days.

 
All external construction materials of KamLAND have been carefully selected
with a $^7$Be program in mind. The KamLAND collaboration believes that
external backgrounds can be managed by means of a fiducial volume cut.
Within the inner scintillator volume KamLAND measures effective Th and U
concentrations of $(5.2\pm 0.8)\times 10^{-17}$ g/g and $(3.5\pm 0.5)\times
10^{-18}$ g/g by means of Bi-Po delayed coincidence, even exceeding the
rigorous requirements for a $^7$Be experiment. For $^{40}$K only a limit
of $<3\times 10^{-16}$ g/g has been determined.  $^{40}$K contained in
the scintillator containment balloon and its holding ropes can again be
countered by an appropriate fiducial volume cut.

Analysis of the low energy background in KamLAND shows $^{85}$Kr and
$^{210}$Pb contaminations at prohibitive concentrations. Some evidence
also points at the presence of $^{39}$Ar. These airborne contaminations
were probably introduced by contact of the scintillator with air. This also
holds for $^{210}$Pb which is a Radon decay product. The singles counting
rate in the solar analysis energy window is now about 500 s$^{-1}$. The
detection of $^7$Be solar neutrinos thus requires a large reduction of
these known contaminants.

The KamLAND scintillator had been purified by means of water extraction and
nitrogen gas stripping during filling. Piping and technical infrastructure
for general scintillator handling exists underground with the appropriate
capacity. The liquid scintillator could thus be re-purified using this
existing infrastructure, augmented by additional purification devices.
Development work toward $^7$Be detection in KamLAND focuses on the
removal of Kr and Pb from the liquid scintillator. 

As very large reduction factors are required the collaboration decided
to conduct laboratory tests to demonstrate technical feasibility, and
repeated application of various techniques did achieve large purification
factors in the lab. Further work will aim at providing proof of principle
which will include the construction of a mid-size pilot device to study
the technical parameters.

To reduce the Radon concentration in the lab the KamLAND collaboration
installed a new fresh air supply system which resulted in factor 10 to 100
reduced environmental Radon concentrations in the KamLAND lab area. It
is further planned to equip all piping and plumbing with external radon
protection.

The R\&D work towards a solar phase of KamLAND is funded on
the Japanese side.  However, no such funding is yet available for the US
collaborators. An R\&D funding proposal is in preparation on the US side.
The KamLAND collaboration hopes to finalize the technical development work
within one year. If technical feasibility can be demonstrated in the lab
then construction of new on-site purification components and purification of
KamLAND's 1000 tons of liquid scintillator are estimated at 2 to 3 years.

\item {\bf Borexino}

Borexino is a liquid scintillator detector with an active mass of 300
tons that is installed at the Gran Sasso Laboratory in Italy~\cite{bx}.
It has an active detector mass of 300 tons and is designed for real time
measurements of low ener gy solar neutrinos.  Neutrino detection is through
the elastic scattering of neu trinos on electrons, a process to which both
charged and neutral currents contri bute.  The rate of neutrino interactions,
thus, depends on neutrino oscillations and flavor conversion. The Borexino
international collaboration includes several European groups and three North
American groups, Princeton University, Virginia Tech, and Queens University.

Although the primary goal of the Borexino experiment is to measure neutrinos
from the solar $^7$Be reaction, if the contamination from $^{14}$C is
low enough, $pp$ neutrinos above the $^{14}$C endpoint may be 
measured~\cite{laura}.  In addition, there is some sensitivity to solar
$pep$, CNO, and $^8$B neutrinos, by tagging and removing cosmic and internal
backgrounds.  The $^8$B neutrinos will produce roughly 100 events a year,
and can be seen down below the energy thresholds other detectors have used
(for example, Super-Kamiokande and SNO).

The installation of the Borexino inner detector at the Gran Sasso Laboratory
was completed in June 2004.  The last major step of the detector assembly was
the installation and inflation of the nested nylon vessels.  The inner steel
sphere was closed on June 9 2004.  The completion of the PMT installation
for the muon veto detector in the water tank is expected in July 2004.
The detector should be fully commissioned and ready to fill with water
by the end of July 2004. A complete description of the detector can be
found in~\cite{bx}.

To minimize radioactive contaminants on the nylon
containment vessels from dust and Radon daughters, the vessels were
made of a specially extruded nylon film that was controlled from the time
of extrusion through the actual fabrication in a class 100 clean room.
The film was pre-cleaned before the fabrication and the surface exposure
to air in each step of the construction process was minimized by providing
protective covers for the film and minimizing the time of exposure to air.
To further reduce Radon daughters during the necessary exposure of the
film, Radon was removed from the make-up air to the clean room by a
pressure swing Radon filter developed for the purpose~\cite{andrea}. The
nylon vessels were recently installed in Borexino, and tests show that
the vessels meet or exceed mechanical design requirements, including
requirements for admissible leak rate.  A final cleaning of the vessels
with a water spray is planned before filling the detector with water,
which will help remove any residual contamination.

Radioactive contaminants from the long-lived chains of U and Th can
also be a problem.  To demonstrate the feasibility of achieving the
required U and Th backgrounds, the collaboration successfully built
and operated the Counting Test Facility (CTF).  The result of the CTF
demonstrated the feasibility of multi-ton detectors with upper limits
on the U and Th impurities of $\sim 10^{-16}$ g/g, as required for solar
neutrino observation~\cite{ctf}.

The commercially-made Borexino scintillator will be pre-purified and tested
in the CTF before filling the detector, to ensure that contamination
levels have met their goals. One other virtue of the initial water
filling is that small quantities of scintillator can be introduced in
Borexino with the full 4$\pi$ water shield as a more sensitive test of
the scintillator background, before the detector is completely filled.
Finally, a purification and liquid handling system is installed that will
enable purification after the detector is filled, if necessary.

The nested structure of the Borexino vessels will allow great reductions
in the diffusion of Radon from the detector periphery to the scintillator
region.  An ultra-high purity liquid nitrogen source will be used in
combination with a stripping column to lower backgrounds from $^{85}$Kr and
$^{39}$Ar inside the the scintillator to below 1 count per day inside the
fiducial region.  To prevent contamination during filling and operation,
the vessels and piping were built with stringent high vacuum tightness
requirements.

Borexino's goal is to determine the $^7$Be flux with a total precision
of 5$\%$.  As noted above, a measurement of $pep$ neutrinos also seems
possible in Borexino.  With an expected $^7$Be neutrino signal rate of
$\sim 30$ counts per day, the statistical precision is $\sim 1\%$ in one
year of counting.  For $pep$ neutrinos, the expected rate is $\sim 1$
count per day, with a statistical uncertainty of $\sim 5\%$ in one year.
In both cases the rates are relatively high and should be sufficient for
a measurement with a statistical uncertainty of few per cent or less in 5 years
of counting.  With the source calibration system, the needed precision in
the fiducial radius (300 cm $\pm \sim 2$ cm) seems possible, given that
the position resolution is expected to be $\sim 10$ cm.  With $\sim 400$ 
detected photoelectrons per MeV, the energy resolution is 5$\%$ at 1 MeV
and 10$\%$ at 250 keV, the lower end of the energy window.

The main issue that will likely determine the final uncertainty are the
backgrounds under the low energy portion of the spectrum where the $^7$Be
neutrinos appear ($< 0.65$ MeV).  Figure~\ref{fig:spectrum}, for example,
\begin{figure}[ht]
\begin{center}
\mbox{
\includegraphics[height=0.4\textheight]{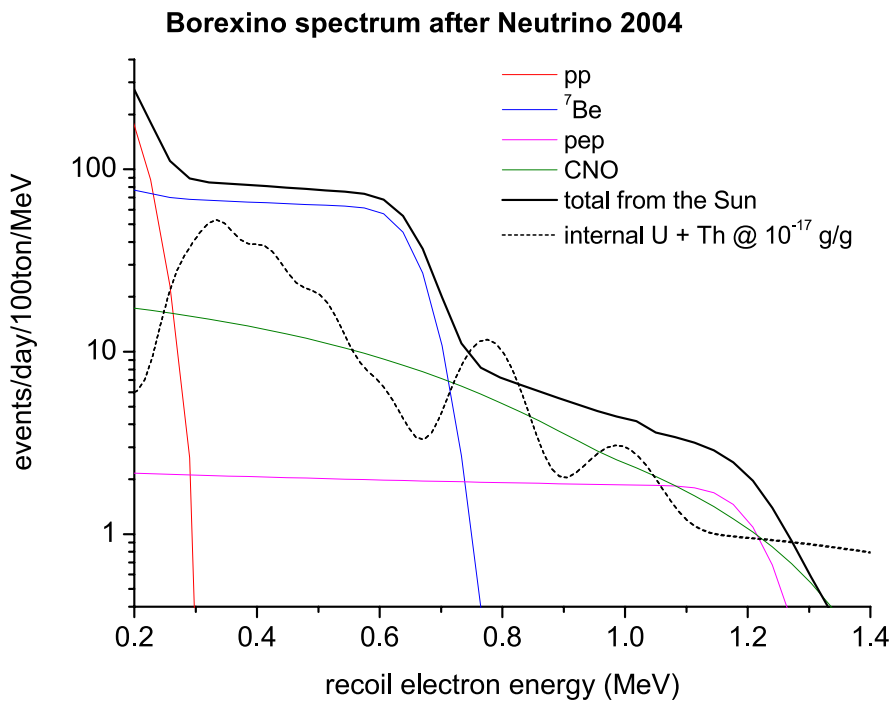}}
\caption{Expected solar neutrino rates in Borexino. The solid line is the
expected total neutrino spectrum between 0.2 and 1.4 MeV, based on the 
MSW LMA solution and BP04 \cite{BP04}. The dotted line is the internal
background from U and Th, assumed at a level of $10 ^{-17}$ g/g, shown as a
reference. Other sources of background are not included.}
\label{fig:spectrum}
\end{center}
\end{figure}
illustrates background from the U and Th chains if their concentrations
are at $10^{-17}$ g/g and no additional cuts are applied.  The background
shown in Figure~\ref{fig:spectrum} can further be reduced by various cuts,
including $\alpha \backslash \beta$ separation which is expected to reduce
the alphas by more than a factor of 10. The U and Th concentrations are
10 times lower than our current limits, but consistent with recent data
from KamLAND. If backgrounds are low enough to measure $^7$Be neutrinos
to 5$\%$, the pep neutrinos should also be measurable, with a precision
better than 10$\%$.

Borexino and other experiments at Gran Sasso were placed under judicial
sequestration following a small spill of scintillator in August 2002.
The sequestration stopped all work underground.  In the spring of 2003
a partial lifting of the sequestration was granted to permit mechanical
construction of the detector to restart.  However, the ban on fluid use
remained in force, owing to the discovery of flaws in the drainage system.
In June 2003 a special commissioner was appointed by the italian government
to assume responsibility for repairing the laboratory infrastructure
and restoring the laboratory to full operations.   As of June 2004, the
commissioner's staff is still implementing the repairs.  A full lifting
of the sequestration is expected late this summer, two years after the
incident. The first operations that will occur after the sequestration is
lifted will be filling the detector with high purity water and studies of
scintillator purification with the Counting Test Facility.

\end{itemize}

\subsubsection{$pp$}

Measurement of the $pp$ neutrino flux will require an experimental
technique that allows very low radioactive backgrounds for energies $<
300$ keV.  In contrast to other low energy solar experiments done so far,
the  proposed experiments aim at measuring the full spectrum below
2 MeV, therefore including the fluxes from all the major sources in the Sun.
Proposed $pp$ experiments fall in two classes: neutrino- electron scattering
(ES) and charge current neutrino absorption (CC).

The ES proposals (CLEAN and HERON) have the advantage of promising very
low internal backgrounds by virtue of the cleanliness of their detection
media. Helium, as a superfluid at 50 mK, is completely free of any activity;
Neon at 27 K can be ultra-purified.  Neutrino-electron elastic scattering
cross-sections are well known from electroweak theory, so ES experiments
do not need to be calibrated with a neutrino source.  CC experiments
are attractive because they exclusively yield  the electron flavor flux
of $^7$Be neutrinos in particular, complementing the NC flavor content
obtainable from present ES experiments such as Borexino and Kamland.
Of course, measuring fluxes {\it both} via ES and CC reactions could allow
a determination of the total active $pp$ neutrino flux independent of
the mixing parameters.

Both types of experiments must be located deep underground  to avoid
backgrounds from muon spallation. CLEAN and HERON have very different
approaches to rejection of gamma ray backgrounds.  CLEAN would use the
relatively dense liquid neon to absorb external gamma rays before they
reach the inner fiducial region, then cut gamma ray events using position
resolution. HERON would have sufficient position resolution to distinguish
point sources (signal) from distributed sources (gamma rays).

The CC proposal (LENS) would use a target of Indium incorporated into
organic scintillator. Because Indium emits a delayed gamma ray after neutrino
absorption, coincidence techniques can be used to greatly reduce radioactive
backgrounds, including bremstrahlung from Indium $\beta$-decay.  The delayed
coincidence signature distinguishes a neutrino event from background, and
allows the simultaneous measurement  of signal and background independently.
Also, because the neutrino energy is entirely captured,  
the shape of the $pp$ and $^7$Be neutrino spectra may be directly measured.  
LENS has only a modest depth requirement ($>$ 2000 m.w.e.). For an 
accurate $pp$ neutrino measurement, the neutrino absorption cross-
section must be calibrated using a MCi neutrino source, probably $\rm ^{37}Ar$.  

	Figure~\ref{fig:spectra} shows the expected reconstructed energy 
spectra from simulations of neutrino interactions and backgrounds in these 
experiments.
\begin{figure}[ht]
\begin{center}
\includegraphics[width=0.5\textwidth]{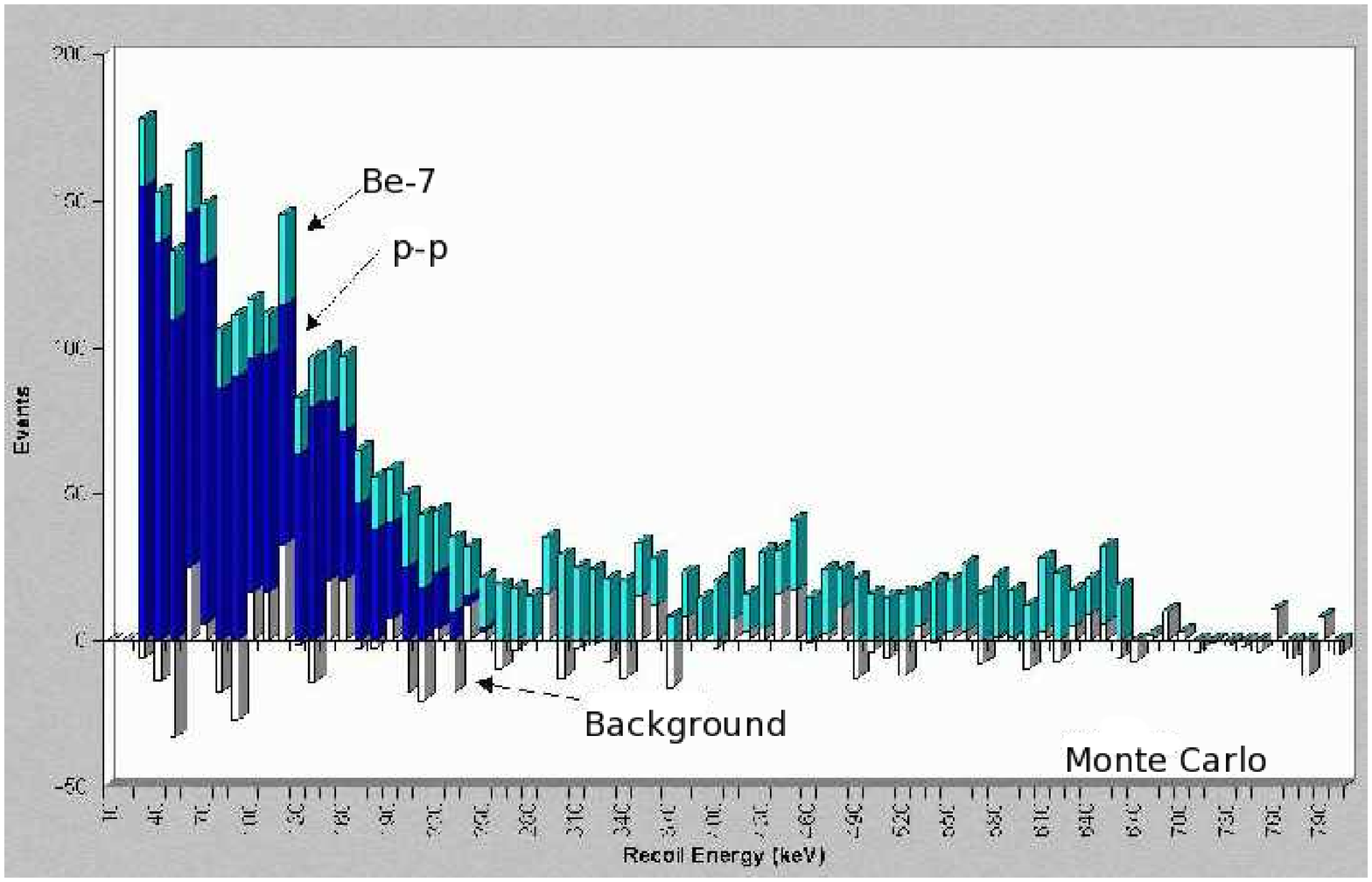}
\includegraphics[angle=-90, width=0.5\textwidth]{./clean_spectrum.epsi}
\includegraphics[width=0.5\textwidth]{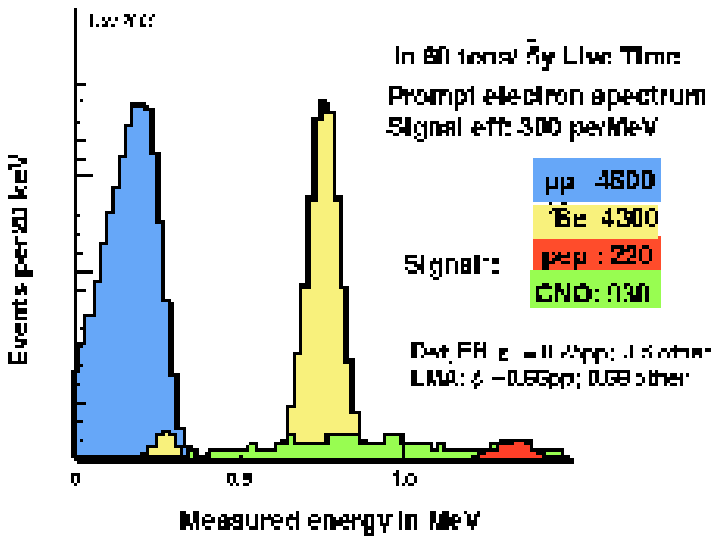}
\caption{Simulations of reconstructed energy spectra for future low energy solar
neutrino experiments. From top to bottom are the simulations for the HERON
experiment, the CLEAN experiment, and the LENS experiment. \label{fig:spectra}}
\end{center}
\end{figure}

	While the technical challenges of these experiments are of high 
order, there has been much progress in overcoming them.  Below we detail
some of the expectations for the precision of the different experiments, and
some of the specific challenges and advantages of each method.

\begin{itemize}

\item {\bf HERON}

  The estimates on the uncertainties for HERON are shown in
Table~\ref{tbl:heron}, and have been made on the basis of extensive prototype
experimentation on the particle detection properties of superfluid helium
and on the wafer detector devices to be used in a full scale device. Also,
detailed simulations of signal and background events from energy deposition
to full event reconstruction in a full-scale detector design have been
used. The detector is assumed to be at a depth of at least 4500 m.w.e.,
externally shielded, and residual backgrounds due to site environmental
sources have been shown to be negligible. The dominant source of background
are the materials of the cryostat and moderator; they have been taken
for the copper cryostat: primordials and cosmogenics (as measured by
double beta decay and dark matter experiments as well as by ICPMS and
NAA measurements). For the moderator N2 : primordials, cosmogenics
and anthropogenics from the Heidelberg LN2 extensive studies. Activity
concentrations for plastics are taken from the large studies at SNO and
KamLAND.

\begin{table}[h!]
\begin{center}
\caption{Uncertainties on $pp$ flux for HERON.\label{tbl:heron}}
\begin{tabular}{|l|r|} \hline
      & {$pp$ uncert.} (\%) \\ \hline
Threshold cut (Energy scale and $\sigma$) &  1.25  \\
Fiducial volume         &  1.3   \\
Efficiency              &  1.5 \\
Signal/backgrounds separation fit & 2.5 \\
Internal background     &  0.0   \\
Density uniformity of target vol. & 0.0 \\
cross section             & 0.25 \\
Deadtime		& 0.04 \\
pp/$^7$Be separation    & 0.25 \\ \hline
{\bf{Total systematic} (quadrature sum) } &  3.4  \\ \hline \hline
{\bf{Statistics} ($pp$)}       &  1.0   \\ \hline \hline
{\bf{Statistics} ($^7$Be)}       &  1.5   \\ \hline \hline
{\bf{Total}}            &  3.5   \\ \hline \hline
\end{tabular}
\end{center}
\end{table}

The energy threshold used to produce the numbers in Table~\ref{tbl:heron}
is set at 45 keV visible electron recoil energy and the energy resolution
(FWHM) ranged from 3.2\% at 600 keV to 10.3\% at 45 keV.  The absolute scale
assumed at 2\%,  and the helium fiducial volume of 68 m$^{3}$ with position
resolution of $\sigma_x=1.62$~cm, $\sigma_y=1.54$~cm,and $\sigma_z=2.46$~cm.
The signal and backgrounds can be separated by their distinct energy and
spatial distributions both inside and outside the fiducial volumes, as well
as the Earth orbital eccentricity variation of the signal neutrino flux.
The superfluid helium itself is entirely free of any activity.

\item {\bf CLEAN}	

Table~\ref{tbl:clean} summarizes the projected uncertainties for solar
neutrino flux measurements with a 300 cm radius liquid neon detector
(CLEAN), assuming both 1 year and 5 year runs.   A fiducial volume defined
by a 200 cm radial cut is assumed, for a total active mass of 40 tonnes.
The detector is assumed to be at a depth of 6000 m water equivalent, where
cosmic-ray induced backgrounds and related uncertainties are negligible.
Dominant sources of backgrounds are assumed to be internal radioactivity,
and radioactivity from the PMT glass found in certain commercially available
phototubes (30 ppb uranium, thorium, 60 ppm potassium).  Above the neutrino
analysis threshold of 35 keV the fiducial volume cut is expected to remove
essentially all background events from PMT activity.  The total event rates
for $pp$ and ${}^{7}$Be neutrino interactions are calculated assuming the
current best-fit LMA solution, and SSM fluxes.  Two analysis windows are
defined:  35-300 keV for $pp$ events, and 300-800 keV for ${}^{7}$Be events.
Fluxes are derived from the event rates in these windows.  Uncertainties
related to the neutrino mixing model are not considered.

\begin{table}[h]
\begin{center}
\caption{Uncertainties on $pp$ and ${}^{7}$Be fluxes for CLEAN.\label{tbl:clean}}
\begin{tabular}{|l|rr|rr|} \hline
      & \multicolumn{2}{|c|}{$pp$ uncert.} & \multicolumn{2}{|c|}{${}^{7}$Be uncert.} \\
      & \multicolumn{2}{|c|}{(\%)} & \multicolumn{2}{|c|}{(\%)} \\
        & 1 y & 5 y & 1 y & 5 y \\ \hline
Energy scale            &  0.34  &  0.34 &  0.87 & 0.87 \\
Fiducial volume         &  0.90  &  0.90 &  0.90 & 0.90 \\
Internal krypton        &  0.25  &  0.25 &  1.87 & 1.87 \\
External backgrounds    &  0.04  &  0.02 &  0.15 & 0.07 \\
${}^{7}$Be $\nu$'s      &  0.25  &  0.11 &  0    & 0 \\ \hline
{\bf{Total systematic}} &  1.03  &  1.00 &  2.26 & 2.25 \\ \hline
{\bf{Statistics}}       &  0.86  &  0.38 &  2.87 & 1.28 \\ \hline
{\bf{Total}}            &  1.34  &  1.07 &  3.65  & 2.59 \\ \hline
\end{tabular}
\end{center}
\end{table}

In Table~\ref{tbl:clean}, the absolute energy scale uncertainty is assumed
to be 1\%, and is assumed to be determined by deploying $\gamma$-ray calibration
sources throughout the detector volume many times.  The dominant uncertainty
is expected to arise from the uncertainty in converting absolute $\gamma$-ray
energies to electron energies with a Monte Carlo model.  The uncertainty in
the neutrino flux arising from the uncertainty on energy resolution is 
negligible.

The dominant uncertainty on the measurement of the $pp$ flux is the 
uncertainty on the fiducial volume.  If CLEAN can do $\sim$ 3 times better
than SNO, then the uncertainty $\frac{\Delta R}{R} = 0.3\%$, leading
to less than a 1\% uncertainty on volume.  Doing as well as 0.3\% will require 
source positioning to be accurate to 0.6~cm, and the positioning system will
need to be able to reach nearly all positions within the detector volume.

Internal background from Krypton,  Uranium, and Thorium are
expected to be small, and in the worst case (Krypton) known to
25\%.  These backgrounds will be measured by assaying the neon, or by measuring them {\it{in-situ}} with the PMT data.  

External backgrounds, dominated by PMT activity, will be removed primarily
by the fiducial volume cut, and can be tested by deploying a very hot
source exterior to the volume and counting the number of events which
reconstruct inside.  The fiducial volume cut is particularly
effective because of the high density of liquid neon (1.2 g/cc).
Position resolution in CLEAN is based on PMT hit pattern and timing,
and is confirmed in detailed Monte Carlo simulations.

\item {\bf LENS}

The following tables summarize preliminary estimates of precision expected
in $pp$ and $^7$Be flux measurements in the LENS-Sol solar detector in
conjunction with the LENS-Cal $^{37}$Ar source calibration. The results
are obtained for two different Indium target masses, 60 and 30 tons to
illustrate the roles of statistical and systematic errors. The latter is
set identical for both target masses.  LENS is planned to have a modular
detector architecture, thus the performance of the full-scale detector
can be closely predicted from bench top tests of individual modules

\begin{table}[h]
\begin{center}
\caption{Uncertainties on $pp$ and ${}^{7}$Be fluxes for LENS. \label{tbl:lens}}
\begin{tabular}{|l|rr|rr|} \hline
      & \multicolumn{2}{|c|}{$pp$ uncert.} & \multicolumn{2}{|c|}{${}^{7}$Be uncert.} \\
      & \multicolumn{2}{|c|}{(\%)} & \multicolumn{2}{|c|}{(\%)} \\
        & 30 t & 60 t & 30 t & 60 t \\ \hline
Signal/Background Statistics &  2.33  &  1.65 &  2.12 & 1.5 \\
Coincidence Detection Efficiency &  0.7  &  0.70 &  0.70 & 0.70 \\
Number of Target Nuclei &  0.3  &  0.3 &  0.3 & 0.3 \\
Cross Section (Q-value) &  0.3  &  0.3 &  0.16 & 0.16 \\
Cross Section (G-T matrix element)  &  1.8  &  1.8 &  1.8    & 1.8 \\ \hline
{\bf{Total Uncertainty }} &  3.05  &  2.57 &  2.87 & 2.46 \\ \hline
\end{tabular}
\end{center}
\end{table}

	The only correlated backgrounds to the triple coincidence arise
from cosmogenic (p,n) reactions on Indium. These are expected to be about
5\% of the solar signal at a depth of 1600 m.w.e., but will be vetoed by
tagging the initiating cosmic.	The triple-coincidence detection efficiency
has been estimated through Monte Carlo simulation, and includes cuts on
energy, time, and the In and In-free parts of the detector.  For the $pp$
neutrinos the efficiency is expected to be $\sim$25\%, and for $^7$Be
neutrinos $\sim$80\%. An experimental determination of the coincidence
efficiency can be made by using a small surface detector and using
cosmic-ray induced products, which can produce the same signals as the
neutrino events.  These measurements are expected to yield the uncertainty
shown in Table~\ref{tbl:lens}.

	The segmentation of the detector will allow the fiducial volume to be
determined by the dimensions of the detector, not on an offline-cut, and
the uncertainty on the number of target nuclei will depend primarily on the
chemical determination of the Indium content in the In-loaded liquid
scintillator.

	To determine uncertainties associated with the cross section (knowledge
of the Q-value as well as the matrix element) a 5-ton calibration detector and
strong (2-MCi) $^{37}$Ar source will be used.

\end{itemize}


\subsection{Supporting Nuclear Physics Measurements}

	As described in Sections~\ref{sec:ssmexp},\ref{sec:npsol}, and~\ref{sec:sterile} comparison of the neutrino fluxes to the predictions of the Standard 
Solar Model allow us to search for both new astrophysics and new particle
physics.  To make the comparison meaningful, we would like for the precision
of the predictions to be comparable to that of the measurements.

A global analysis of all solar experiments and KamLAND yields the total
$^8$B solar neutrino flux with a precision of $\pm$ 4\% \cite{roadmap}.
We would therefore hope to reduce the uncertainties on the Standard Solar Model
prediction to a level of 5\% or smaller.

Recently, new measurements have been made of the C, N, O, Ne, and
Ar abundances on the surface of the Sun \cite{Allende}.  The current
uncertainty in solar composition (Z/X) leads to a large 8\% and 20\%
uncertainty ~\cite{Bah04,Ricci,Couv} in the predicted $^7$Be and $^8$B
solar neutrino flux, respectively, and therefore to improve the precision
on the prediction, we need new measurements.



Nuclear inputs to the SSM, in particular the cross sections \sos and \stf, as
defined by Adelberger {\em et al.} \cite{Adel}, need to be known with a precision
better than 5\%.  High precision (3-5\%) measurements of \xsos are now available
from experiments using very different methods or experimental procedures
\cite{Seatt,Weiz,Iw99,Sch03}.  The mean of the modern direct measurements below the
630-keV $1^+$ resonance gives S$_{17}$(0) to 4\%.  Significant differences
are apparent, however, between the indirect (Coulomb dissociation and heavy-ion
transfer) and direct determinations of S$_{17}$(0), which merit further
exploration---see Fig.~\ref{Extrap}. An additional high-precision direct
measurement is in progress.


The most recent evaluation
of \stf was performed in 1999 by Adelberger {\em et al.} \cite{Adel} and
unfortunately no new data on \stf were reported in the intervening time
period. A 13\% discrepancy between the low and high values of \stf was
found by Adelberger {\em et al.}, who quote \stf with a 9\% accuracy. 
Additional direct experimental measurements are necessary to reduce the
uncertainty on S$_{34}$(0) below 5\%.

As a particular example we note that the precision value measured
by the Seattle group \sos = 22.1 $\pm$ 0.6 \cite{Seatt}, together with
the larger value of \stf = 572 $\pm$ 26 eV-b deduced from $^7$Be activity
measurements \cite{Adel,Ham} yield a predicted total $^8$B neutrino flux
that is 20\% larger than measured by SNO. The smaller value of \stf = 507
$\pm$ 16 eV-b \cite{Adel} on the other hand reduces the discrepancy to 9\%.
Currently the $^8$B solar neutrino flux is predicted with 23\% uncertainty
\cite{BP04} with the main uncertainty due to the solar composition (Z), as
discussed above.  Such a discrepancy can only be considered significant
with an improved precision of the prediction of the SSM, and it may for
example provide evidence of oscillation into sterile neutrino.

\begin{figure}
\begin{center}
  \includegraphics[width=3in]{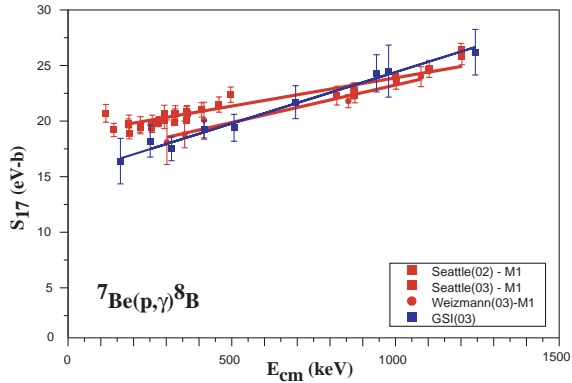}
  \caption{\label{Extrap} 
A comparison of a selection of the recent Seattle(02) \cite{Seatt},
Weizmann(03) \cite{Weiz}, and GSI \cite{Iw99,Sch03} data. 
An M1 contribution due to the resonance at 632 keV is subtracted from the 
direct capture data. The slope of the indirect Coulomb breakup results appear
to disagree somewhat with that of the direct cross section measurements. A
detailed theoretical re-evaluation is in progress.}
\end{center}
\end{figure}

\subsection{Other Physics with Solar Neutrino Detectors}
\label{sec:solop}

	Nearly all detectors described in the previous sections are capable
of doing other physics besides solar neutrino observations.  The detection
of neutrinos from supernovae can be done particularly well by large-scale
water Cerenkov experiments, but also by many of the other detectors as well.
The detection of the constituents of dark matter are an explicit goal of
some of the $pp$ experiments.  In addition, many of these experiments can
serve as antineutrino detectors as well, perhaps able to observe (or limit)
the flux of geoneutrinos originating within the Earth.	Although the focus
of this report is solar and atmospheric neutrino detection, we do want
to emphasize the fact that the detectors themselves have a justification
that includes a range of physics outside that focus.

\pagebreak

\section{Atmospheric Neutrino Experiments}

	The study of atmospheric neutrinos---neutrinos produced by the
interactions of cosmic rays with the atmosphere---should have been
straightforward. While the detection of such neutrinos was by themselves
interesting, and while some envisioned the possibilities for these neutrinos
to help us discover leptonic flavor transformation, the primary physics we
expected to learn was how cosmic rays interact in the atmosphere, what the
fluxes of different types of neutrinos were, and what the neutrino energy
spectra were.

	It was clear almost from the outset, however, that something
was wrong. The two largest experiments capable of detecting atmospheric
neutrinos---IMB~\cite{IMB1,IMB2} and Kamiokande~\cite{KIIatm}---both saw that
the ratio of $\nu_{\mu}$'s to $\nu_e$'s was significantly
different than expectations.  One possibility which explained such an
observation was that upward $\nu_{\mu}$'s were disappearing due to
neutrino oscillations, but other possibilities were still considered.
In addition, if oscillations were the explanation, the data indicated
that the mixing between the neutrino mass states was maximal or nearly
so---something which contradicted the prejudices based on the small quark
mixing angles.

	It was not until 1998, when the Super-Kamiokande collaboration
published a high statistics plot of the number of detected neutrinos as
a function of zenith angle, that the oscillation hypothesis was clearly
demonstrated~\cite{SKatm}.  Figure~\ref{fig:coszatm} shows an updated
\begin{figure}[ht!]
\begin{center}
\includegraphics[width=0.8\textwidth]{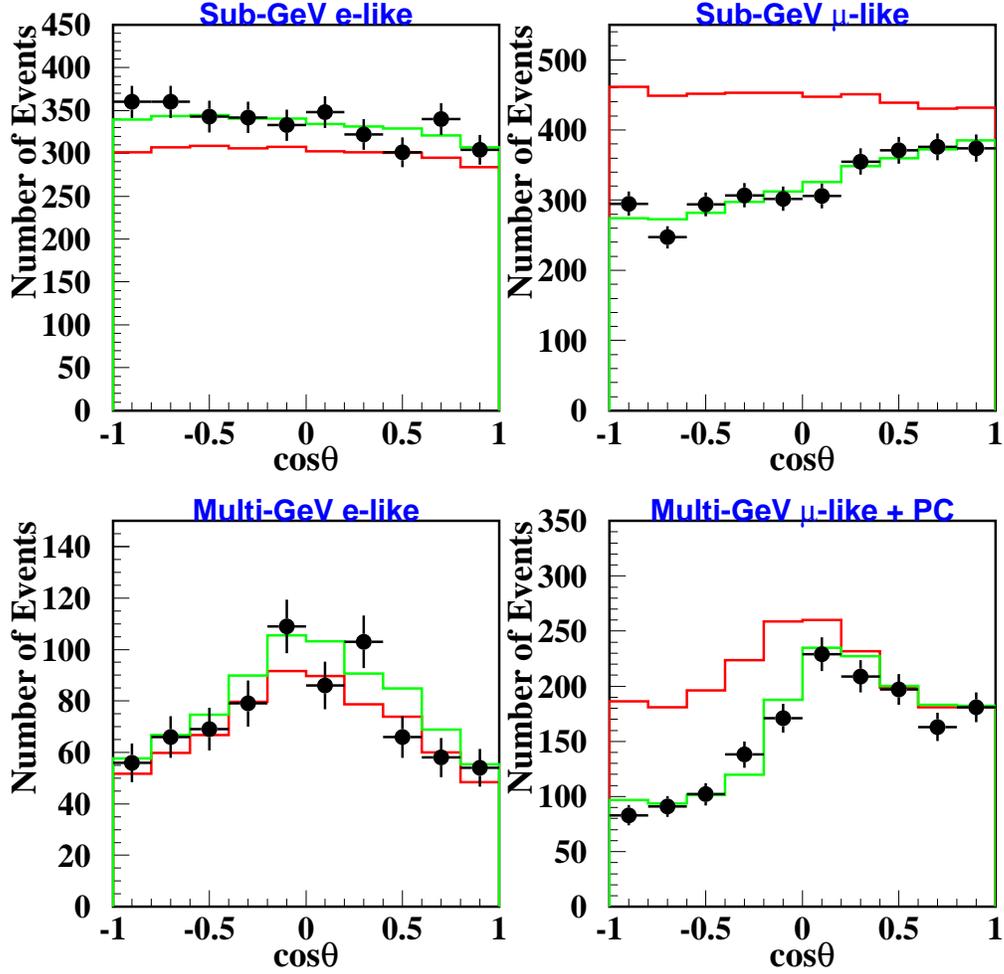}
\caption{Zenith angle distributions for atmospheric neutrino interactions
in Super-Kamiokande~\cite{SKatm04}.
\label{fig:coszatm}}
\end{center}
\end{figure}
version of the $\cos{\theta_z}$ distribution for both $\nu_e$'s and
$\nu_{\mu}$'s, compared to the oscillation hypothesis.	We can see that
the fit to the data for an oscillation of $\nu_{\mu}$'s is extremely good,
and that the null hypothesis of no transformation is not possible.  Other
experiments---MACRO~\cite{macro} and Soudan 2~\cite{soudan2}---using very
different methods subsequently confirmed Super-Kamiokande's measurements.
Fig.~\ref{fig:gallagher}~\cite{gallagher} shows the summary of the results
on the mixing parameters determined by all the atmospheric experiments.

	In the following sections, we describe both the role that future
atmospheric experiments can play in testing the three-flavor oscillation
model, as well as the potentially critical role they may play in resolving
the mass hierarchy.  We finish with a short discussion of some of the
non-oscillation physics which can be done with these experiments.

\subsection{Testing the Neutrino Oscillation Model}

	As discussed in Section~\ref{sec:soltests}, our model of neutrino
flavor transformation requires the addition of (at least) seven new
fundamental parameters to the Standard Model of particle physics: three
mixing angles, a complex phase, and three neutrino masses.  With these new
parameters, the Model predicts all transformation phenomena regardless of
energy, baseline, lepton number, flavor, or intervening matter.  To test
the Model, we therefore need to measure the parameters and compare them
across experimental regimes such as energy, baseline, etc., verify some
of the explicit predictions of the Model such as the oscillatory nature
of the transformation, look for the predicted sub-dominant effects, 
and search for some of the possible non-Standard Model transformation
signatures.

	The enormous experimental regime covered by the atmospheric
measurements means that they are particularly sensitive tests, and
in many ways the atmospheric sector is far ahead of the solar sector in
verifying some of the finer details of the transformation model.

\begin{figure}
\begin{center}
\includegraphics[width=0.6\textwidth]{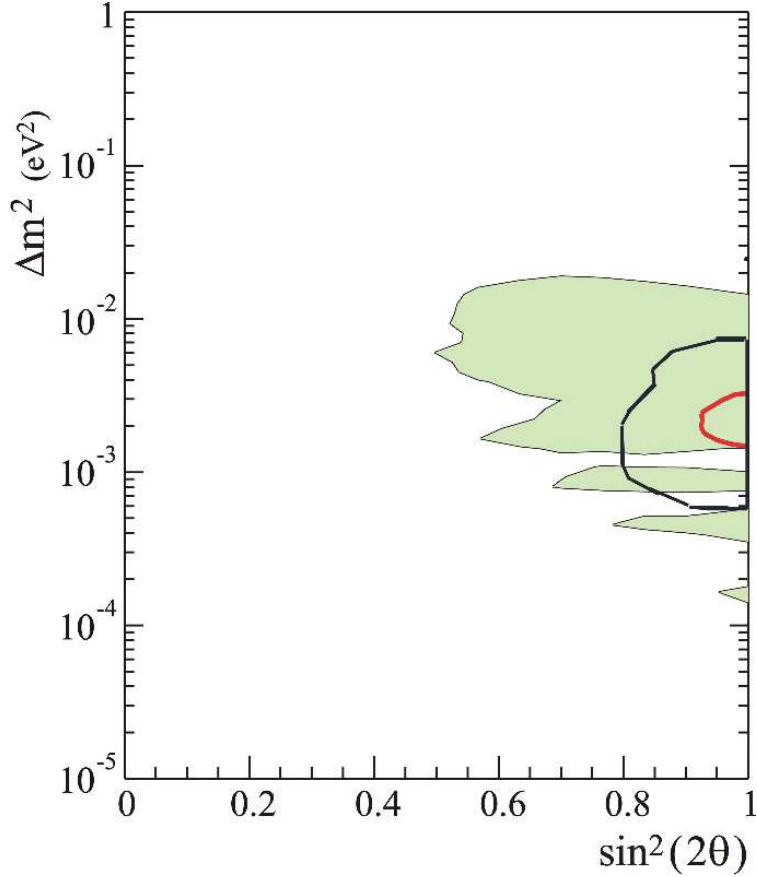}
\caption{The shaded region shows the allowed region from the latest
Soudan 2 results; the outer unfilled black contour shows the allowed
region from MACRO upward-going muons; and the inner solid red contour
shows the results from Super-Kamiokande..
Figure taken from Ref.~\cite{gallagher}.
\label{fig:gallagher}}
\end{center}
\end{figure}


\subsubsection{Precision Measurements in the (2,3) Sector}

	The wide dynamic range of neutrino energies and baselines in the
atmospheric sector mean that atmospheric experiments provide their own tests
of the oscillation model---the predictions can be shown to hold across all 
the accessible experimental regimes.  Improved precision in these experiments
thus provide interesting tests of the oscillation model even in the
absence of other experimental approaches.  

\begin{figure}
\begin{center}
\includegraphics[height=2.7in]{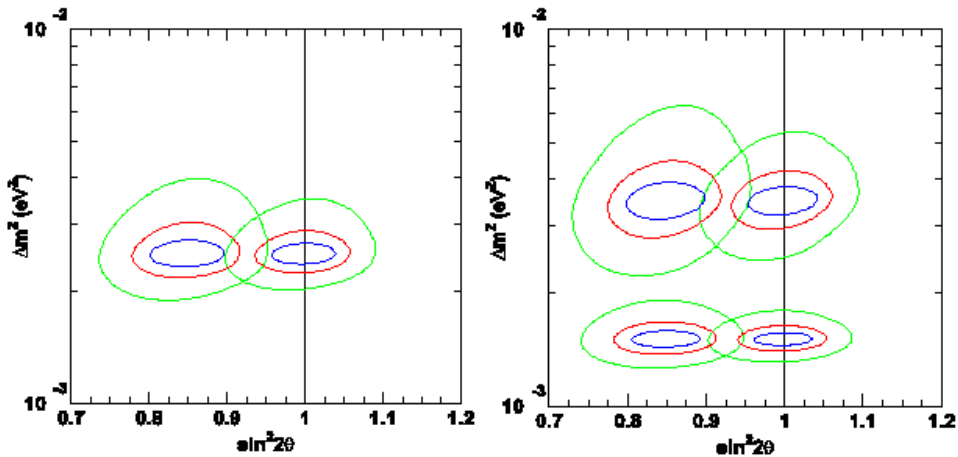}
\caption{Allowed regions showing expected sensitivity to dominant oscillation
parameters for Super-Kamiokande (or similar water Cerenkov detector). Each
set of contours corresponds to a particular assumption of true oscillation
parameters.  For each set of contours, the outer (green) one represents a
113 kton-year exposure (5 years of SK); the red represents 450 kton-years
(20 years of SK); the inner (blue) one represents 1800~kton-years (80 years
of SK).  Figures taken from Ref.~\cite{kajitanoon04}.  \label{fig:kajita}}
\end{center}
\end{figure}

Figure~\ref{fig:kajita} shows the expected sensitivity to the dominant
oscillation parameters from atmospheric neutrinos, for various exposures
of Super-Kamiokande and various assumed true values of the parameters.
These results assume use of an analysis similar to the high resolution
L/E analysis of Ref.~\cite{SKdip04} (and see Section~\ref{sec:loverE}).
The size of the regions shrink with the square root of the exposure,
as expected, but it is clear from the plot that ultimate sensitivity
to $\Delta m^2$ depends also on the actual values of the parameters.
Of course, the sensitivities of the next generation of long baseline
accelerator experiments are competitive with the measurements made
by the atmospheric experiments.

The first long baseline accelerator neutrino experiment K2K~\cite{K2K,
K2Knu2004} has confirmed this picture, thus providing the first test of
the oscillation model in the atmospheric sector.
Further data from Super-K, K2K and MINOS will enable
more precise determination of these two-flavor mixing parameters, and
comparison of these experiments provide more stringent tests of the
three-flavor oscillation model.

\subsubsection{Direct Observations of the Oscillatory Behavior}
\label{sec:loverE}

	One of the most explicit tests of the oscillation model is to observe
the oscillations themselves---up until recently, no measurement of flavor
transformation could show the expected oscillation with baseline and energy
(L/E) which is a fundamental prediction of the model.  That has changed
now that the Super-Kamiokande (SK) collaboration has shown an ``oscillation
dip'' in the $L/E-$dependence, of the $\mu-$like atmospheric neutrino events
(see Fig.~\ref{fig:l-over-e}, Ref. ~\cite{SKdip04}) \footnote{The sample
used in the analysis of the $L/E$ dependence consists of $\mu-$like events
for which the relative uncertainty in the experimental determination of the
$L/E$ ratio does not exceed 70\%.}, $L$ and $E$ being the distance traveled
by neutrinos and the neutrino energy.  As is well known, the SK atmospheric
neutrino data are best described in terms of dominant two-neutrino $\nu_{\mu}
\rightarrow \nu_{\tau}$ ($\bar{\nu}_{\mu} \rightarrow \bar{\nu}_{\tau}$)
vacuum oscillations with maximal mixing, $\sin^22\theta_{23} \cong 1$
with $\deltaatm$ being the neutrino mass squared difference responsible
for the atmospheric $\nu_{\mu}$ and $\bar{\nu}_{\mu}$ oscillations.
This result represents the first ever observation of a direct oscillatory
dependence of $L/E$.

\begin{figure}[h!t]
\begin{minipage}{3in}
\begin{centering}
\includegraphics[height=2.8in,bb=22 189 498 609]{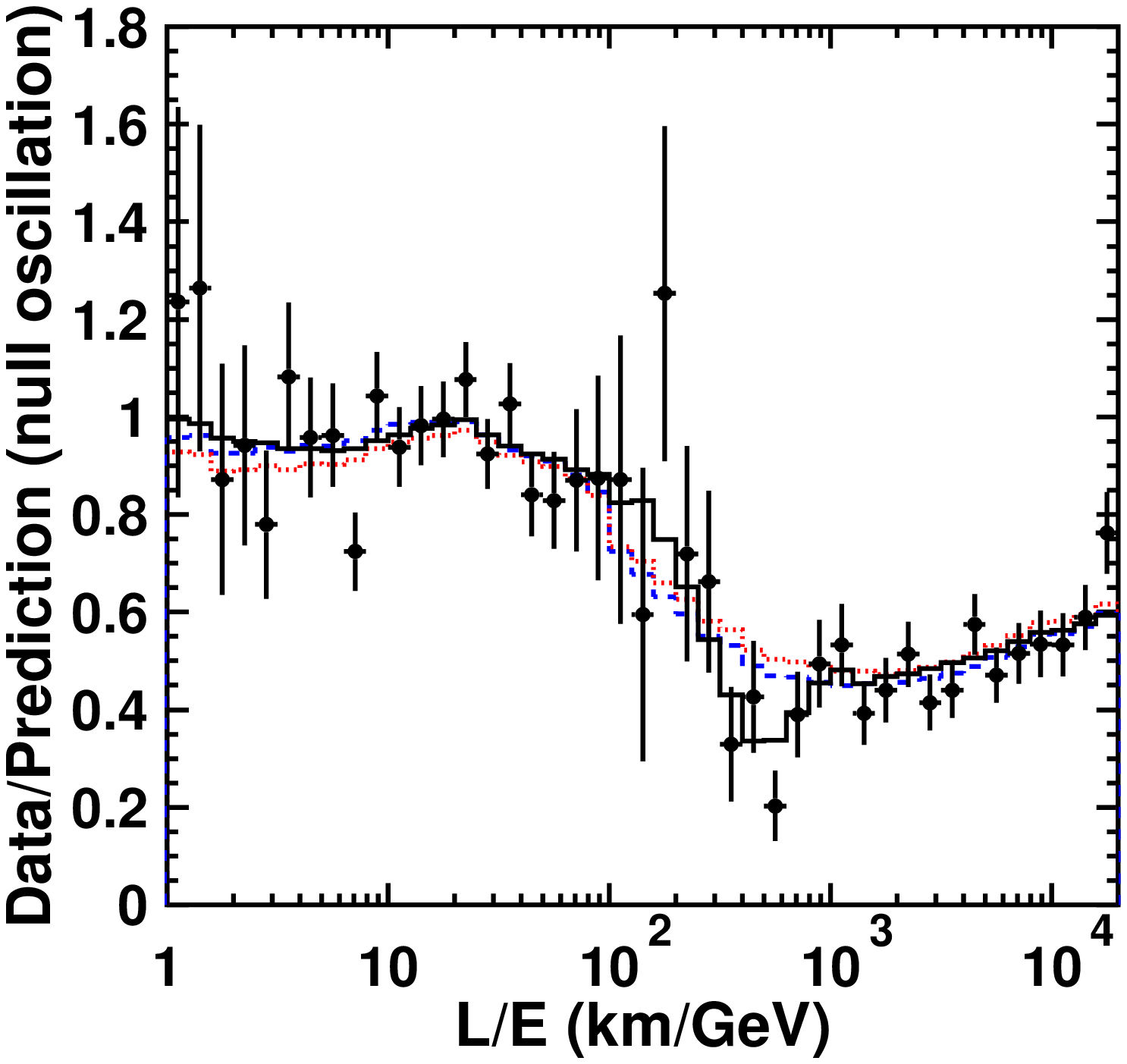}
\end{centering}
\end{minipage}
\begin{minipage}{3in}
\begin{centering}
\includegraphics[height=2.7in,bb=22 189 498 609]{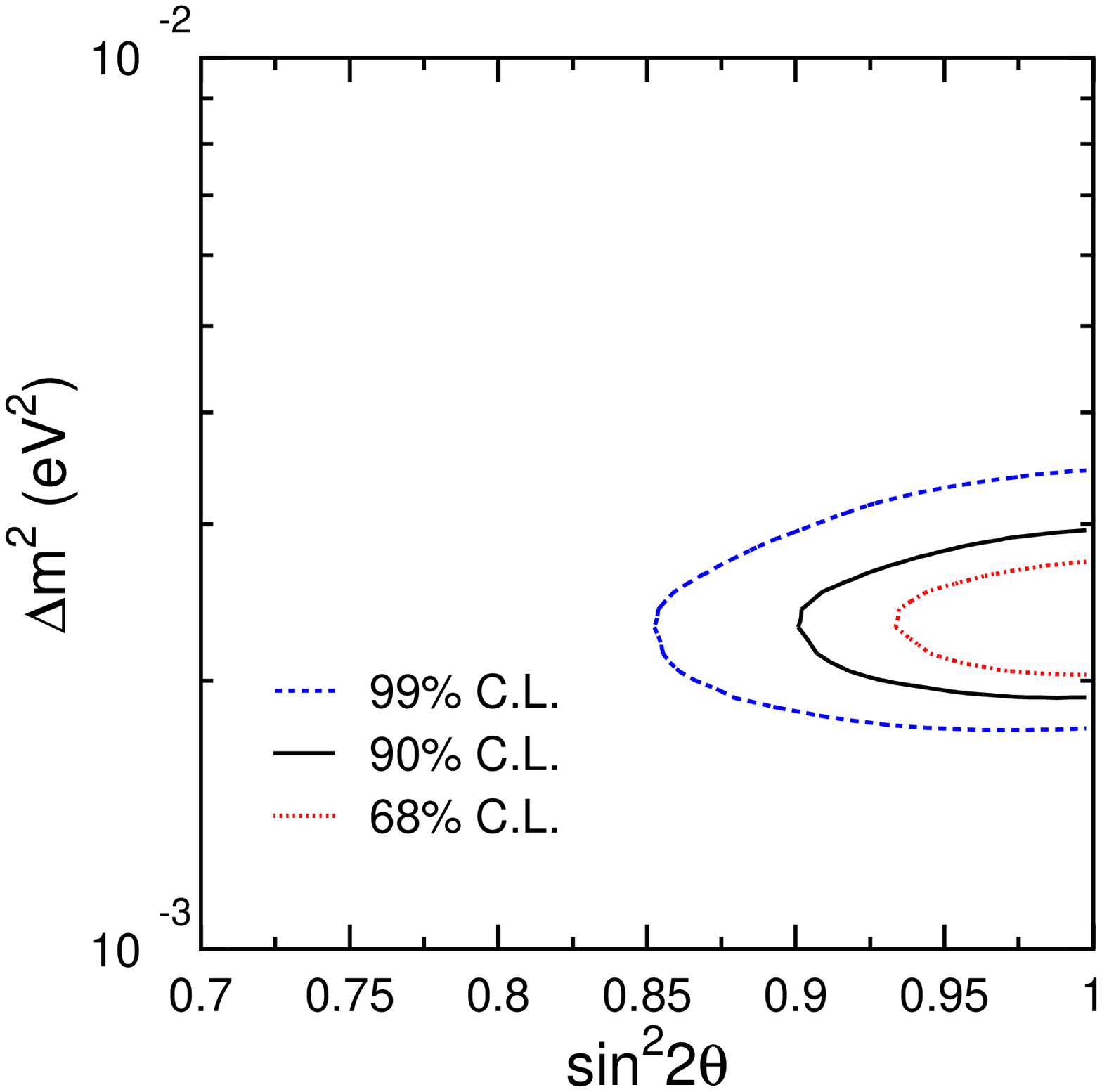}
\end{centering}
\end{minipage}
\caption{\label{fig:l-over-e} Left: ratio of data to Monte Carlo events
without neutrino oscillation (points) as a function of reconstructed
L/E, with best-fit expectation for two-flavor 
 $\nu_{\mu} \rightarrow \nu_{\tau}$ 
oscillations (solid line).  The dashed and dotted lines show
disfavored decay and decoherence models.  Right: corresponding
allowed oscillation parameter regions.
Figures taken from Ref.~\cite{SKdip04}.
}
\end{figure}

Future, larger-scale experiments such as UNO~\cite{uno} or
Hyper-Kamiokande~\cite{hyperk} should be able to see this kind of effect
with far greater signficance.  Figure~\ref{fig:loeuno} shows the oscillation
pattern which could be observed by the UNO detector.

\begin{figure}[h]
\begin{center}
\includegraphics[height=2.7in]{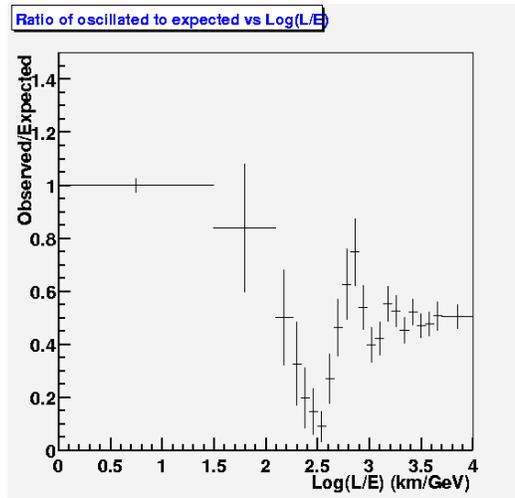}
\caption{Simulated oscillation pattern observable by UNO~\cite{uno}.  
\label{fig:loeuno}}
\end{center}
\end{figure}

\subsubsection{Searches for Sub-Dominant Effects}


   The $\nu_{\mu} \rightarrow \nu_{e}$ ($\bar{\nu}_{\mu} \rightarrow
\bar{\nu}_{e}$) and $\nu_{e} \rightarrow \nu_{\mu (\tau)}$ ($\bar{\nu}_{e}
\rightarrow \bar{\nu}_{\mu (\tau)}$) subdominant oscillations of
atmospheric neutrinos should exist and their effects could be observable
if genuine three-flavor-neutrino mixing takes place in vacuum, i.e., if
$\sin^22\theta_{13} \neq 0$, and if $\sin^22\theta_{13}$ is sufficiently
large~\cite{SP3198,AkhDig,core}.  The subdominant $\deltasol$ effects
depend crucially on the value of $\theta_{23}$, i.e. whether it is larger or
smaller than 45$^o$\footnote{It turns out that if $\theta_{13}$ is very small
($\sin^{2}{ 2 \theta_{13}} \ltap 0.01$), the only effect able to discriminate
the octant of $\theta_{23}$ is the one related to subdominant atmospheric
$\deltasol$ oscillations, as will be described later in the section.}.
These effects, as those associated with $\delta$ (CP-violating phase),
only show up at sub-GeV energies, for which the oscillation length due to
$\deltasol$ becomes comparable to the typical distances for atmospheric
neutrinos crossing the Earth.

In addition, if $\sin^22\theta_{13}$ is sufficiently large,
subdominant effects should exist in the multi-GeV range too. In this
case, $\nu_{\mu} \rightarrow \nu_{e}$ 
($\bar{\nu}_{\mu} \rightarrow \bar{\nu}_{e}$)
and $\nu_{e} \rightarrow \nu_{\mu (\tau)}$
($\bar{\nu}_{e} \rightarrow \bar{\nu}_{\mu (\tau)}$)
transitions of atmospheric neutrinos are amplified by Earth matter
effects. But matter affects neutrinos and
antineutrinos differently, and thus the study of these subdominant effects can
provide unique information (see Section~\ref{sec:masshier}).

The analytic analyses of references~\cite{3nuSP88,SPNu98} imply that in
the case under study the effects of the $\nu_{\mu} \rightarrow \nu_{e}$,
$\bar{\nu}_{\mu} \rightarrow \bar{\nu}_{e}$, and $\nu_{e} \rightarrow
\nu_{\mu (\tau)}$, $\bar{\nu}_{e} \rightarrow \bar{\nu}_{\mu (\tau)}$,
oscillations i) increase with the increase of $s^2_{23}$ and are maximal
for the largest allowed value of $s^2_{23}$, ii) should be considerably
larger in the multi-GeV samples of events than in the sub-GeV samples, iii)
in the case of the multi-GeV samples, they lead to an increase of the rate
of $e-$like events and to a slight decrease of the $\mu-$like event rate.
This analysis suggests that in water-\v{C}erenkov detectors, the quantity
most sensitive to the effects of the oscillations of interest should be
the ratio of the $\mu-$like and $e-$like multi-GeV events (or event rates),
$N_{\mu}/N_{e}$.

  The magnitudes of the effects we are interested in depend also on
the 2-neutrino oscillation probabilities.  
In the case of oscillations in vacuum we have
$P_{2\nu}(\deltaatm, \theta_{13};E,\theta_{n}) = \bar{P}_{2\nu}(\deltaatm,
\theta_{13};E,\theta_{n}) \sim \sin^22\theta_{13}$. Given the existing limits
on $\sin^22\theta_{13}$, the probabilities $P_{2\nu}$ and $\bar{P}_{2\nu}$
cannot be large if the oscillations take place in vacuum. However, $P_{2\nu}$
or $\bar{P}_{2\nu}$ can be strongly enhanced by the Earth matter effects 
(see Section~\ref{sec:masshier}).


For small $\theta_{13}$ the only effect able to discriminate the octant
of $\theta_{23}$ is associated with subdominant $\Delta m^2_{21}$ 
oscillations which are neglected in the hierarchical approximation
used in the standard three-neutrino oscillation analysis.
This effect can be understood in terms of approximate analytical expressions
developed in Ref.~\cite{alexei,ournohier}.  

The sensitivity for future experiments can be determined by
constructing a $\chi^2$ which is a function of the oscillation
parameters and the data (see Ref.~\cite{ournohier}), and evaluating
the difference in $\chi^2$ for ``false'' and ``true'' $\theta_{23}$
minima.  Figure~\ref{fig:disc} summarizes the results of
ref~\cite{ournohier}: it shows that unless $\theta_{23}$ is very close
to maximal mixing, there is good discrimination power from a high
statistics future atmospheric neutrino experiment. This effect is much
increased if the theoretical uncertainties on the atmospheric fluxes
and the interaction cross section as well as the experimental
systematic uncertainties are reduced.

\begin{figure}
\begin{center}
\includegraphics[height=8cm]{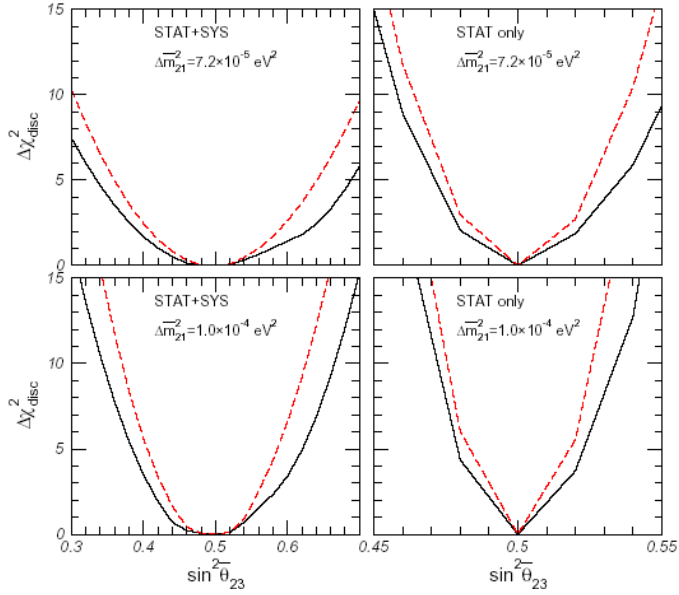}
\caption{$\Delta\chi^2$ between the
``false'' and ``true'' $\theta_{23}$ minima 
for future ATM (+OTHERS)  experiments as a function of the simulated 
$\overline{\sin^2\theta_{23}}$. The upper (lower) panels are for 
the simulated parameters :
$\overline{\Delta m^2_{21}}=7.2\; (10) \; \times 10^{-5}\;{\rm eV}^2$, 
$\overline{\tan^2\theta_{21}}=0.42$, 
$\overline{\Delta m^2_{31}}=2.2\times 10^{-3}\; {\rm eV}^2$  
$\overline{\sin^2\theta_{13}}=0$, 
$\overline{\delta}=0$.
The left panel assumes 20 times SK statistics but the same
theoretical and systematic errors. In the right panel, no theoretical
and systematic errors are included.  \label{fig:disc}}
\end{center}
\end{figure}

\subsubsection{Other Transformation Hypotheses}

	As discussed in Section~\ref{sec:npsol}, any new physics which
leads to differences in neutrino propagation may lead to neutrino
transformation effects.  Although these effects are now all excluded
as the dominant source of flavor transformation in both the solar
and atmospheric sectors, in the atmospheric sector they are severely
restricted even at the sub-dominant level.

The atmospheric neutrino data span several decades in neutrino energy
and distance. As a consequence it is very sensitive to these forms of
new physics.  The question arises, therefore, at what point the possible
presence of these forms of new physics, even if sub-dominant, may affect
the derived ranges of masses and mixing from the oscillation analysis
of the atmospheric data. Or in other words, to what level our present
determination of the neutrino masses and mixing is robust under the presence
of phenomenologically allowed new physics effects.

This question has been recently  answered in Ref.~\cite{atmnp} in 
which  a global analysis
of the atmospheric is performed with $\nu_\mu \to \nu_\tau$
transitions driven by neutrino masses and mixing in the presence of
these forms of new physics. 

In general when both neutrino masses and mixing and some of these
generic forms of new physics coexist, the evolution of the neutrinos
is governed by the equation
\begin{equation} \label{eq:hamil}
    \mathbf{H}_\pm \equiv
    \frac{\Dmq}{4 E}
    { \mathbf{U}_\theta}
    \left(\begin{array}{cc}
	-1 & ~0 \\
	\hphantom{-}0 & ~1
    \end{array}\right)
    { \mathbf{U}_\theta^\dagger}
    +\sigma_n^\pm \frac{{\Dlt_n}\, { E^n}}{2}
    { \mathbf{U}_{\xi_n}}
    \left(\begin{array}{cc}
	-1 & ~0 \\
	\hphantom{-}0 & ~1
    \end{array}\right)
    {\mathbf{U}_{\xi_n}^\dagger} \;,
\end{equation}
where $\Dmq$ is the mass--squared difference between the two neutrino
mass eigenstates, $\sigma_n^\pm$ accounts for a possible relative
sign of the NP effects between neutrinos and antineutrinos and
$\Dlt_n$ parametrizes the size of the NP terms. 
$\mathbf{U}_\theta$  ($\mathbf{U}_{\xi_n,\pm\eta_n}$) 
is the rotation matrices between the flavor states and the 
mass eigenstates (NP eigenstates).  In general
a non-vanishing relative phase $\eta_n$ is also possible. 

For Violation of Equivalence Principle
\begin{displaymath}    
{
 \Dlt_1 = 2 |\phi|(\gamma_1- \gamma_2) 
\equiv 2 |\phi| \Delta\gamma} \,,
    \qquad \sigma_1^+ = \sigma_1^- \,.
\end{displaymath}
{For Violation of Lorentz Invariance}: 
\small 
\begin{displaymath}    
    { \Dlt_1 = (v_1- v_2)\equiv\,\delta v} \,,
    \qquad \sigma_1^+ = \sigma_1^- \,.
\end{displaymath}
{\normalsize For Coupling to a space-time torsion field}
\small 
\begin{displaymath}
    { \Dlt_0= Q (k_1- k_2)\equiv Q \,\delta k} \,,
    \qquad \sigma_0^+ = \sigma_0^- \,.
\end{displaymath}
{\normalsize For Violation of Lorentz Invariance via CPT violation}
\small
\begin{displaymath}
    { \Dlt_0 = b_1-b_2 \equiv \delta b} \,,
    \qquad 
\sigma_0^+ = -\sigma_0^-
\end{displaymath}
{\normalsize For NSNI}
\begin{displaymath}
 { \Dlt_0} = 2\sqrt{2}\, G_F \, N_f(\vec{r})
{ \sqrt{\varepsilon_{\mu\tau}^2+
\frac{(\varepsilon_{\mu\mu}-\varepsilon_{\tau\tau})^2}{4}}}
    \qquad 
{ \sin^2 2\xi=\frac{\varepsilon_{\mu\tau}}
{\sqrt{\varepsilon_{\mu\tau}^2+\frac{(
\varepsilon_{\mu\mu}-\varepsilon_{\tau\tau})^2}{4}}}}
    \qquad \sigma_0^+ = -\sigma_0^-
\end{displaymath}
\normalsize

In all these scenarios the oscillation probabilities can
be written as:
\begin{equation} \label{eq:prob}
    P_{\nu_\mu \to \nu_\mu} = 1 - P_{\nu_\mu \to \nu_\tau} =
    1 - \sin^2 2\Theta \, \sin^2 \left( 
    \frac{\Dmq L}{4E} \, \mathcal{R} \right) \,.
\end{equation}
where the correction to the $\Dmq$-OSC wavelength, $\mathcal{R}$, and
to the global mixing angle, $\Theta$, verify
\begin{eqnarray}
    {\mathcal R} \cos 2\Theta \,
    & = \cos 2\theta + R_n\, \cos 2\xi_n \,,
    \\ 
    {\mathcal R} \sin 2\Theta \
    & =  |\sin 2\theta 
    + R_n \,\sin 2\xi_n \, e^{i\eta_n} | \,,
\end{eqnarray}
with $R_n$ being the ratio between the NP--induced and 
$\Dmq$--induced contributions to the oscillation wavelength.
For $P_{\bar{\nu}_\mu \to \bar{\nu}_\mu}$ the same expressions hold
with the exchange $\sigma_n^+ \to \sigma_n^-$ and $\eta_n \to
-\eta_n$.

Ther results of the analysis of the present atmospheric data is
shown in Fig.~\ref{fig:atmnp} and ~\ref{fig:nsireg}.  
\begin{figure}
\begin{center}
    \includegraphics[width=4.in]{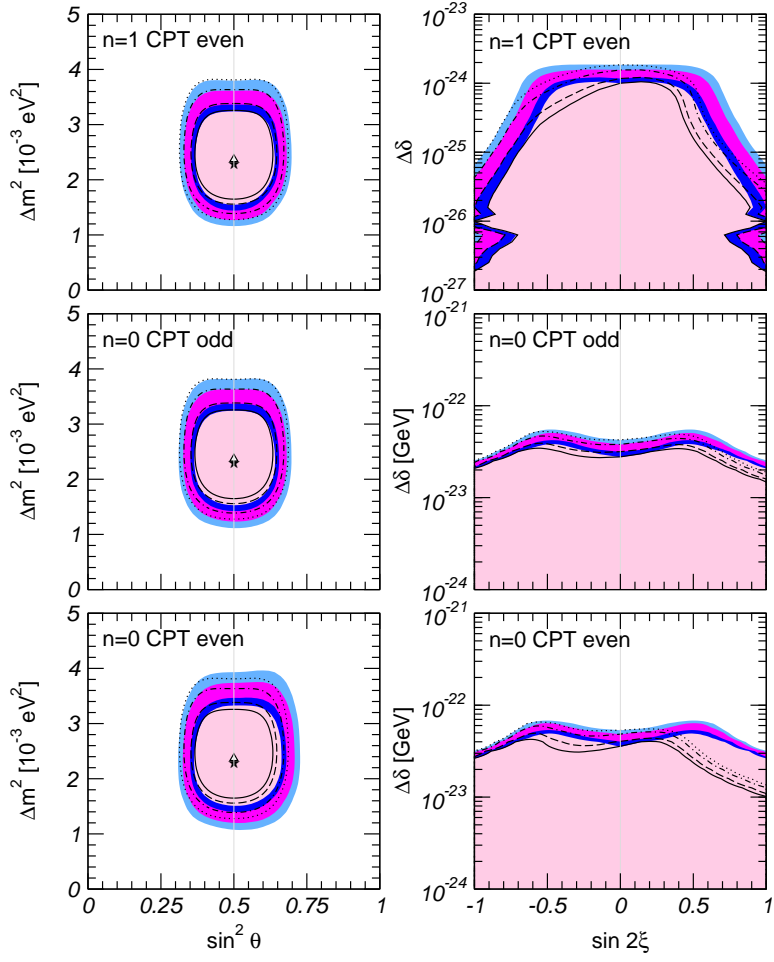}
    \caption{\label{fig:atmnp}%
      Allowed parameter regions for the analysis of atmospheric and
      K2K data in presence of $\nu_\mu \to \nu_\tau$ oscillations and
      different NP effects as labeled in the figure. 
The different contours correspond to the
      allowed regions 
at 90\%, 95\%, 99\% and
      $3\sigma$ CL. The filled areas in the left panels show the
      projected two-dimensional allowed region on the oscillation
      parameters $\Dmq$--$\sin^2\theta$ plane. The best fit point is
      marked with a star. We also show the
      lines corresponding to the contours in the absence of new
      physics and mark with a triangle the position of the best fit
      point. 
      The regions on the right panels show the allowed values for the
      parameters characterizing the strength and mixing of the NP. The
      full regions corresponds to arbitrary values of the phase
      $\eta_n$ while the lines correspond to the case $\eta_n \in
      \{0,\,\pi\}$.}
\end{center}  
\end{figure}

The figure demonstrates that the data does not show any evidence of 
presence of NP even as a sub-dominant effect and  
the robustness of the allowed ranges of
mass and mixing derived from the analysis of atmospheric and K2K data
under the presence of these generic NP effects. Thus the analysis
allow us to derive well-defined upper bounds on the NP strength. 
At 95\% CL
\begin{eqnarray}
&&    |\delta v| \leq 8.1\times 10^{-25} \nonumber \\[+0.05cm]
&&   |\phi\, \Delta \gamma| \leq 4.0\times 10^{-25} \nonumber \\[+0.05cm]
&&    |\delta b| \leq 3.2\times 10^{-23} \nonumber \\[+0.05cm]
&&    |Q \,\delta k| \leq 4.0\times 10^{-23} \nonumber \\[+0.05cm]
&&    |\varepsilon_{\mu\mu}-\varepsilon_{\tau\tau}|\leq 0.013~\nonumber \\ [+0.05cm]
&&     |\varepsilon_{\mu\tau}|\leq 0.034~
\nonumber 
\end{eqnarray}

These limits are among the strongest constraints on these forms of NP.

Next we illustrate the attainable sensitivity at a future atmospheric
neutrino experiment. 
In order to show this we have assumed a SK-like detector with 20 times the
present SK statistics with the same theoretical and systematic uncertainties
as SK. The results are shown in Figs.~\ref{fig:atmnpfut1} and 
~\ref{fig:atmnpfut2}. 
In Fig~\ref{fig:atmnpfut1} we have assumed that 
the observed rates will correspond to the present SK data. The figure
illustrates that within the fluctuations in the existing data there 
is still room for NP. In Fig~\ref{fig:atmnpfut2} we have assumed that 
the observed rates will correspond to the present best fit point for
pure $\Delta m^2$ oscillations. The figure
illustrates the possible improvement in the constraints on the NP
strength due to the improved statistics.
\begin{figure}
\begin{center}
    \includegraphics[width=5in]{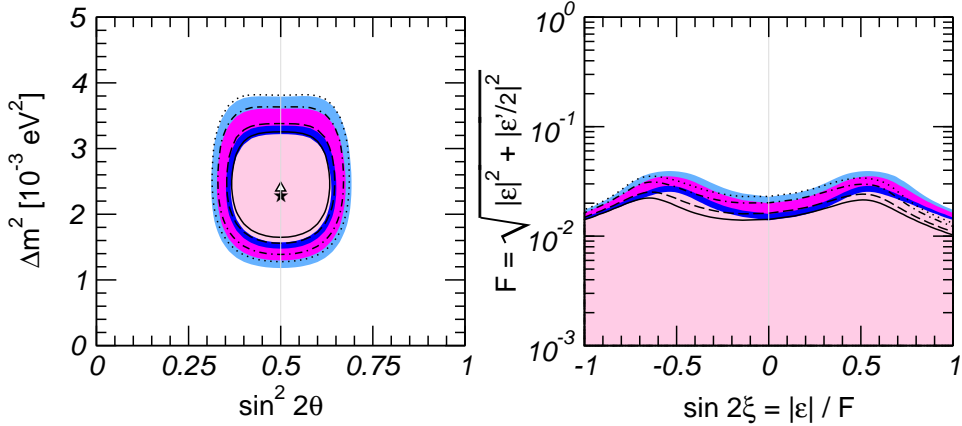}
    \caption{\label{fig:nsireg}%
      Same as Fig.~\ref{fig:atmnp} for the case of $\Delta m^2$-OSC+NSI.
For the sake of concretness we  have assumed NSI to $d$-quarks} 
\end{center}
\end{figure}
\begin{figure}
\begin{center}
    \includegraphics[width=5in]{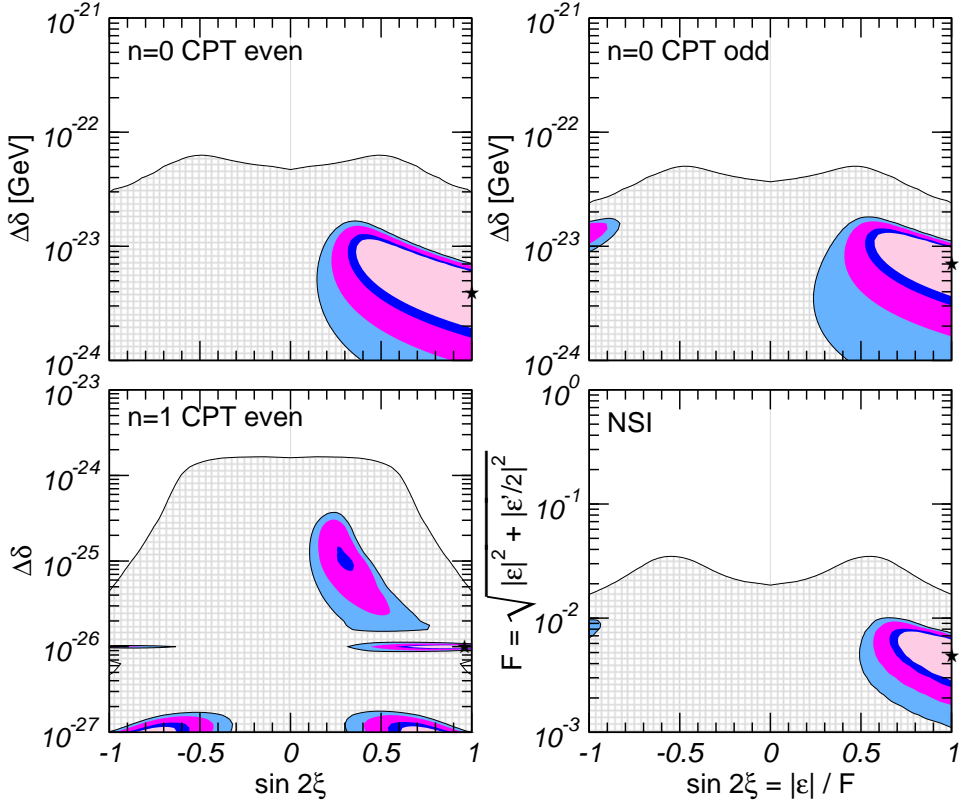}
    \caption{\label{fig:atmnpfut1}
Allowed regions for the NP parameters for a future atmospheric neutrino
experiment with 20 SK exposure and 
observed rates corresponding to the
present SK data. 
The filled contours show the allowed region 
at 90\%, 95\%, 99\% and
      $3\sigma$ CL. The hatched area is the presently excluded region at 
$3\sigma$.}
\end{center}
\end{figure}
\begin{figure}
\begin{center}
    \includegraphics[width=5in]{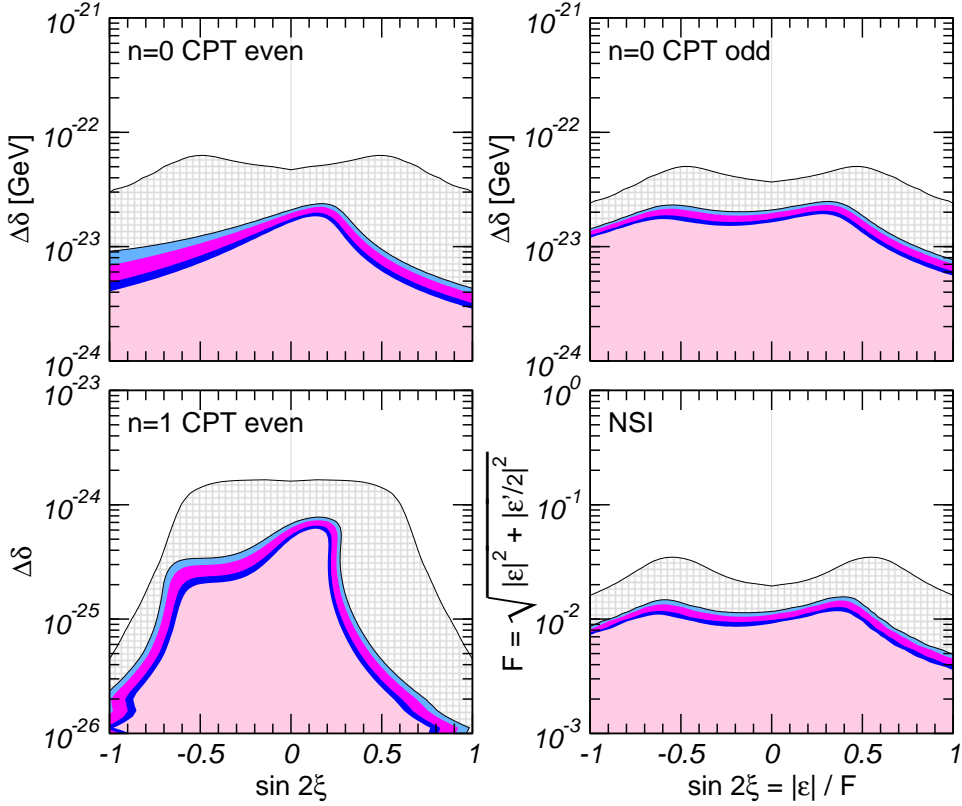}
    \caption{\label{fig:atmnpfut2}
Allowed regions for the NP parameters for a future atmospheric neutrino
experiment with 20 SK exposure and observed rates corresponding to the
simulated $\Delta m^2$ oscillation best fit point.
The filled contours show the allowed region 
at 90\%, 95\%, 99\% and  $3\sigma$ CL. The hatched area is the 
presently excluded region at $3\sigma$.}
\end{center}
\end{figure}
where the correction to the $\Dmq$-OSC wavelength, $\mathcal{R}$, and
to the global mixing angle, $\Theta$, verify
\begin{eqnarray}
    {\mathcal R} \cos 2\Theta \,
    & = \cos 2\theta + R_n\, \cos 2\xi_n \,,
    \\ 
    {\mathcal R} \sin 2\Theta \
    & =  |\sin 2\theta 
    + R_n \,\sin 2\xi_n \, e^{i\eta_n} | \,,
\end{eqnarray}
with $R_n$ being the ratio between the NP--induced and 
$\Dmq$--induced contributions to the oscillation wavelength.
For $P_{\bar{\nu}_\mu \to \bar{\nu}_\mu}$ the same expressions hold
with the exchange $\sigma_n^+ \to \sigma_n^-$ and $\eta_n \to
-\eta_n$.

A more recent analysis~\cite{fried}, which included the effects of all three 
flavors found that non-standard interactions could play a role as large
as Standard Model oscillations. If so, the mixing angle $\theta_{23}$ could be 
non-maximal, and $\Delta m_{23}^2$ somewhat higher than its present value.

\subsection{The Matter Effect and the Mass Hierarchy}
\label{sec:masshier}

If $\sin^2\theta_{13}\neq 0$, the Earth matter effects can resonantly
enhance either the  $\nu_{\mu} \rightarrow \nu_e$ and $\nu_{e}
\rightarrow \nu_{\mu}$, or the $\bar{\nu}_{\mu} \rightarrow \bar{\nu}_e$
and $\bar{\nu}_{e} \rightarrow \bar{\nu}_{\mu}$ transitions, depending on
the sign of $\deltaatm$. For $\deltaatm > 0$, the $\nu_{\mu} \rightarrow
\nu_e$ and $\nu_{e} \rightarrow \nu_{\mu}$ transitions in the Earth lead to
a reduction of the rate of the multi-GeV $\mu^{-}$ events, observable in
detectors with charge discrimination, with respect to the case of absence
of these transitions~\cite{SP3198,SPNu98,core,107}.  If $\deltaatm <
0$, the $\mu^{+}$ event rate will be reduced.  Correspondingly, as an
observable which is sensitive to the Earth matter effects, and thus to
the value of $\sin^2\theta_{13}$ and the sign of $\deltaatm$, as well as
to $\sin^2\theta_{23}$, we can consider the nadir-angle distributions of
the $N(\mu^{-})-N(\mu^+)$ asymmetry of the multi-GeV $\mu^-$ and $\mu^+$
event rates.  

   {\it We note that atmospheric neutrino experiments are the only
method other than long baseline experiments of determining the mass hierarchy,
and for some values of the mixing parameters, they may be the only method.}

In a water-Cerenkov detector the distinction between neutrino and
antineutrino events is not possible (on an event-by-event basis), and in
principle determining the type of neutrino mass hierarchy does not seem
feasible. However, due to the difference of cross sections for neutrinos
and antineutrinos, approximately 2/3 of the total rate of the $\mu-$like
and $e-$like multi-GeV atmospheric neutrino events in a water-Cerenkov
detector, i.e., $\sim 2N_{\mu}/3$ and $\sim 2N_e/3$, are due to neutrinos
$\nu_{\mu}$ and $\nu_e$, respectively, while the remaining $\sim 1/3$ of the
multi-GeV event rates, i.e., $\sim N_{\mu}/3$ and $\sim N_e/3$, are produced
by antineutrinos $\bar{\nu}_{\mu}$ and $\bar{\nu}_e$.  This implies that
the Earth matter effects in the multi-GeV samples of $\mu-$like events
will be smaller and that of $e-$like will be larger if $\deltaatm >
0$, i.e., if the neutrino mass spectrum is with normal hierarchy, than
if $\deltaatm < 0$ and the spectrum is with inverted hierarchy.  Thus,
the ratio $N_{\mu}/N_e$ of the multi-GeV $\mu-$like and $e-$like event
rates measured in the SK experiment could be sensitive, in principle,
to the type of the neutrino mass spectrum.

   It follows from the simple analysis of reference~\cite{mantle} that
Earth matter effects can amplify $P_{2\nu}$ significantly when the neutrinos
cross only the mantle i) for $E \sim (5 - 10)$ GeV, i.e., in the multi-GeV
range of neutrino energies, and ii) only for sufficiently long neutrino
paths in the mantle, i.e., for $\cos\theta_n \gtap 0.4$.  The magnitude of
the matter effects in the ratio $N_{\mu}/N_e$ of interest increases with
increasing $\sin^2\theta_{13}$.  
The same conclusions are valid for
the antineutrino oscillation probability $\bar{P}_{2\nu}$ in the case of
$\deltaatm < 0$. As a consequence, the ideal situation for distinguishing
the type of mass hierarchy would be a detector with charge discrimination,
such that neutrino interactions can be distinguished from those due to
antineutrinos.

In the case of atmospheric neutrinos crossing the Earth core, new resonant
effects become apparent. For $\sin^2\theta_{13} < 0.05$ and $\deltaatm
> 0$, we can have $P_{2\nu} \cong 1$ {\it only} due to the effect of
maximal constructive interference between the amplitudes of the the
$\nu_{e} \rightarrow \nu'_{\tau}$ transitions in the Earth mantle and in
the Earth core~\cite{SP3198,107,106}.  The effect differs from the MSW
effect~\cite{SP3198} and the resonances happen at lower energies, between the
resonance energies corresponding to the density in the mantle and that
of the core.  The {\it mantle-core enhancement effect} is caused by the
existence (for a given neutrino trajectory through the Earth core) of {\it
points of resonance-like total neutrino conversion}, $P_{2\nu} = 1$, in
the corresponding space of neutrino oscillation parameters~\cite{107,106}.
The location of these points determines the regions where $P_{2\nu}$ is
large, $P_{2\nu} \gtap 0.5$.  These regions vary slowly with the nadir
angle; they are remarkably wide in the nadir angle and are rather wide
in the neutrino energy~\cite{107}, so that the transitions of interest
produce noticeable effects: we have $\delta E/E \cong 0.3$ for the values
of $\sin^2\theta_{13}$ of interest.

  The effects of the mantle-core enhancement of $P_{2\nu}$ (or
$\bar{P}_{2\nu}$) increase rapidly with $\sin^2\theta_{13}$ as long as
$\sin^2\theta_{13}\ltap 0.01$, and should exhibit a rather weak dependence on
$\sin^2\theta_{13}$ for $0.01 \ltap \sin^2\theta_{13} < 0.05$.  If 3-neutrino
oscillations of atmospheric neutrinos take place, the magnitude of the
matter effects in the multi-GeV $e-$like and $\mu-$like event samples,
produced by neutrinos crossing the Earth core, should be larger than in
the event samples due to neutrinos crossing only the Earth mantle (but
not the core).  This is a consequence of the fact that in the energy range
of interest the atmospheric neutrino fluxes decrease rather rapidly with
energy (approximately as $E^{-2.7}$), while the neutrino interaction cross
section rises only linearly with $E$, and that the maximum of $P_{2\nu}$
(or $\bar{P}_{2\nu}$) due to the these new resonance-like effects takes
place at approximately one half the energy than that due to the MSW
effect for neutrinos crossing only the Earth mantle (e.g., at $E \cong (4.2 -
4.7)~{\rm GeV}$ and $E \cong 10~{\rm GeV}$, respectively, for $\deltaatm =
3\times 10^{-3}~{\rm eV^2}$).

Thus, summarizing, from the study of the Earth matter effects on atmospheric
neutrinos one can conclude that: i) the medium effects, which discriminate
between neutrino and antineutrino propagation, can help to determine the
sign of the atmospheric $\deltaatm$~\cite{LBL}; ii) for $\sin\theta_{13}
= 0$ electron neutrinos decouple from the oscillations of the atmospheric
neutrinos in matter, whereas they mix with the third (heaviest) mass
eigenstate neutrino and take part in the atmospheric neutrino oscillations
if $\sin\theta_{13} \not= 0$, although their mixing with the first
(lightest) mass eigenstate neutrino still vanishes; iii) non-resonant
medium effects are already apparent in the sub-dominant channels $\nu_e
\rightarrow \nu_{\mu}$ and $\bar{\nu}_e \rightarrow \bar{\nu}_{\mu}$,
for baselines $L \sim 3000$ km, in both the mixing and oscillation phase
shift (see also refs.~\cite{mantle,AMMS99}); iv) in order for the medium
effects in the muon neutrino survival probability to be observable, the
resonant MSW effect in the $\nu_{e(\mu)} \rightarrow \nu_{\mu(e)}$ or
$\bar{\nu}_{e(\mu)} \rightarrow \bar{\nu}_{\mu(e)}$ transitions must be
operational, which requires baselines larger than $L \sim 7000$ km, the
optimal baseline being a function of the value of $\sin\theta_{1 3}$; v)
taking into account the initial atmospheric $\nu_{\mu}$, $\bar{\nu}_{\mu}$,
and $\nu_{e}$, $\bar{\nu}_{e}$ fluxes and the relevant charged current
neutrino-nucleon deep inelastic scattering cross-sections, it was shown that
the matter-induced CPT-odd~\cite{Lang87} and CP-odd~\cite{LBL,mantle,AMMS99}
asymmetries are observable.

\noindent
\textbf{Observables: Water Cerenkov Detectors}\label{waterobs}
\hskip 0.6truecm
\begin{figure}
\begin{center}
\includegraphics[height=6cm]{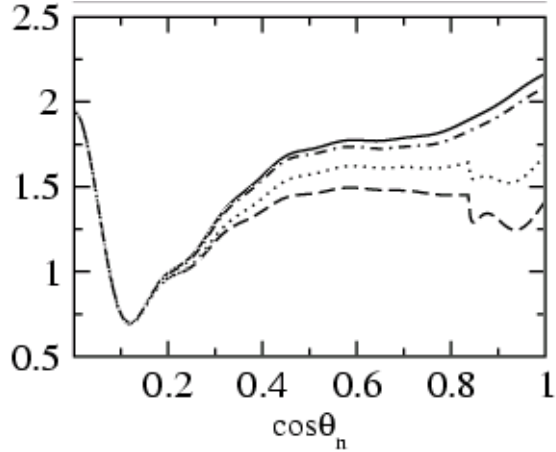}
\caption{\label{nadir} The dependence on $\cos\theta_n$ of the ratios of the
multi-GeV $\mu^-$ and $e^-$ like events (or event rates), integrated over
the neutrino energy in the interval $E = (2.0 - 10.0)$ GeV, in the cases
i) of 2-neutrino $\nu_{\mu} \rightarrow \nu_{\tau}$ and $\bar{\nu}_{\mu}
\rightarrow \bar{\nu}_{\tau}$ oscillations in vacuum and no $\nu_e$ and
$\bar{\nu}_e$ oscillations, $N^{2\nu}_{\mu}/N^{0}_{e}$ (solid lines), ii)
3-neutrino oscillations in vacuum of $\nu_{\mu}$, $\bar{\nu}_{\mu}$, $\nu_e$
and $\bar{\nu}_e$, $(N^{3\nu}_{\mu}/N^{3\nu}_{e})_{vac}$ (dash-dotted
lines), iii) 3-neutrino oscillations of $\nu_{\mu}$, $\bar{\nu}_{\mu}$
$\nu_e$ and $\bar{\nu}_e$ in the Earth and neutrino mass spectrum with
normal hierarchy $(N^{3\nu}_{\mu}/N^{3\nu}_{e})_{\rm NH}$ (dashed lines),
or with inverted hierarchy, $(N^{3\nu}_{\mu}/N^{3\nu}_{e})_{\rm IH}$
(dotted lines).  The results shown are for $|\deltaatm| = 3\times
10^{-3}~{\rm eV^2}$, $\sin^2\theta_{23} =~0.64$, and $\sin^22\theta_{13}
= 0.05$. Figure taken from Ref.~\cite{BerPalPet03}.  } 
\end{center}
\end{figure}


The predicted dependence on $\cos\theta_n$ of the ratios of the
multi-GeV $\mu-$ and $e-$ like events (or event rates), integrated
over the neutrino energy from the interval $E = (2.0 - 10.0)$ GeV, for
various cases, and for a particular choice of parameters, are shown in
Fig.~\ref{nadir}~\cite{BerPalPet03}\footnote{For an
analysis of the resonant effects in terms of the up-down asymmetry see
refs.~\cite{AkhDig,PalPetProc}.}.

\begin{figure}
\begin{center}
\includegraphics[height=6.3cm]{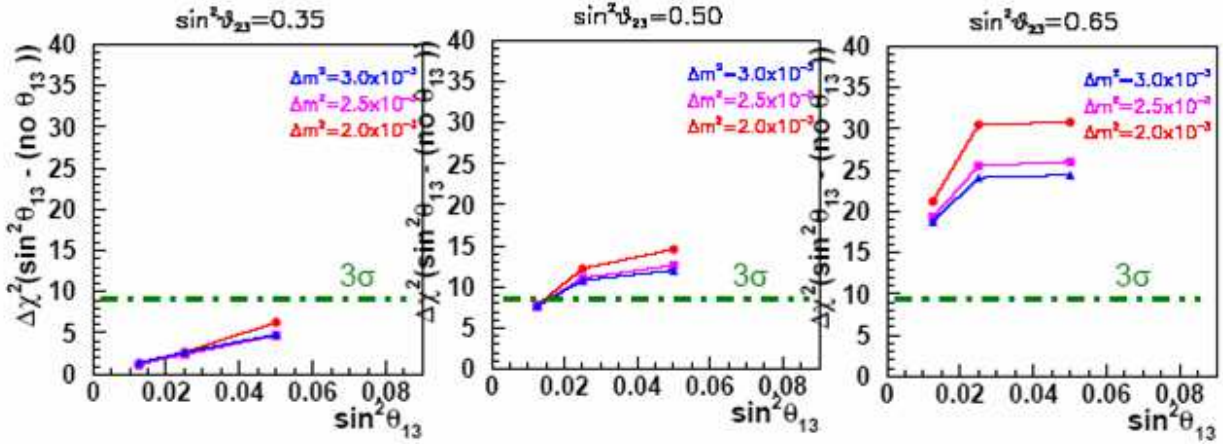}
\caption{\label{20SK} Statistical significance of a non-zero
  $\sin^2\theta_{13}$ with a water-Cerenkov detector with an
  exposure of 450 kton yrs (assuming normal hierarchy, i.e.,
  $\deltaatm > 0$) for $\deltaatm = 2.0 \times 10^{-3}~{\rm
  eV}^2~{(\rm red~line;)}~2.5 \times 10^{-3}~{\rm eV}^2~{\rm
  (pink~line); and}~3.0 \times 10^{-3}~{\rm eV}^2~{\rm (blue~line)}$
  and $\sin^2\theta_{23} = 0.35~{\rm (left~panel);}~0.50~{\rm (middle~
  panel)~and}~0.65~{\rm (right~panel)}$. Figure taken from
  Ref.~\cite{kajitanoon04}. 
}
\end{center}
\end{figure}

\begin{figure}
\begin{center}
\includegraphics[height=6.1cm]{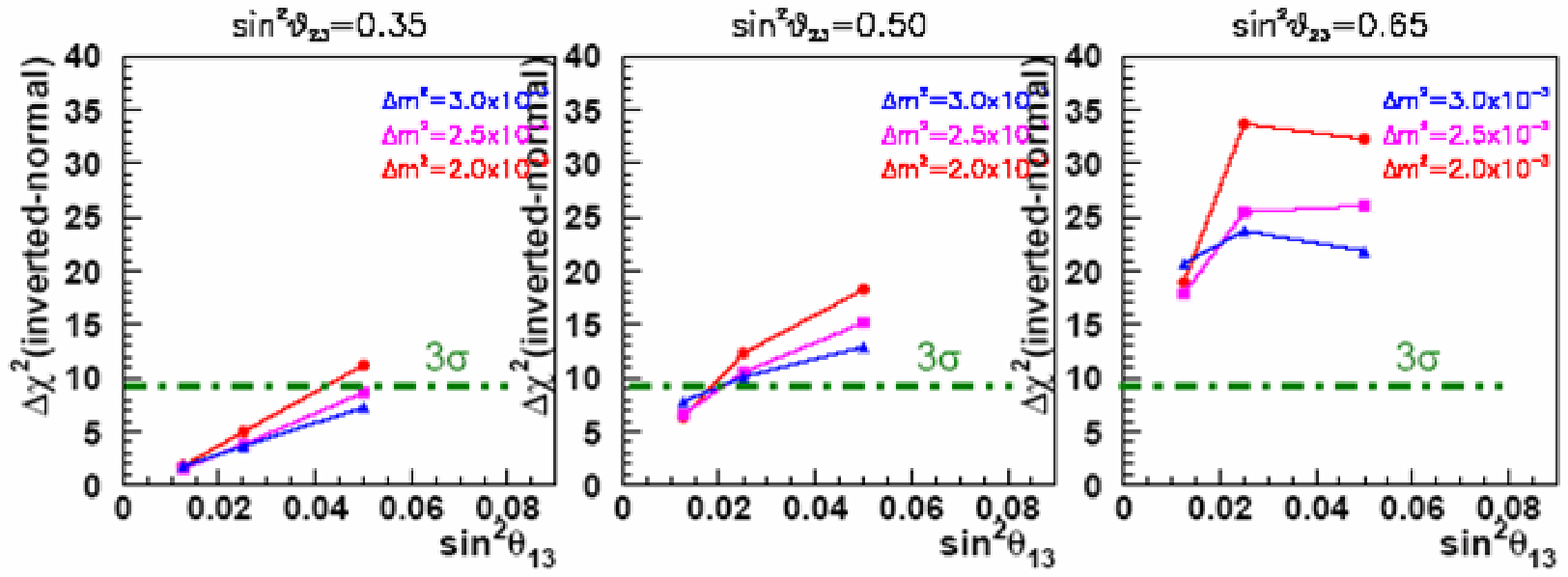}
\includegraphics[height=6.37cm]{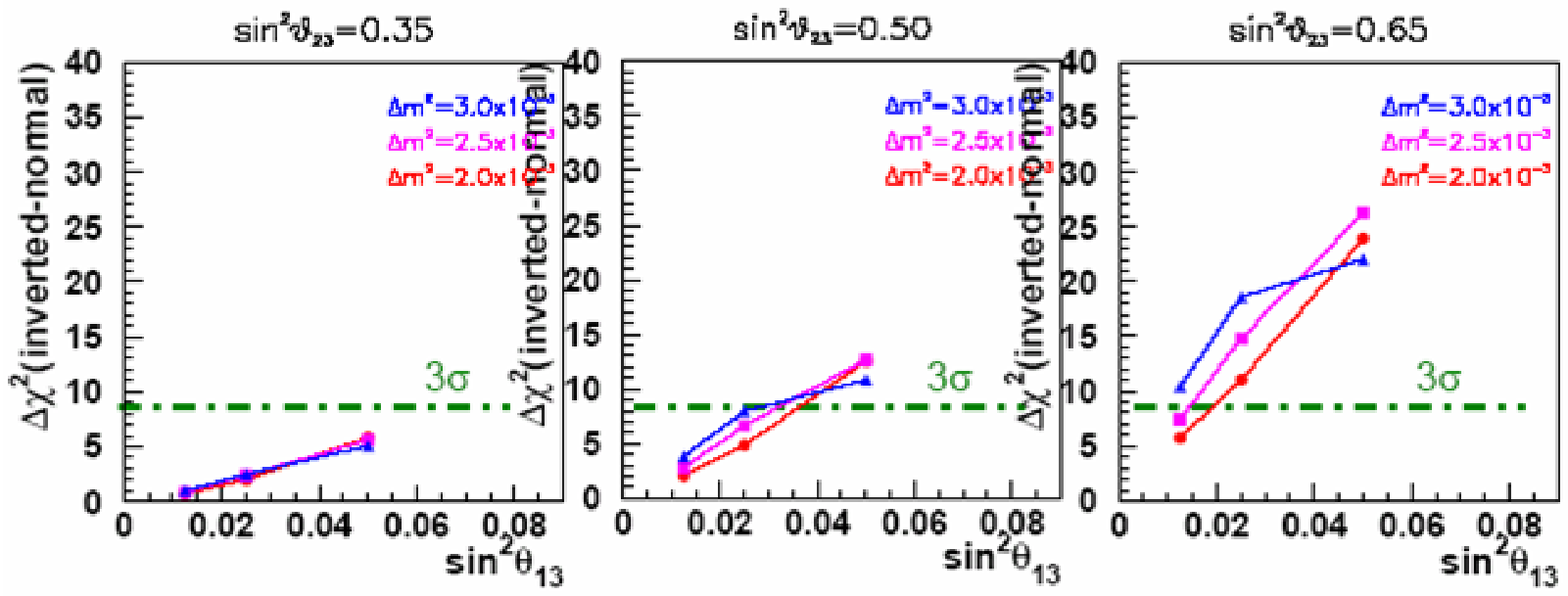}
\caption{\label{hier80SK} Statistical significance for
  measuring the sign of $\deltaatm$ for an exposure of 1.8 Mton yrs if
  the true hierarchy turns out to 
  be normal (upper plots) or inverted (lower plots), for different
  values of $|\deltaatm|$ and $\sin^2\theta_{23}$, as a function of
  $\sin^2\theta_{13}$ (same as in Fig.~\ref{20SK}). Figure taken from
  Ref.~\cite{kajitanoon04}.  
}
\end{center}
\end{figure}

In Fig.~\ref{20SK}, we show the statistical significance (assuming normal
hierarchy) for a non-zero $\sin^2{\theta_{13}}$ with 20 years of data taking
by Super-Kamiokande (450 kton yrs) for different values of $\deltaatm$
\cite{kajitanoon04}.  We can see that not only a large enough value of
$\theta_{13}$ is needed, but also $\sin^2{\theta_{23}} \gtap 0.5$. We
see that for $\sin^2{2\theta_{13}} \ltap 0.025$ there is an increase
of sensitivity with $\theta_{13}$, whereas for larger values a constant
sensitivity is obtained.  

In Fig.~\ref{hier80SK}, we show the possibilities to disentangle 
the type of mass hierarchy with 80 years of data from 
Super-Kamiokande (1.8 Mton yrs), but just about four for a detector
like UNO or less than two for HyperKamiokande. The upper plot shows
the case if the true neutrino mass hierarchy is normal and the lower
plot if it is inverted. The same set of parameters as in
Fig.~\ref{20SK} are used~\cite{kajitanoon04}. We see again that, even
with this huge exposure, $\sin^2{\theta_{23}} \gtap 0.5$ is needed and
that the case of inverted hierarchy is the most difficult to
distinguish, because just 1/3 of the (multi-GeV) events are being
affected by matter, in comparison to 2/3 in the case of normal
hierarchy.

\noindent
\textbf{Observables: Magnetized Detectors}

\begin{figure}
\begin{center}
\includegraphics[height=6cm]{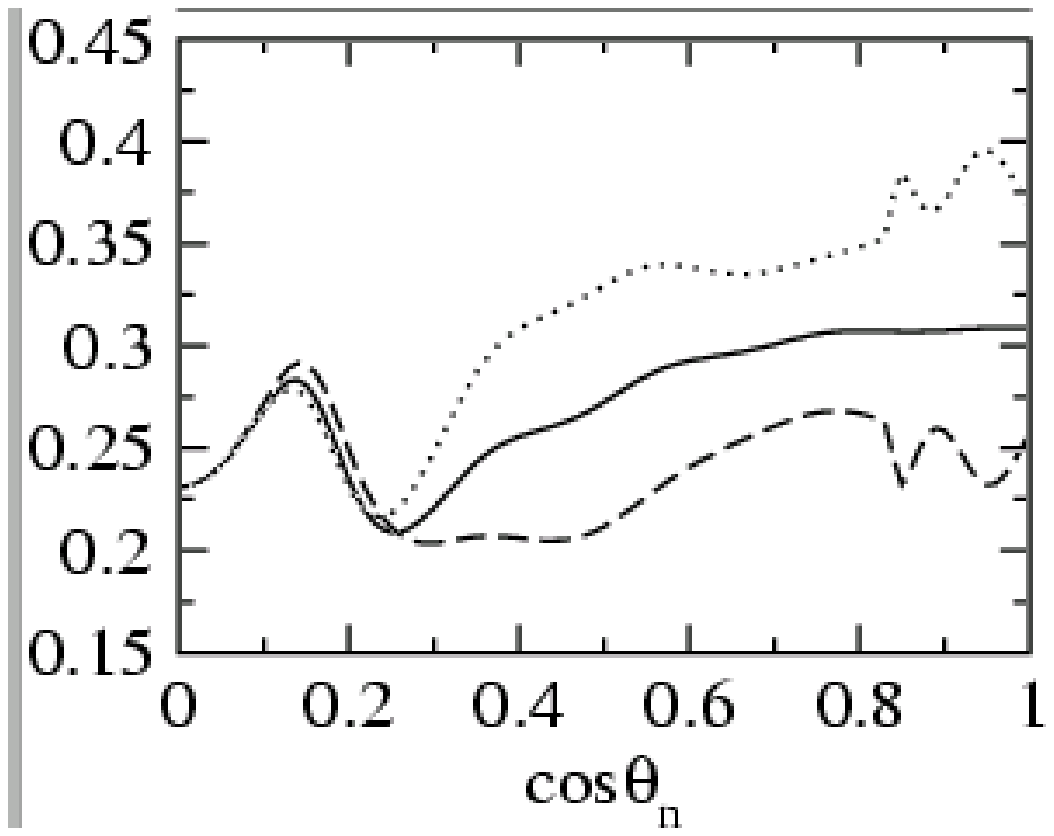}
\caption{\label{asymnadir}
The nadir angle distribution of the charge asymmetry, $A_{\mu^-\mu^+}
\equiv \frac{N(\mu^{-})-N(\mu^+)}{N(\mu^{-})+N(\mu^+)}$, measurable
in magnetized calorimeters, of the multi-GeV $\mu^--$ and $\mu^+-$
like events (or event rates), integrated over the neutrino (and muon)
energy in the interval $E = (2.0 - 10.0)$ GeV, in the cases i) of
two-neutrino $\nu_{\mu} \rightarrow \nu_{\tau}$ and $\bar{\nu}_{\mu}
\rightarrow \bar{\nu}_{\tau}$ oscillations in vacuum and no $\nu_e$ and
$\bar{\nu}_e$ oscillations, $A_{\mu^-\mu^+}^{2\nu}$ (solid lines), ii)
three-neutrino oscillations of $\nu_{\mu}$, $\bar{\nu}_{\mu}$ $\nu_e$
and $\bar{\nu}_e$ in the Earth and neutrino mass spectrum with normal
hierarchy $(A_{\mu^-\mu^+}^{3\nu})_{\rm NH}$ (dashed lines), or with
inverted hierarchy, $(A_{\mu^-\mu^+}^{3\nu})_{\rm IH}$ (dotted lines).
The results shown are for $|\deltaatm| = 3\times 10^{-3}~{\rm eV^2}$,
$\sin^2\theta_{23} =~0.64$, and $\sin^22\theta_{13} = 0.05$. Figure taken
from Ref.~\cite{PalPet}.  } 
\end{center}
\end{figure}


\begin{figure}
\begin{minipage}{3.2in}
\begin{centering}
\includegraphics[height=6.5cm]{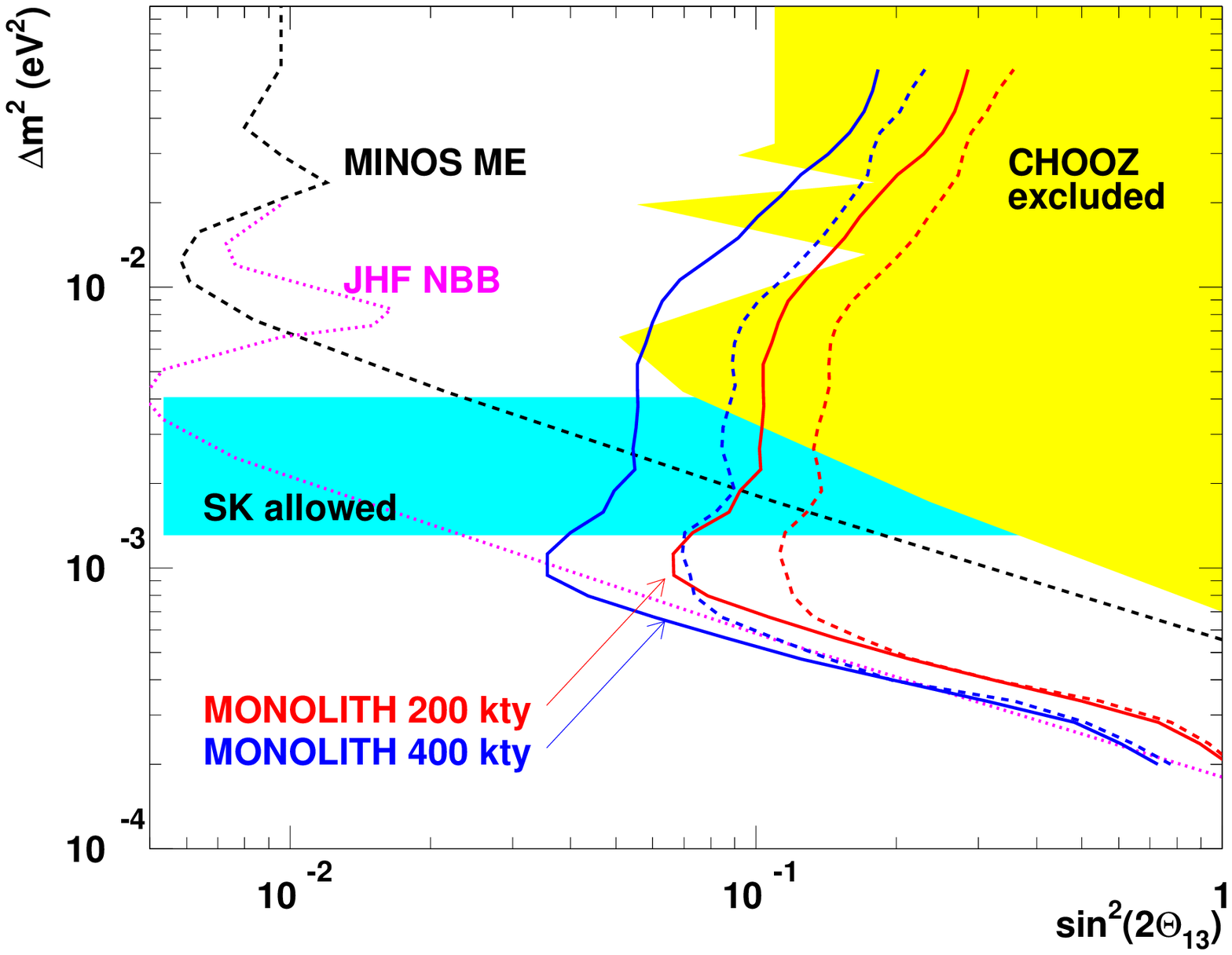}
\end{centering}
\end{minipage}
\begin{minipage}{3.2in}
\begin{centering}
\includegraphics[height=6.5cm]{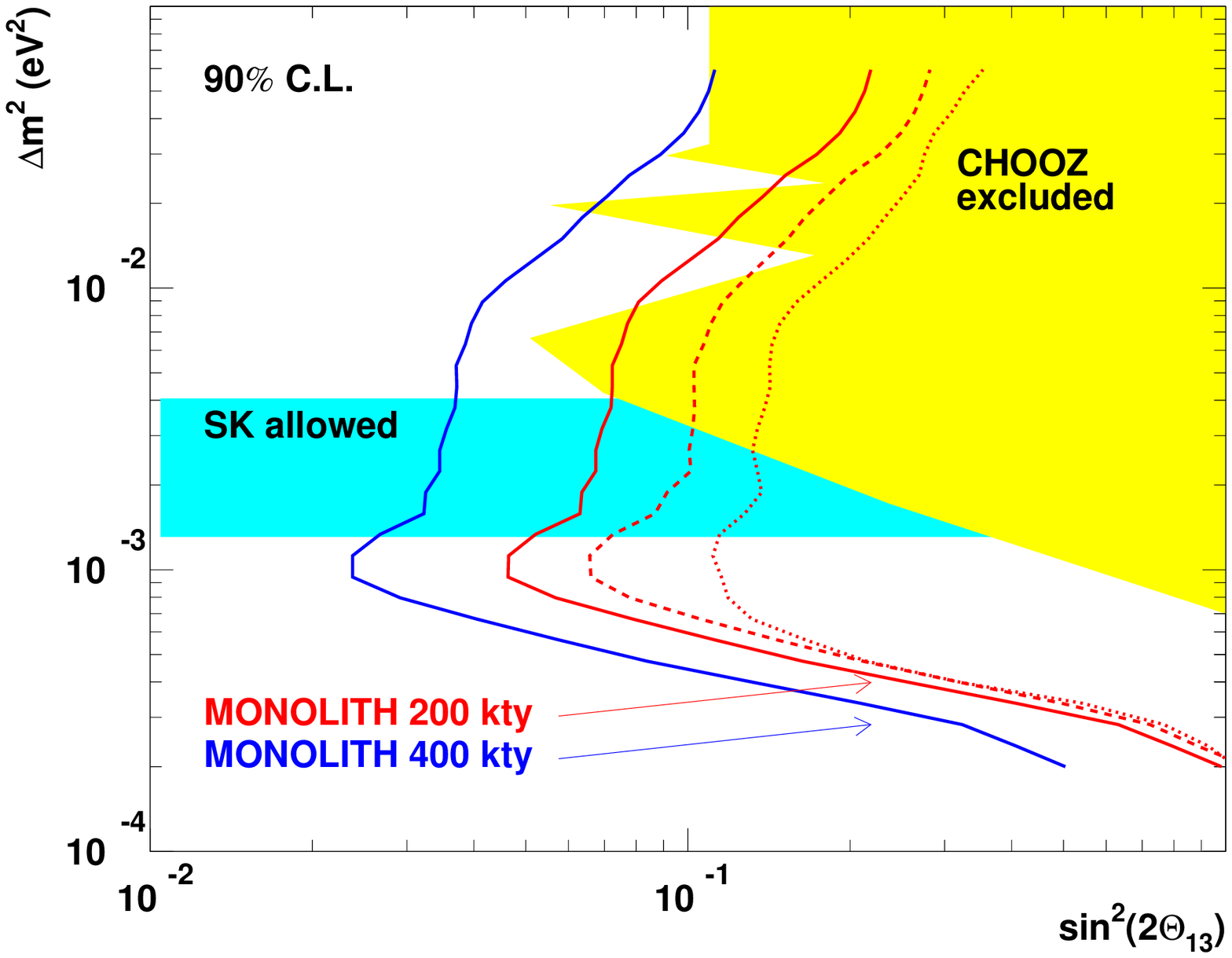}
\end{centering}
\end{minipage}

\caption{\label{mono13} Left: Sensitivity regions at 90\% C.L. in a
  MONOLITH-like iron calorimeter for 200 kton yrs (red lines) and 400
  kton yrs (blue lines), for $\deltaatm > 0$ (solid lines) and
  $\deltaatm < 0$ (dashed lines). The regions excluded by CHOOZ and
  allowed by Super-Kamiokande (at the time this analysis was done),
  together with the expected sensitivity regions of MINOS (Medium
  Energy) and JHF (Low Energy) are also shown. Right: Sensitivity
  regions at 90\% C.L. in a MONOLITH-like iron calorimeter for 200
  kton yr (red solid line) and 400 kton yrs (blue line), for which the
  sign of $\deltaatm$ can be determined, assuming that
  $\sin^2{2\theta_{13}}$ is known with a 30\% accuracy. If no prior
  knowledge of $\sin^2{2\theta_{13}}$ is assumed, the regions over
  which the sign of $\deltaatm$ can be determined, if it is
  positive (red dashed line) or negative (red dashed line), with an
  exposure of 200 kton-yr, are also depicted.
  Figures taken from
  Ref.~\cite{Tabarelli}. 
}
\end{figure}


Having the capability of discriminating the charge of the
neutrino-induced muon, the observable (for magnetized calorimeters)
which is the most sensitive to the value of $\sin^2{\theta_{13}}$ and
the sign of $\deltaatm$ is the charge asymmetry, $A_{\mu^-\mu^+} \equiv
\frac{N(\mu^{-})-N(\mu^+)}{N(\mu^{-})+N(\mu^+)}$ for multi-GeV $\mu^-$
and $\mu^+$ event rates~\cite{PalPet}.  The qualitative behavior of this
observable can be understood from the already explained matter effects for
neutrinos crossing the Earth.  In Fig.~\ref{asymnadir} we show an example
of the nadir angle distribution of $A_{\mu^-\mu^+}$.  From this figure we
see that the distinction of the type of neutrino mass spectrum is in this
case much more clear than in the case of water-Cerenkov detectors,
each type having a different sign with respect to the case of 2-neutrino
oscillations (see Ref.~\cite{PalPet}).

An analysis of the sensitivity for an iron calorimeter like the
proposed MONOLITH~\cite{Tabarelli} is shown in Fig.~\ref{mono13}; 
INO should be able to achieve
similar results.


\label{monolith}

From these considerations and
Figs.~\ref{20SK},~\ref{hier80SK},and ~\ref{mono13}, we
learn that a  
magnetized detector is the preferred experiment to measure 
the sign of $\deltaatm$ and the value of $\sin^2{2\theta_{13}}$ 
with atmospheric neutrinos. 

\subsection{Non-Oscillation Atmospheric Neutrino Studies}

Aside from the dramatic physics discussed in previous sections,
there are many related topics and many intimately intertwined areas
of physics which need to be carried out to facilitate maximum
exploitation of the solar and atmospheric neutrino research program.
Here we briefly review some of the other physics topics which have
been neglected up to this point. We remind the reader of the
fantastic physics of supernova neutrinos studies, which lies in
energy between solar and atmospheric studies and so is related in
technology and backgrounds in physical detectors.  We give an
overview of the thorny issue of precision atmospheric neutrino
calculations, which constrain our ability to use the atmospheric
neutrino beam.  We mention the necessity of updated accelerator
measurements of hadronic and neutrino interactions as input to all
this work, and finally we point out the necessity of ongoing support
for the computer simulation tools which are mainstays of all these
works.

\subsubsection{Other Fundamental Physics}

As discussed in Section~\ref{sec:intro}, atmospheric neutrino detectors
were designed originally to do other physics.  The generality of the
detector technology has allowed searches for a
variety of particles such as magnetic monopoles, quark nuggets,
Q-balls, WIMPs, free quarks, mirror particles, etc.  In addition, we have
studied the cosmic radiation in terms of content, spectrum, particle
clustering in space and time, astronomical uniformity, temporal
variation (correlation with solar cycles, etc.), sensitivity to
atmospheric conditions (season, temperature) and correlations with
events such as gamma ray bursts, solar outbreaks and bursting
activity from distant quasars.  


\subsubsection{Supernova Neutrino Studies}

As we mentioned in Section~\ref{sec:solop} for the planned solar neutrino
experiments., many of the future atmospheric experiments can serve
as detectors of  neutrinos from supernovae.  


The rate of galactic SN Types Ib/c and II (namely gravitational
collapse) events in our galaxy is observationally only once per 250
years from historical records.  Of course only a small fraction of
the galaxy is visible by eye, perhaps 1/6.  The rate from external
galaxies (averaging over types) is about 1/60 years.  We have had
almost totally continuous neutrino monitoring now for about 20
years, but waiting for the next galactic SN requires patience.
A detector like Super-Kamiokande can see of order 100k events in
10 s for a Type II supernova at the galactic center.  Future, 
large-scale atmospheric experiments may be able to see 50 times more,
hopefully allowing us to better understand the still ill-understood
physics of stellar collapse.

	In addition to seeing a burst of neutrinos from a supernova collapse
in `real-time', it may be possible to  observe isolated neutrinos arriving
from distant and dim SN, the so-called ``Supernova relic neutrinos''.
Those arriving from great distances (eg. z$>$1) will have significantly
down-shifted spectra, encroaching upon the solar neutrino energy regime.
Detection of such neutrinos could provide much interesting information on
stellar origins and evolution.	Antineutrino detectors offer the least
background, until one gets down to below about 8 MeV and starts picking up
background from terrestrial reactors. SuperK results come within about a
factor of three of the most optimistic and credible models, and are limited
by cosmic ray neutrino induced backgrounds.  A large-scale water Cerenkov
detector could see these, if it is located deep enough.

We suggest that the recommendation be that all detectors with SN detection
capability should be encouraged, and that they should be encouraged to
cooperate on a world wide basis to enable wide and prompt notification of
the next supernova.  A megaton detector with adequate energy
sensitivity (for single neutrons) could provide coverage of the Milky Way.

\subsubsection{Atmospheric Neutrino Flux}


``Atmospheric neutrinos'' arise from interactions of the incoming
primary cosmic rays with the Earth's atmosphere, which being not too
dense permits many secondaries to decay, resulting in substantial
neutrino fluxes. In fact the dominant neutrino flux on (and below)
the earth's surface between a few tens of MeV and perhaps 10 PeV
arises from this halo of cosmic ray interactions typically 20km high
in the atmosphere.  In the first few GeV energy range, the ratios of
neutrino types are well determined by the decay kinematics of pions
and muons, of both signs.  The ratio of $\pi^+$ to $\pi^-$ is about
5:4.  Hence at such lower energies we have flux ratios known a
priori to around 5\%. 


Calculations of atmospheric neutrino fluxes are generally carried out
starting from the flux of primary cosmic rays, including non-trivial
amounts of nuclei heavier than hydrogen, hitting air nuclei.  Secondary
particles may interact and cascade in the atmosphere, particularly at
higher energies.  Aside from knowing the incoming primary cosmic ray
spectrum and compositions, one needs to know nucleus-nucleus cross-sections
and partial cross-sections for all secondaries.  Much of this is not well
known.  The most egregious deficiency is perhaps in the knowledge of the
pion to kaon ratio in the forward direction. As one goes to higher
energies the flux is dominated at first by pion decays, but beyond around
100 GeV by kaon decays.  

Further complications in the atmospheric neutrino flux calculations are
the earth's magnetic field, which cuts off the incoming primary cosmic
rays (up to a few GeV) depending upon magnetic latitude and which bends
the secondary tracks of particles so that a straight line approximation is
not adequate below a few GeV.  Moreover when one is attempting to do 
better than tens of percent in these calculations, one needs to use a 
realistic atmosphere model, and perhaps even include seasonal effects (a 
warmer atmosphere is less dense and makes more neutrinos).  There are also 
effects due to the state of solar activity, which are at present only 
crudely modeled.

The atmospheric neutrino flux even today remains frustratingly imprecise
in absolute magnitude.  The most recent calculations claim the accuracy
has improved from around 25\%  to $<$10\% below 10 GeV.
Above a TeV it is perhaps uncertain to a factor of two!  Even in the latest
reduction of data from SuperK I (2004), the absolute normalization of the
neutrino flux is allowed to float and they find it to be off (including
oscillations) by 14.4\% (more events seen than expected).  Now one may be
tempted to ignore such nuisances, but this lack of knowledge has effects on
oscillations studies and every other physics analysis for which atmospheric
neutrinos pose a background (eg. proton decay searches and searches for
extraterrestrial neutrinos).

It is worth remarking that while variations in the atmospheric flux model
have not changed the conclusions at all about muon neutrino oscillations
taking place and being nearly maximal, the flux model can indeed move the
derived $\Delta m^2$ around on the order of a factor of two.  There has
seemed to be a discrepancy between earlier and smaller experiments which
inferred larger $\Delta m^2$ values based upon $\mu / e$ ratios, and the
results of SuperK, which is pulled to lower values by shape (energy and
angle) fits.  Thus the influence of atmospheric neutrino flux calculations 
is significant when we start moving to the era of precise oscillation 
parameter measurements. 
Atmospheric neutrino flux measurements are also the limiting factor in searches 
for sub-dominant processes, and any new physics which might be showing up 
at second order.  

The limited precision of the atmospheric flux calculations means that we
remain unable to fully exploit the great dynamic range of the atmospheric
natural beam.  Our recommendation is that strong support be given to
efforts to update hadronic interaction measurements (for which there is
also need by those making neutrino beams) at accelerators (HARP, MIPP,
NA-49), for better cosmic ray studies into the TeV regime, and for further
improvements in flux modeling.

\subsubsection{Neutrino Cross-sections}

As stated above, improved hadronic cross-sections are needed for
atmospheric (and accelerator) neutrino flux calculations.
Interaction cross-section improvements are needed on the other end
as well, in measurements of the neutrino partial cross sections.  A
key problem for atmospheric neutrino studies is predictions of
interactions in the energy range of a few GeV.  This is just the
region where the transition from quasi-elastic to deep inelastic
takes place, and where the theory is murky.  Particularly central is
the production of single $\pi^o$'s.  This is because asymmetric
$\pi^o$ decay mimics a single (CC) electron (positron), which is
background for searches for muon to electron neutrino oscillations.
Some work on this is underway or planned at the various long
baseline experiments (eg. K2K, MINERVA, T2K). 



%

\subsection{Future Atmospheric Experiments}

In this section we will briefly outline the future of
existing and proposed experiments which may have the potential
to carry out the atmospheric neutrino measurements discussed above.

\subsubsection{Future of Super-K}

Super-K's current configuration is Super-K~II with 47\% of inner
detector photomultiplier tubes; in this configuration 
atmospheric neutrino data quality is scarcely diminished from
that of Super-K~I\cite{SKatm04}.
Super-K will be replenished to its full complement of tubes
during 2005/06, and Super-K~III is expected
to collect high quality
atmospheric neutrino data for many years beyond that.
The plots in section~\ref{waterobs} give a sense of
Super-K's long term reach for $\theta_{13}$ and the
sign of $\deltaatm$, via atmospheric neutrinos.

\subsubsection{SNO}

	Although SNO is a relatively small detector, it has two advantages
as an atmospheric neutrino detector.  The first is its depth and overburden: 
at over 6000 m.w.e, it is the deepest operating atmospheric neutrino 
detector.  This depth, and the flat overburden, mean that SNO can measure
the atmospheric neutrino flux using throughgoing muons even above the
detector horizon (up to $\cos \eta = 0.4$) without any contamination from
cosmic ray muons.  These data above the horizon are important, because they
will tell us the unoscillated flux of neutrinos and therefore reduce the
reliance on atmospheric neutrino flux models.  In addition, the charged current
interactions of antineutrinos within SNO should produce additional 
neutrons compared to neutrinos, and it may be possible to make a crude 
measurement of the relative rates ot these.  SNO will continue running
until the end of 2006, at which point it is likely to have over 1000
live-days of data.

\subsection{Next Generation Water Cherenkov Detectors}

Large next generation underground water Chrenkov detectors are proposed in US
(UNO~\cite{NNN}~\cite{uno}), in Japan (Hyper-Kamiokande~\cite{hyperk})
and in Europe. These large megaton class detectors are proposed as
multi-purpose detectors that probe  physics  beyond the  sensitivities
of  the highly successful  Super-Kamiokande detector  utilizing  a  well-
tested technology.  The physics goals of these detectors include: nucleon
decay searches, observation of neutrinos from supernova explosions,
observation of supernova relic neutrinos, and precision measurements of
neutrino oscillation parameters using atmospheric, solar and accelerator
produced neutrinos.

The UNO detector is current proposed to be built in the Henderson mine
located at Empire, Colorado. The optimal depth of the detector is
considered to be about ~4,000 mwe. The Hyper-Kamiokande detector is
proposed to be build in the Tochibora mine, about 8 km away from the
Kamioka mine in the west coast of the main island, Japan. The proposed
detector site has a depth of 1400 ~ 1900 mwe. In Europe, an UNO-like
detector is poposed to be built in the Frejus tunnel, which is located
about 130 km from CERN. The detector can be located as deep as ~4800 mwe.

Because of their sheer size that provides high statistics data and larger
energy containment than Super-K, these detectors have capabilities to
observe multiple oscillation minima and possibly determine mass hierarchy
as described in the previous sections (see Sections~\ref{sec:loverE}
and \ref{sec:masshier}) and to observe $\nu_{\tau}$ appearance in the
atmospheric neutrinos.  For the scale of detector discussed here, we would
expect about 1 tau event per kton-year exposure, yielding about 400 tau
events in one year.

\subsubsection{MINOS}

\hskip 0.6truecm Among the operating detectors, MINOS is an iron
magnetized calorimeter and thus, has muon charge identification
capabilities for multi-GeV muons~\cite{MINOS}.  The MINOS experiment
is currently collecting atmospheric neutrino data.  The detector has
relatively small mass, but after 5 years of data-taking it is expected
to collect about 440 atmospheric $\nu_{\mu}$ and about 260 atmospheric
$\bar{\nu}_{\mu}$ multi-GeV events (having the interaction vertex
inside the detector).  Reference~\cite{minosveto} gives a summary of
physics reach for MINOS from atmospheric neutrinos (in particular what
can be learned from $\nu$ versus $\bar{\nu}$ tagging).

\subsubsection{INO}

There are also plans to build a 30-50 kton magnetized tracking iron
calorimeter detector in India within the India-based Neutrino
Observatory (INO) project~\cite{INO}. The primary goal is to study the
oscillations of atmospheric $\nu_{\mu}$ and $\bar{\nu}_{\mu}$.
This detector is planned to have efficient muon charge identification,
high muon energy resolution ($\sim 5\% $) and muon energy threshold of
about 2 GeV.  It will accumulate sufficiently high statistics of
atmospheric $\nu_{\mu}$ and $\bar{\nu}_{\mu}$ induced events over
several years, which would permit a search for the effects of the
subdominant $\nu_{\mu} \rightarrow \nu_e$ ($\nu_{e} \rightarrow
\nu_{\mu}$) and $\bar{\nu}_{\mu} \rightarrow \bar{\nu}_e$
($\bar{\nu}_{e} \rightarrow \bar{\nu}_{\mu}$) transitions.
The sensitivity to new physics should be comparable to
that evaluated for a similar detector, MONOLITH (section~\ref{monolith}
and Ref.~\cite{Tabarelli}).  A 30\% effect---roughly the size we would
expect for the best values of the mixing parameters---could be observed
by INO with 5 years of running.

We have seen that both water Cerenkov and magnetized detectors
have sensitivity to these effects; however a
magnetized detector is a superior atmospheric neutrino experiment for
measurement of the value of $\sin^2{2\theta_{13}}$ and the sign of
$\deltaatm$,
due to ability to distinguish neutrinos and antineutrinos.

In conclusion, due to their large range of L/E and to
the propagation through matter, $\theta_{13}-$driven resonant effects
can show up in atmospheric neutrino experiments. Therefore, these
experiments have direct access to a fundamental parameter in neutrino
physics: the type of neutrino mass hierarchy.

\section{Facilities}

	We note that for nearly all the experiments discussed here,
underground facilities are a common element.  We therefore recommend the
development of underground space for use in both solar and atmospheric
neutrino experiments.


\pagebreak


\begin{thebibliography}{99}


\bibitem{davis} R. Davis Jr., {\it Solar neutrinos. II. Experimental}, {\it
Phys. Rev. Lett.} {\bf 12} (1964) 303.
\bibitem{IMB1} T.J. Haines {\it et al}., {\it Phys. Rev. Lett.} {\bf 57}, (1986).
\bibitem{IMB2} D. Casper et al. Phys. Rev. Lett. 66, 2561 (1991)
\bibitem{KIIatm} K.S. Hirata {\em et al.}, Physics Letters {\bf B205} 
pp. 416-420, 1988.
\bibitem{KIIsol}  K.S. Hirata {\it et al}., {\it Phys. Rev. Lett.} {\bf 63}, 16 
(1989).
\bibitem{SAGE} V.N. Gavrin, Proc. Int. Conf. on Topics in Astroparticle and Underground Physics, Seattle WA, Sept. 5 -9, 2003.
\bibitem{GALLEX}GALLEX collaboration, W. Hampel et al., {\it GALLEX
solar neutrino observations: results for GALLEX IV}, {\it Phys. Lett.} {\bf B
447} (1999) 127.
\bibitem{sksol}  S. Fukuda {\it et al}., {\it Phys. Lett.} {\bf B539}, 179
(2002); {\it Phys. Rev. Lett.} {\bf 86}, 5651 (2001).
\bibitem{snocces} SNO Collaboration: Q.R. Ahmad {\em et al.}, Phys. Rev. Lett. {\bf 87}, 071301 (2001)  
\bibitem{GNO} E. Bellotti, Proc. Int. Conf. on Topics in Astroparticle and Underground Physics, Seattle WA, Sept. 5 -9, 2003.
\bibitem{snoccnc} SNO Collaboration: Q.R. Ahmad {\em et al.}, Phys. Rev. Lett. {\bf 89}, 011301 (2002)  
\bibitem{snodn} SNO Collaboration: Q.R. Ahmad {\em et al.}, Phys. Rev. Lett. {\bf 89}, 011302 (2002).  
\bibitem{snosalt} SNO Collaboration: S. N. Ahmed {\em et al.} nucl-ex/0309004.
\bibitem{jnbweb} {\tt http://www.sns.ias.edu/~jnb/}
\bibitem{superkdn}  M. B.. Smy {\it et al}., {\it Phys. Rev.} {\bf D69},
011104, (2004). 
\bibitem{bahcall64} J.N. Bahcall, Phys. Rev. Lett. {\bf 12}, pp. 300-302, 1964.
\bibitem{roadmap} J.N. Bahcall and C. Pe\~{n}a-Garay, JHEP {\bf 0311}:004,
2003,({\em hep-ph}/0305159v3).
\bibitem{chlorine} B.T. Cleveland {\em et al.}, Astrophys. J. {\bf 496}, 505 (1998).
\bibitem{sage02}
V.~Gavrin, {\it Results from the Russian American gallium experiment
(SAGE)}, talk at VIIIth International Conference on Topics in
Astroparticle and Underground Physics (TAUP03), Seattle, Sept. 5--9, 2003;
SAGE collaboration, J.N. Abdurashitov et al., {\it J. Exp. Theor. Phys.} {\bf 95} (2002) 181 [astro-ph/0204245].
\bibitem{kamland} KamLAND Collaboration: K. Eguchi {\em et al.} Phys. Rev. Lett {\bf 90} 021802 (2003) [hep-ex/0212021].
\bibitem{jnbhist} J. N. Bahcall {"Solar Models: An Historical Overview"}, {\em
Nucl. Phys. B (Proc. Suppl.)}, {\bf 118}, 77 (2003).
\bibitem{VEP}
M.~Gasperini, Phys.\ Rev.\ D {\bf 38} (1988) 2635;
Phys.\ Rev.\ D {\bf 39}, 3606 (1989);
%
\bibitem{torsion} V.\ De Sabbata and M.\ Gasperini,
Nuovo Cimento A {\bf 65}, 479 (1981).
%
\bibitem{VLI} S.\ Coleman and S.L.\ Glashow,
Phys.\ Lett.\ B {\bf 405}, 249 (1997);
S.L.\ Glashow, A.\ Halprin, P.I.\ Krastev,
C.N.\ Leung, and J.\ Pantaleone,
Phys.\ Rev.\ D {\bf 56}, 2433 (1997).
%
\bibitem{NSI} L. Wolfenstein, Phys. Rev. {\bf D17}, 236 (1978)
%
\bibitem{VLICPT}
D.~Colladay and V.A.~Kostelecky, 
{\it Phys. Rev.} {\bf D55}, 6760 (1997); 
S.\ Coleman and S.L.\ Glashow,
Phys.\ Rev.\ D {\bf 59}, 116008 (1999).
%
\bibitem{solnp}
A.~M.~Gago, M.~M.~Guzzo, P.~C.~de Holanda, H.~Nunokawa,
O.~L.~G.~Peres, V.~Pleitez and R.~Zukanovich Funchal,
Phys.\ Rev.\ D {\bf 65}, 073012 (2002)
%
\bibitem{barger}
V. Barger, D. Marfatia, K. Whisnant, B. Wood,
hep-ph/0204253.

\bibitem{mavans}
David B. Kaplan, Ann E. Nelson, and Neal Weiner, hep-ph/0401099.
%
\bibitem{carlos}
J.~N.~Bahcall and C.~Pena-Garay,
JHEP {\bf 0311}, 004 (2003)
%
\bibitem{orlando}
M.~M.~Guzzo, P.~C.~de Holanda and O.~L.~G.~Peres,
hep-ph/0403134.
\bibitem{smirnovdn} P.C. de Holanda and A.Yu. Smirnov, Astropart. Phys.{\bf 21}, 287-301,
(2004).
\bibitem{uno} {\it Physics Potential and Feasibility of UNO (UNO Whitepaper)},
ed. D. Casper, C.K. Jung, C. McGrew, and C. Yanagisawa, June 2001, SBHEP01-3.
\bibitem{3m} M.V. Diwan {\it et al}, hep-ex/0306053, June 2003.
\bibitem{dnuno} J.N. Bahcall and C. Pe\~{n}a-Garay, New Journal of Physics {\bf
6} (2004), 63.
\bibitem{kamland2} KamLAND Collaboration: T. Araki {\em et al.} [hep-ex/0406035],
2004.
\bibitem{SKatm}
Y.~Fukuda {\it et al.}  [Super-Kamiokande Collaboration],
Phys.\ Rev.\ Lett.\  {\bf 81}, 1562 (1998).
\bibitem{skatm01} T. Toshito, for the Super-Kamiokande Collaboration,
hep-ex/0105023
\bibitem{macro} MACRO Collaboration: M. Ambrosio {\em et al}, Phys. Lett. {\bf B
434}, 451 (1998).
\bibitem{soudan2} Soudan-2 Collaboration: W.W.M. Allison {\em et al}, Phys. Lett. {\bf B 449}, 137 (1999). 
\bibitem{saji} C. Saji, Proc. Conf. NOON04, http://www-sk.icrr.u-tokyo.ac.jp/noon2004/.
\bibitem{alexei}
    O.~L.~Peres and A.~Y.~Smirnov,
    Phys.\ Lett.\ B {\bf 456}, 204 (1999);
    O.~L.~Peres and A.~Y.~Smirnov,
    Nucl.\ Phys.\ Proc.\ Suppl.\  {\bf 110} (2002) 355.
%
\bibitem{ournohier}
M.~C.~Gonzalez-Garcia and M.~Maltoni,
Eur.\ Phys.\ J.\ C {\bf 26}, 417 (2003)
%
\bibitem{thomas}
P.~Huber, M.~Lindner, M.~Rolinec, T.~Schwetz and W.~Winter,
arXiv:hep-ph/0403068.
%
\bibitem{msw}L. Wolfenstein, {\it Neutrino oscillations in matter},
{\it Phys. Rev.} {\bf D 17} (1978) 2369; S.P. Mikheyev and A.Y. Smirnov, {\it
Resonance enhancement of
oscillations in matter and solar neutrino spectroscopy}, {\it Sov. Jour. Nucl.
Phys.} {\bf 42} (1985)
913.
\bibitem{bethe}
H.A.~Bethe, {\it Possible explanation of the solar-neutrino puzzle},
{\it Phys.\ Rev.\ Lett.}\  {\bf 56} (1986) 1305.
\bibitem{Mikheev:ik}
S.P.~Mikheev and A.Y.~Smirnov, {\it Neutrino oscillations in matter with
varying density}, Proc. of the
6th Moriond Workshop on {\it Massive Neutrinos
in Astrophysics and Particle Physics}, eds.
O. Fackler and J. Tran Thanh Van (Editions Fronti\`eres 1986), p. 355.
\bibitem{fogli} G.~L.~Fogli, E.~Lisi, A.~Marrone and G.~Scioscia,
Phys.\ Rev.\ D {\bf 60}, 053006 (1999)
\bibitem{messiah}
A.~Messiah, {\it Treatment of electron-neutrino oscillations in solar matter:
the MSW Effect}, Proc. of the
6th Moriond Workshop on {\it Massive Neutrinos in Astrophysics and Particle
Physics}, eds. O. Fackler and J. Tran Thanh Van (Editions Fronti\`eres
1986), p. 373.
\bibitem{neutrinoastrophysics}J.N. Bahcall, {\it Neutrino Astrophysics},
Cambridge
University Press, Cambridge, 1989.
\bibitem{conchayossi}M.C.~Gonzalez-Garcia and Y.~Nir,
{\it Neutrino masses and mixing: evidence and implications}, {\it Rev.
Mod. Phys.} {\bf 75} (2003) 345 [hep-ph/0202058].
\bibitem{maltoni} M.Maltoni, T. Schwetz, M.A. T\'{o}rtola, and J.W.F. Valle, hep-ph/0309130v2.
\bibitem{atmnp} 
M.C.Gonzalez-Garcia, M. Maltoni hep-ph/0404085. 
\bibitem{concha} M.C. Gonzalez-Garcia and C. Pe\~{n}a-Garay, hep-ph/0111432v2. 
\bibitem{K2K} K2K Collaboration: M.H. Ahn {\em et al.} Phys. Rev. Lett. {\bf 90}, 041801 (2003); hep-ex/0212007.
\bibitem{fogli1} G.L. Fogli, E. Lisi, and A. Marrone, Phys. Rev. {\bf D64}, 093005 (2001); hep-ph/0105139
\bibitem{fogli2} G.L. Fogli, E. Lisi, D. Montanino, and A. Palazzo, Phys. Rev. {\bf D62}, 013002 (2000);hep-ph/9912231
\bibitem{heeger} K.M. Heeger and R.G.H. Robertson, Phys. Rev. Lett. {\bf 77}, 3720 (1996).
\bibitem{hamish}, R.G.H. Robertson, ``Solar Neutrinos and $\theta_{13}$'',
in preparation, 2004.
\bibitem{postkam} J.N. Bahcall, M.C. Gonzalez-Garcia, and C. Pe\~{n}a-Garay,
hep-ph/0406294v1 (2004).
\bibitem{spiro} M. Spiro and D. Vignaud, Phys. Lett. {\bf B 242}, 279 (1990).
\bibitem{jnblum} J.N. Bahcall, Phys. Rev. {\bf C 65}, 025801 (2003). 
\bibitem{BP} J.N. Bahcall and M.H. Pinsonneault, Revs. Mod. Phys. {\bf 67}, 781(1995).
\bibitem{Bahc3} J.N. Bahcall, Phys. Rev. {\bf 135}, B137 (1964); {\em ibid.}Nucl. Phys. B (Proc. Suppl.) {\bf 38}, 98 (1995).
\bibitem{BU} J.N. Bahcall and R.K. Ulrich, Revs. Mod. Phys. {\bf 60}, 297(1988).\bibitem{Bahc5} J.N. Bahcall, Revs. Mod. Phys. {\bf 50}, 881 (1978).
\bibitem{Bahc6} J.N. Bahcall, E. Lisi, D.E. Alburger, L. deBraeckeleer, S.J.Freedman, and J. Napolitano, Phys. Rev. C {\bf 54}, 411 (1996).
\bibitem{Trinder} W. Trinder {\em et al.}, Phys.  Lett. {\bf B349}, 267 (1995).
\bibitem{Hampel} W. Hampel and R. Schlotz, Proc. Intern. Conf. Atomic Massesand Fundamental Constants 7, Darmstadt-Seeheim (TH Darmstadt Lehrdruckerei,1984), p. 89.
\bibitem{Haxton} N. Hata and W.C. Haxton, Phys. Lett. {\bf B353}, 422 (1995).
\bibitem{BP04} J.N. Bahcall and M. H. Pinsonneault; Phys. Rev. Lett.
{\bf 92}(2004)121301.

\bibitem{hyperk} Y. Itow {\it et al}, hep-ex/0106019v1, 2001.

\bibitem{bx} {The Borexino Collaboration}, G. Alimonti et al., {``Science and
technology of Borexino: a real-time detector for low energy solar neutrinos''},
{\em Astrop. Phys.}, {\bf 16}(3), 205 (2002).

\bibitem{laura} L. Cadonati, {``The Borexino Solar Neutrino Experiment and its
Scintillator Containment Vessel''}, Ph.D. Thesis, Princeton University, Jan.
2001.

\bibitem{andrea} A. Pocar, {``Low Background Techniques and Experimental
Challenges for Borexino and its Nylon Vessels''}, Ph.D. Thesis, Princeton
University, Nov. 2003.

\bibitem{ctf} G. Alimonti et al., {``Ultra-low background measurements in a
large volume underground detector''}, {\em Astrop. Phys.}, {\bf 8}(3), 141
(1998).

\bibitem{Allende} C. Allende Prieto, D.L. Lambert, and M. Asplund;
ApJ {\bf 556}(2001)L63.
C. Allende Prieto, D.L. Lambert, and M. Asplund; ApJ {\bf 573}(2002)L137. 
M. Asplund, N. Grevesse, A.J. Sauval, C. Allende Prieto, and D. Kiselman, A\&A,
in press, astro-ph/0312290;

\bibitem{Bah04} J.N. Bahcall, and A.M. Serenelli; astro-ph/0403604.

\bibitem{Ricci} G. Fiorentini and B. Ricci; astro-ph/0310753.

\bibitem{Couv} S. Couvidat,S. Turck-Chie`ze, and A. G. Kosovichev;
bitem{Adel} E.G. Adelberger {\em et al.}; Rev. of Modern Phys.
{\bf 70}(1998)1265.

\bibitem{Adel} E.G. Adelberger {\em et al.}; Rev. of Modern Phys.
{\bf 70}(1998)1265.

\bibitem{Gai} M. Gai; nucl-ex/0312003.

\bibitem{Seatt} A.R. Junghans {\em et al.}; Phys. Rev. Lett {\bf
88}(2002)041101,
  ibid Phys. Rev. {\bf C68}(2003)065803.

\bibitem{Weiz} L.T. Baby {\it et al.}, Phys. Rev. Lett. {\bf 90}(2003)022501,
  ibid Phys. Rev. {\bf C67}(2003)065805.

\bibitem{Iw99} N. Iwasa {\em et al.}; Phys. Rev. Lett. {\bf 83}(1999)2910.

\bibitem{Sch03} F. Schumann {\em et al.}; Phys. Rev. Lett. {\bf 90}(2003)232501.

\bibitem{Ham} R.G.H. Robertson {\em et al.}; Phys. Rev. {\bf C27}(1983)27.

\bibitem{gallagher}Hugh Gallagher, Neutrino 2004 talk

\bibitem{K2Knu2004} T. Nakaya, Neutrino 2004 talk

\bibitem{SKdip04} [Super-Kamiokande Collaboration], hep-ex/0404034

\bibitem{ConchaNOON04} M.C. Gonz\'alez-Garc\'{\i}a,
Talk given at the Int. Workshop on Neutrino Oscillations and their Origin
(NOON2004), February 11 - 15, 2004, Tokyo, Japan.

\bibitem{MINOS} D.~Michael (MINOS Collaboration),
Talk at the Int. Conf. on Neutrino Physics and
Astrophysics ``Neutrino'02'', May 25 - 30, 2002, Munich, Germany.

\bibitem{OPICA03} M. Komatsu, P. Migliozzi and F. Terranova,
J.\ .Phys.\ G {\bf 29} (2003) 443.

\bibitem{MSpironu02} M.~Spiro, Summary talk at the
Int. Conf. on Neutrino Physics and Astrophysics ``Neutrino'02'',
May 25 - 30, 2002, Munich, Germany.

\bibitem{LBL} A.~De R\'{u}jula, M.~B.~Gavela and P.~Hern\'{a}ndez,
  Nucl.\ Phys.\ B {\bf 547}, 21 (1999);
   V.~Barger {\it et al.},
    Phys.\ Rev.\ D {\bf 62}, 013004 (2000).

\bibitem{AMMS99} M.~Freund {\it et al.},
       Nucl.\ Phys.\ B {\bf 578}  (2000) 27.

\bibitem{HLM} 
V.~Barger, D.~Marfatia and K.~Whisnant,
Phys.\ Lett.\ B {\bf 560} (2003) 75;
P.~Huber, M.~Lindner and W.~Winter,
Nucl.\ Phys.\ B {\bf 654} (2003) 3

\bibitem{BPP1}  S.~M.~Bilenky, S.~Pascoli  and S.~T.~Petcov,
Phys.\ Rev.\ D {\bf 64} (2001) 053010;
S.~M.~Bilenky {\it et al.}, Phys.\ Lett.\ B
{\bf 465} (1999) 193.

\bibitem{PPSNO23bb} S.~Pascoli and  S.~T.~Petcov,
      Phys.\ Lett.\ B {\bf 544} (2002) 239, and
      Phys.\ Lett.\ B {\bf 580} (2004) 280;
      S.~Pascoli, S.~T.~Petcov and W.~Rodejohann,
      Phys.\ Lett.\ B {\bf 558} (2003) 141.

\bibitem{kajitanoon04} T.~Kajita,
Talk given at the Int. Workshop on Neutrino
Oscillations and their Origin
(NOON2004), February 11 - 15, 2004, Tokyo,
Japan.

\bibitem{fried} A. Friedland, C. Lunardini, and Michele Maltoni,
Phys.\ Rev.\ D {\bf 70}, p. 111301, (2004).

\bibitem{SP3198}  S.~T.~Petcov,
Phys.\ Lett.\ B {\bf 434} (1998) 321,
(E) {\it ibid.} B {\bf 444} (1998) 584.

\bibitem{AkhDig}
E.~K.~Akhmedov {\it et al.},
Nucl.\ Phys.\ B {\bf 542}, 3 (1999).

\bibitem{core}
J.~Bernab\'{e}u, S.~Palomares-Ruiz, A.~P\'{e}rez and S.~T.~Petcov,
Phys.\ Lett.\ B {\bf 531}, 90 (2002)

\bibitem{3nuSP88} S.~T.~Petcov, Phys.\ Lett.\ B {\bf 214}, 259 (1988).

\bibitem{SPNu98} S.~T.~Petcov,
Nucl.\ Phys.\  B (Proc. Suppl.) {\bf 77} (1999) 93,
hep-ph/9809587, hep-ph/9811205 and hep-ph/9907216;
M.~V.~Chizhov, M.~Maris
and S.~T.~Petcov, hep-ph/9810501.

\bibitem{107} M.~V.~Chizhov and S.~T.~Petcov,
Phys.\ Rev.\ D {\bf 63} (2001) 073003.

\bibitem{mantle}
M.~C.~Ba\~nuls, G.~Barenboim and J.~Bernab\'eu,
Phys.\ Lett.\ B {\bf 513}, 391 (2001);
J.~Bernab\'eu and S.~Palomares-Ruiz,
hep-ph/0112002, and
Nucl.\ Phys.\ Proc.\ Suppl.\  {\bf 110}, 339 (2002),
hep-ph/0201090.
\bibitem{106} M.~V.~Chizhov and S.T.~Petcov,
Phys.\ Rev.\ Lett.\ {\bf 83} (1999) 1096.


\bibitem{Lang87} P. Langacker et al., {\sl Nucl.\ Phys.} {\bf B282}
                  (1987) 589.

\bibitem{BerPalPet03}
J.~Bernab\'eu, S.~Palomares Ruiz and S.~T.~Petcov,
Nucl.\ Phys.\ B {\bf 669}, 255 (2003);
S.~Palomares-Ruiz and J.~Bernab\'eu,
hep-ph/0312038.
\bibitem{PalPetProc} S.~T.~Petcov and S.~Palomares-Ruiz, hep-ph/0406106.

\bibitem{PalPet} S.~Palomares-Ruiz and S.~T.~Petcov,
hep-ph/0406096.

\bibitem{Tabarelli}
T.~Tabarelli de Fatis,
Eur.\ Phys.\ J.\ C {\bf 24} (2002) 43.

\bibitem{SKatm04} E. Kearns, Neutrino 2004 talk

\bibitem{NNN} C.K. Jung, Feasibility of a Next Generation Underground
Water Cherenkov Detector: UNO, (hep-ex/0005046); Next Generation Nucleon
Decay and Neutrino Detector, AIP Conf. Proc. 533, edited by M.V. Diwan and
C.K. Jung (2000).

\bibitem{minosveto} MINOS collaboration document, \\
http://www-numi.fnal.gov/numinotes/public/ps/numi1037/numi1037.ps.gz


\bibitem{INO} See http://www.imsc.res.in/$\sim$ino \\
and working reports and talks therein.




\end{thebibliography}
\end{document}